\documentclass[10pt,a4paper]{article}
 \usepackage{titlesec}

 \newenvironment{dedication}
        {\vspace{3ex}\begin{quotation}\begin{flushright}
        \begin{em}}
        {\par\end{em}\end{flushright}\end{quotation}}

\newcounter{subsubsubsection}[subsubsection]
\def\subsubsubsectionmark#1{}

\def\subsubsubsection{\@startsection
     {subsubsubsection}{4}{\z@} {-3.25ex plus -1
     ex minus -.2ex}{1.5ex plus .2ex}{\normalsize\bf}}
\def\l@subsubsubsection{\@dottedtocline{4}{4.8em}
     {4.2em}}

\usepackage{amsmath}
\numberwithin{equation}{section}

\usepackage{authblk}
 \usepackage[utf8]{inputenc}
 \usepackage{amsmath}
 \usepackage{amsfonts}
 \usepackage{amssymb}
 \usepackage{makeidx}
 \usepackage{graphicx}
 \usepackage{lmodern}
 \usepackage{kpfonts}
\numberwithin{equation}{section}
\usepackage{fullpage}
\usepackage{cite}
\usepackage[T1]{fontenc}
\usepackage{mathtools}
\newcounter{mysubequations}
\usepackage{pict2e}

\makeatletter
\DeclareRobustCommand{\loplus}{\mathbin{\mathpalette\dog@lsemi{+}}}
\DeclareRobustCommand{\lotimes}{\mathbin{\mathpalette\dog@lsemi{\times}}}
\DeclareRobustCommand{\roplus}{\mathbin{\mathpalette\dog@rsemi{+}}}
\DeclareRobustCommand{\rotimes}{\mathbin{\mathpalette\dog@rsemi{\times}}}

\newcommand{\dog@rsemi}[2]{\dog@semi{#1}{#2}{-90,90}}
\newcommand{\dog@lsemi}[2]{\dog@semi{#1}{#2}{270,90}}
\newcommand{\dog@semi}[3]{%
  \begingroup
  \sbox\z@{$\m@th#1#2$}%
  \setlength{\unitlength}{\dimexpr\ht\z@+\dp\z@\relax}%
  \makebox[\wd\z@]{\raisebox{-\dp\z@}{%
    \begin{picture}(1,1)
    \linethickness{\variable@rule{#1}}
    \roundcap
    \put(0.5,0.5){\makebox(0,0){\raisebox{\dp\z@}{$\m@th#1#2$}}}
    \put(0.5,0.5){\arc[#3]{0.5}}
    \end{picture}%
  }}%
  \endgroup
}
\newcommand{\variable@rule}[1]{%
  \fontdimen8  
  \ifx#1\displaystyle\textfont3\else
    \ifx#1\textstyle\textfont3\else
      \ifx#1\scriptstyle\scriptfont3\else
        \scriptscriptfont3\relax
  \fi\fi\fi
}
\makeatother

\newcounter{my-counter} 

\title{Generalized   $ \left\{ h (1) \oplus h(1) \right\} \roplus u(2)  $ commensurate anisotropic Hamiltoninan and  ladder operators; energy spectrum, eigenstates and associated coherent and squeezed states}

 \author{Nibaldo Edmundo Alvarez Moraga\footnote{Email address: nibaldo.alvarez.m@mail.pucv.cl} }
 
  \affil{ Autonomous Center of Theoretical Physics and Applied Mathematics, \\
2805 Place de Darlington, Montreal (Quebec), H3S1L4, Canada}

\begin{document}

  \maketitle

\begin{abstract}
   In this article a study was made of the conditions under which a Hamiltonian which is an element of the complex $ \left\{ h (1) \oplus h(1) \right\} \roplus u(2) $ Lie algebra admits ladder operators which are also elements of this algebra. The  algebra eigenstates of the lowering operator  constructed in this way are computed and from them both the energy spectrum and the energy eigenstates of this Hamiltonian are generated in the usual way with the help of the corresponding raising operator. Thus, several families of generalized Hamiltonian systems are found, which, under a suitable similarity transformation, reduce to a basic set of systems, among which we find the 1:1, 2:1, 1:2, $su(2)$ and some other non-commensurate and commensurate anisotropic  2D quantum oscillator systems. Explicit expressions for the normalized eigenstates of the Hamiltonian and its associated lowering operator are given, which show the classical structure of two-mode separable and non-separable generalized coherent and squeezed states. Finally, based on all the above results, a proposal for new ladder operators for the $p:q$ coprime commensurate anisotropic quantum oscillator is made, which leads us to a class of Chen $SU(2)$ coherent states.
\end{abstract}

\begin{dedication}
\hspace{1.0cm}
\vspace*{0.0cm}{This work is dedicated to my mother, 

Mrs. Norma Moraga Alzamora,

}

\end{dedication}
\section{Introduction}
\setcounter{equation}{0}
  In quantum mechanics, the Hamiltonian operator representing the energy of a system is directly linked to the time evolution of the physical states of this system by the Schr\"odinger\cite{Schro} equation or its mathematical equivalent pictures \cite{Heisenberg}--\cite{Weyl}. This operator is part of the complete set of commuting observables of the system, i.e. a minimal set of commuting Hermitian operators whose common eigenstates, under special conditions, generate the Hilbert space of the physical states of the system. Therefore, it is of crucial importance to determine this set, which in turn is closely related to the group of symmetries of the Hamiltonian or, going further, to the Lie algebras of  symmetry of the system. In the study of a quantum system, the inverse process is also possible, i.e., starting from a Lie  symmetry algebra,  whose realization  of  its generators in a given   representation space is known, we can generate the conditions for the existence of a physical system which is  compatible with this symmetry algebra.  

 The usefulness of beginning the study of a general quantum system by first constructing the dynamical observables or essential operators of that system from the elements of a symmetry algebra has been well demonstrated in the literature. This allows us, among other things, to obtain a general overview of all possible realizations of such a system and to determine the scope and limits that this algebra can offer us in a given context. This technique has been implicitly used by Klauder\cite{Klauder} to redefine the canonical coherent states as eigenstates of the annihilation operator associated to the oscillator algebra, by Barut and Giraldello \cite{Ba-Gi} to define a class of generalized coherent states based on the concept of ladder operators, by C. Brif, \cite{Cbr-96},\cite{Cbr-97}, when introducing the concept of algebra eigenstates and applying it to the Lie algebras $su(1,1)$ and $su(2)$, and by many other authors, see for example \cite{NAM-VH} and \cite{NEAM-2023} and references therein.

    In this article we will deal with the two-boson realization of the semidirect sum   $ \left\{ h (1) \oplus h(1) \right\} \roplus u(2)  $ Lie algebra,  where $u(2) = u(1) \oplus su(2)$ and $h(1)$ is the Heisenberg Lie algebra. Our goal is to take a Hermitian operator as a Hamiltonian, which  is a general element of this algebra and determine the necessary conditions for the existence of associated ladder operators that are also elements of this algebra. Some examples of physical systems that would be part of the set of systems generated in this way would be the system of two interacting oscillators with external linear coupling\cite{SuKu-Mehta}, the   commensurate isotropic  2D quantum oscillator and the $1:2$ and  $2:1$  commensurate anisotropic  2D  quantum oscillator \cite{Louck-Moshinsky-Wolf},\cite{SuKu-Dutta}, \cite{JM-VH}. It is precisely this last system that has caught our attention because we think that its study within a general framework could give us some clues to better understand Cheng's conjecture\cite{Chen-1},\cite{Chen-2} on the structure of the $SU(2)-$type coherent states  associated to the  $p:q$ commensurate anisotropic  2D quantum oscillator, where $p$ and $q$ are mutually prime integers.

 This article is organized as follows: In section \ref{sec-one} we present the generators of the extended algebra $ \left\{ h (1) \oplus h(1) \right\} \roplus u(2), $ along with their commutation relations and their realization in the two-mode boson number representation.  There we also establish the notation that we will use in the rest of the article. In section \ref{sec-two}, we define a generalized  $u(2)$ Hamiltonian and determine the conditions for such a Hamiltonian to possess associated ladder operators which also belong to this algebra.  We classify the various physical systems represented by these  operators in terms of the values of suitably chosen parameters, which give rise to a class of basic $u(2)$ systems.     We then compute the algebra eigenstates of the lowering operator thus obtained,  using the technique of the algebra of operators  together with the technique of partial differential equations, which appears in a  natural way  when we  express the Hamiltonian and the ladder  operators in the Fock-Bargmann representation space of two-dimensional  complex analytic functions. At this level, a rich variety of two-mode eigenstates emerges,  which have the structure of coherent and squeezed separable and non-separable  states. The separable structures arise not only due to the effect of the non-separability quality of the fundamental state of the system, but also due to the presence of a squeezing unitary operator of the type $su(2)$  which is able to mix the states. The  same type of squeezing operator  has been   used, for example, to show the equivalence between the $1:2$ and $2:1$  commensurate anisotropic quantum oscillator systems or to find  the energy eigenstates and the truncated spectrum of pure  $su(2)$ type  two-mode Hamiltonian systems. Then, requiring that the eigenstates of the lowering operator are in turn eigenstates of the Hamiltonian, and appealing to the commutation relations satisfied by the Hamiltonian and its associated  ladder operators, we derive the energy spectrum. Furthermore, by applying the  raising operator on the zero eigenvalue eigenstates of  the lowering operator, we are able to generate  chains of energy eigenstates of the system, which appear automatically expressed in terms of  right linear combinations of the states that form the subspace of degeneracy of a given energy eigenvalue. This procedure, which is slightly different in form but not in substance from that  used in reference \cite{JM-VH-IM}, will  allow us to vizualize more easily  the structure presented by the  algebra eigenstates of the lowering operator, in particular of those states associated to  the zero eigenvalue. In fact, it is in this way, that at the end of this article, more precisely in section \ref{sec-four}, that we will take advantage of this to propose an alternative set of  energy eigenstates for the $p:q$ coprime  commensurate anisotropic  2D  quantum oscillator, which have a similar structure to the Cheng $SU(2)$ coherent states\cite{Chen-1},\cite{Chen-2}.  In section  \ref{sec-three}, we  add  $h(1) \oplus h(1)$ terms to the $u(2)$ Hamiltonian, i.e., we add linear external coupling to our system. In this case, we show  that the Hamiltonian system can be brought into one of the $u(2)$ basic forms  by  appropriate unitary similarity transformations. These unitary transformations  give rise to  the mixing  $su(2)$-type squeezing operator and the well-known  one-mode displacement operator associated with the canonical quantum oscillator.   Finally, in the appendices  \ref{appa} and \ref{appb-0},   we give a list of all possible quantum systems that can be generated from   the requirement that the Hamiltonian together with its associated  ladder operators belong to the algebra   $ \left\{ h (1) \oplus h(1) \right\} \roplus u(2) , $ for suitably chosen values of the parameters.

   \section{Two-boson realization of the $ \left\{ h (1) \oplus h(1) \right\} \roplus u(2)  $ algebra  }
 \label{sec-one}
 The algebra $ \left\{ h (1) \oplus h(1) \right\} \roplus u(2)  $ is a extended Lie algebra constructed from the Heisenberg-Weyl $h(1)$ Lie algebra and the unitary $u(2)$ algebra.  It has the structure of a semi-direct sum between the extended direct sum $h(1) \oplus h(1)$ algebra and $u(2).$ By denoting  $\hat{a}_i, \hat{a}^\dagger_i, \hat{I}_i, \; i=1,2,$ the generators of $h(1) \oplus h(1)$ and  $\hat{N}, \hat{J}_+, \hat{J}_-, \hat{J}_3$ the corresponding generators of  $u(2),$ the characteristic   commutation relations of this extended Lie algebra are given by 
 
$$
[\hat{a}_i , \hat{a}^\dagger_j ] =  \delta_{ij} \hat{I}_i, \quad [\hat{a}_i , \hat{a}_j ] = [\hat{a}^\dagger_i , \hat{a}^\dagger_j]= [\hat{a}_i, \hat{I}_j] =[\hat{a}^\dagger_i, \hat{I}_j]=0 ,\quad i,j=1,2,
$$

$$  [\hat{N}, \hat{J}_{\pm}] =  [\hat{N}, \hat{J}_{3}] = 0, \quad [\hat{J}_3 , \hat{J}_\pm ] = \pm \;  \hat{J}_{\pm}, \quad \text{and} \quad [\hat{J}_+ , \hat{J}_-] = 2 \hat{J}_3 ,$$
  
$$
[\hat{N}, \hat{a}_i] = -  \frac{\hat{a}_i}{2}, \quad [\hat{N}, \hat{a}^\dagger_i] =  \frac{\hat{a}^\dagger_i}{2}, \quad[\hat{J}_3, \hat{a}_i] = (- 1)^i \; \frac{\hat{a}_i}{2},  \quad[\hat{J}_3, \hat{a}^\dagger_i] = (- 1)^{(i-1)} \; \frac{\hat{a}^\dagger_i}{2}, \quad i=1,2,
$$

$$
[\hat{J}_+ , \hat{a}_1 ] = - \hat{a}_2, \quad [\hat{J}_+ , \hat{a}^\dagger_2 ] =  \hat{a}^\dagger_1, 
\quad 
[\hat{J}_- , \hat{a}_2 ] = - \hat{a}_1, \quad [\hat{J}_- , \hat{a}^\dagger_1] =   \hat{a}^\dagger_2,
$$
 and
$$
[\hat{J}_+ , \hat{a}^\dagger_1] = [\hat{J}_+ , \hat{a}_2] = [\hat{J}_- , \hat{a}^\dagger_2] = [\hat{J}_- , \hat{a}_1] = [\hat{J}_{\pm} , \hat{I}_{i}] =0.\quad i=1,2.
$$

In  the two-boson realization of this extended algebra, the representation space is spanned by the set of orthonormal number  states  $\left\{ \mid n_1 , n_2 \rangle \right\}_{n_1=0,n_2=0}^{\infty,\infty}, $ where $\mid n_1 , n_2 \rangle \equiv \mid n_1 \rangle \otimes \mid n_2 \rangle.$ In order to simplify the notation,  it is convenient to redefine the generators of the $h(1) \oplus h(1)$ algebra in the following form  $$\hat{a}_1 \equiv \hat{a}_1 \otimes \hat{I}_2, \quad   \hat{a}^\dagger_1  \equiv \hat{a}^\dagger_1 \otimes \hat{I}_2,  \quad  \hat{a}_2 \equiv  \hat{I}_1  \otimes \hat{a}_2,  \quad  \hat{a}^\dagger_2  \equiv \hat{I}_1 \otimes \hat{a}^\dagger_2 \quad   \text{and} \quad \hat{I} \equiv \hat{I}_1 \otimes \hat{I}_2. $$  Thus, the action of the basic creation and annihilation  generators on the two-mode oscillator number representation states  is given by

\begin{eqnarray}
\hat{a}_1 \mid n_1 , n_2 \rangle &=& \sqrt{n_1} \mid n_1 -1 , n_2 \rangle, \quad  \hat{a}^\dagger_1 \mid n_1   , n_2 \rangle = \sqrt{n_1 + 1} \mid n_1 + 1 , n_2 \rangle   \nonumber \\
  \hat{a}_2 \mid n_1 , n_2 \rangle = \sqrt{n_2} \mid n_1  , n_2 -1\rangle & =& \quad \text{and } \quad   \hat{a}^\dagger_2 \mid n_1    , n_2  \rangle = \sqrt{n_2 +1 } \mid n_1  , n_2 +1\rangle. \nonumber
\end{eqnarray}

On the other hand,   the generator of the $u(1)$ algebra is  represented by  $$\hat{N}= \frac{\hat{a}^\dagger_1  \hat{a}_1   + \hat{a}^\dagger_2  \hat{a}_2}{2}, $$ and those of the $su(2)$ algebra are represented by  two-boson  Schwinger \cite{Schwinger} realization  $$\hat{J}_3 = \frac{\hat{a}^\dagger_1  \hat{a}_1  -  \hat{a}^\dagger_2  \hat{a}_2}{2},  \quad \hat{J}_+ = \hat{a}^\dagger_1    \hat{a}_2 \quad \text{and} \quad  \hat{J}_- = \hat{a}_1   \hat{a}^\dagger_2.$$  
 
We notice that if we define $\hat{F}_i \equiv  F_i (\hat{a}_i, \hat{a}^\dagger_i, \hat{I}_i), \hat{G}_i \equiv  G_i (\hat{a}_i, \hat{a}^\dagger_i, \hat{I}_i), \; i=1,2 $ according with with the general  commutation rule for  bosonic operators 
$$[\hat{A} \hat{B}, \hat{C} \hat{D}] = \hat{A} [\hat{B}, 
\hat{C}  ]\hat{D} + \hat{A} \hat{C} [\hat{B}, \hat{D}] + \hat{C} [\hat{A}, 
\hat{D}  ]\hat{B} +  [\hat{A}, \hat{C}] \hat{D} \hat{B},$$
we have

\begin{equation}
[\hat{F}_1 \otimes \hat{F}_2, \hat{G}_1 \otimes \hat{G}_2 ] = \hat{F}_1  \hat{G}_1  \otimes  [\hat{F}_2, \hat{G}_2, ] +   [\hat{F}_1, \hat{G}_1, ] \otimes \hat{G}_2 \hat{F}_2  ,
\end{equation}
for example
\begin{eqnarray}
[\hat{J}_+ , \hat{J}_-]& =& [\hat{a}^\dagger_1\hat{a}_2,\hat{a} _1  \hat{a}_2^\dagger ] 
\nonumber \\ &=& [\hat{a}^\dagger_1 \otimes \hat{a}_2, \hat{a}_1 \otimes \hat{a}^\dagger_2]  \nonumber \\
&=& \hat{a}^\dagger_1 \hat{a}_1 \otimes [\hat{a}_2 ,\hat{a}_2^\dagger ]  +  [\hat{a}^\dagger_1 ,\hat{a}_1 ] \otimes   \hat{a}^\dagger_2 \hat{a}_2 \nonumber \\ &=&  
  \hat{a}^\dagger_1 \hat{a}_1 \otimes \hat{I}_2 - \hat{I}_1 \otimes \hat{a}^\dagger_2 \hat{a}_2 \nonumber \\ 
  &=&  (\hat{a}^\dagger_1 \otimes \hat{I}_2)  (\hat{a}_1 \otimes \hat{I}_2)  - (\hat{I}_1 \otimes \hat{a}^\dagger_2 ) (\hat{I}_1 \otimes \hat{I}_2) \nonumber \\  &=&
 \hat{a}_1^\dagger  \hat{a}_1 -   \hat{a}_2^\dagger  \hat{a}_2 = 2 \hat{J}_3.
 \end{eqnarray}

\section{Two interacting oscillators without external coupling: \\ 
Basic $u(2)$ Hamiltonian systems}
\setcounter{equation}{0}
\label{sec-two}
Let us start with a general two dimensional Hamiltonian, $\hat{H}_{2D} ,$   which is an  Hermitian element of the algebra generated by the set of  the  $u(2)$ operators  $\{\hat{N}, \hat{\vec{J}}  \} $ and the identity $\hat{I},$ that is,

\begin{equation}
\hat{H}_{2D} =  \beta_{0} \hat{N} + \vec{\beta} \cdot \hat{\vec{J}} =
\beta_{0} \hat{N} +   \beta_{+} \hat{J}_-  +   \beta_{-} \hat{J}_+  + \beta_{3} \hat{J}_3  + h_0 \hat{I},  \label{H-2D}
\end{equation} 
where $h_0, \beta_0 $ and $\beta_3$ are real numbers whereas $\beta_{+} =\beta_{-}^\ast$ is a complex number.  Let us now define another element of algebra  $ \left\{ h (1) \oplus h(1) \right\} \roplus u(2)  $  as follows

\begin{equation}
 \hat{A}_{(\vec{u} , \vec{\alpha} )}= \mu_1 \hat{a}_1 +  \mu_2  \hat{a}_2 + \alpha_- \hat{J}_+  +  \alpha_+  \hat{J}_-  + \alpha_{3} \hat{J}_3, \label{A-lowering-basic}
\end{equation}
 where all coefficients  are complex numbers, and require that the following  commutation  relation be satisfied:
 
 \begin{equation}
 \left[ \hat{H}_{2D} ,  \hat{A}_{(\vec{u} , \vec{\alpha} )} \right]= - \hat{A}_{(\vec{u} , \vec{\alpha} )}, \label{general-commutator-H-2D-A}
 \end{equation}
 which automatically  implies that
 \begin{equation}
 \left[ \hat{H}_{2D} ,  \hat{A}^\dagger_{(\vec{u} , \vec{\alpha} )} \right]= \hat{A}^\dagger_{(\vec{u} , \vec{\alpha} )}. 
 \label{general-commutator-H-2D-A-dagger}
 \end{equation} 
 
 Then equation (\ref{general-commutator-H-2D-A})  imposes the following restrictions on the constants
 
 \begin{equation}
 \begin{pmatrix}
 1- \frac{\beta_0+ \beta_3}{2} & - \beta_+ \\
\beta_-  &  \frac{\beta_0 - \beta_3}{2} -1  
 \end{pmatrix}
\begin{pmatrix}
\mu_1 \\ \mu_2  \end{pmatrix}= \begin{pmatrix}
0 \\ 0 
\end{pmatrix} 
 \quad \text{and } \quad
 \begin{pmatrix}
 2 \beta_-  & - 2 \beta_+ &  1 \\
1- \beta_3 & 0 &   \beta_+ \\
0& 1 + \beta_3 & - \beta_-  
 \end{pmatrix}
\begin{pmatrix}
\alpha_+ \\ \alpha_-  \\ \alpha_3 \end{pmatrix}= \begin{pmatrix}
0 \\ 0  \\0 
\end{pmatrix} . \label{eq-parameters}
\end{equation}
Each one of these  systems has non-trivial solutions if and only if 

\begin{equation}
b^2= 4 \beta_+ \beta_- + \beta_3^2 = (2 - \beta_0)^2 \quad \text{ and} \quad  b^2= 1,  \label{conditions}
\end{equation} 
respectively. 

Thus, if $b \neq 1,$    all parameters $\alpha$ are equal to zero and then for the lowering operator $\hat{A}_{(\vec{u} , \vec{\alpha} )}$ in (\ref{A-lowering-basic}) not to be null we must have $b= |2 -\beta_0|.$   On the other hand,   if  $b=1,$ we have the possibility that $\alpha_+ , \alpha_-$ or $\alpha_3$ be different from $0,$ whereas under the same condition, if we want to $\mu_1$ or $\mu_2$ be different from $0,$ we also need that  $\beta_0 \in \{1,3\}.$   Thus, three classes of lowering operators can arise, that is, one composed by only generator of the  $h(1) \oplus h(1)$ algebra, one composed by only generators of the $su(2)$ algebra and  a mixed one. These different cases  will be analyzed in the next subsections.
\subsection{The case $b^2 \neq 1 :$ Isotropic and  fractional  commensurate anisotropic  2-D oscillator systems}
\label{sec-fractional}
When $b^2 \neq 1,$ then  from (\ref{eq-parameters}) we have  $\vec{\alpha}=\vec{0}$ and to have  $\mu \neq 0$ we need $|2 - \beta_0|= b= \sqrt{4 \beta_+  \beta_- +  \beta_3^2}.$ Therefore, the following situations can occur:

\begin{itemize}
\item When  $\beta_{\pm} =0,$

\begin{equation}
\hat{H}_{2-D} = (2  +  \beta_3) \hat{N}  + \beta_3  \hat{J}_3 = (1 + \beta_3) \hat{a}_1^\dagger \hat{a}_1 + \hat{a}_2^\dagger \hat{a}_2, \label{H-2D-beta-3-plus} \end{equation} and
\begin{equation} \hat{A}_{(\vec{u} , \vec{0} )} = \mu_2 \hat{a}_2  \label{A-2D-beta-3-plus} \end{equation}
where $\beta_3 \neq 0$,

\begin{equation}
\hat{H}_{2-D} = (2  -  \beta_3) \hat{N}  + \beta_3  \hat{J}_3 =  \hat{a}_1^\dagger \hat{a}_1 +   (1-\beta_3) \hat{a}_2^\dagger \hat{a}_2\label{H-2D-beta-3-minus}  \end{equation} 
and 
\begin{equation} \hat{A}_{(\vec{u} , \vec{0} )} =\mu_1 \hat{a}_1  \label{A-2D-beta-3-minus} 
\end{equation}
where  $\beta_3 \neq 0$ and
\begin{equation}
\hat{H}_{2-D} = \hat{a}_1^\dagger \hat{a}_1 +  \hat{a}_2^\dagger \hat{a}_2\label{H-2D-b-zero}
\end{equation}
and
 \begin{equation} 
 \hat{A}_{(\vec{u} , \vec{0} )} = \mu_1 \hat{a}_1  + \mu_2 \hat{a}_2.  \label{A-2D-b-zero}
\end{equation}
when $\beta_3=0.$

\item When $\beta_{\pm} \neq 0,$
\begin{equation}
\hat{H}_{2-D} = \beta_0 \hat{N} +   \vec{\beta} \cdot \vec{\hat{J}} \label{H-2D-beta-0-vec-beta} 
\end{equation}
and
 \begin{equation} 
 \hat{A}_{(\vec{u} , \vec{0} )} = \mu_1  \left(  \hat{a}_1  + \frac{2 \beta_-}{ 2 - \beta_0 + \beta_3} \hat{a}_2\right). \label{A-2D-beta-0-vec-beta} 
\end{equation}
\end{itemize}
Let us notice that the Hamiltonian operators shown in  (\ref{H-2D-beta-3-plus}) and in (\ref{H-2D-beta-3-minus}) represent to a $(1 + \beta_3): 1$ and $1 : (1+ \beta_3)$ 2-D anisotropic oscillator  system, respectively. Furthermore, when $\beta$ is an integer or a  fractional number we are in presence of  commensurate  anisotropic oscillator systems. Moreover, looking at  (\ref{H-2D-b-zero}) and (\ref{A-2D-b-zero}), we observe that when  $b=0,$ i.e., when $\beta_0=2,$ we are in presence of the isotropic  2D  quantum oscillator system.   

\vspace{1.0cm}

Let us now write $\beta_\pm= R e^{i \theta}$ and  define  the  $su(2)$ unitary mixing operator
\begin{equation}
\hat{T}_{(\epsilon, b, \beta_3,\theta)} = \exp\left[ - \arctan \left(  \epsilon \sqrt{\frac{b -\epsilon \beta_3}{b + \epsilon \beta_3}} \right)  
\left( e^{-i \theta} \hat{J}_+ - e^{i \theta}  \hat{J}_- \right)
\right], \quad \epsilon=\pm 1, \label{T-unitary-epsilon-b} 
\end{equation}
where $e^{i \theta}= \sqrt{\frac{\beta_+}{\beta_-}}, $ which disentangled form, among others,  is
\begin{equation}
\hat{T}_{(\epsilon,b, \beta_3,\theta)} = \exp\left[ -  \epsilon \sqrt{\frac{b -\epsilon \beta_3}{b+ \epsilon \beta_3}}  e^{-i \theta} \hat{J}_+ \right]
\exp \left[ \ln\left(   \frac{2b}{b + \epsilon \beta_3}     \right) \hat{J}_{3}  \right]
\exp\left[   \epsilon \sqrt{\frac{b -\epsilon \beta_3}{b+ \epsilon \beta_3}}  e^{i \theta} \hat{J}_- \right], \quad \epsilon=\pm 1,
\end{equation} and whose action on the one mode annihilation operators is given by

\begin{equation}
\hat{T}^\dagger_{(\epsilon,b, \beta_3,\theta)} \hat{a}_1 \hat{T}_{(\epsilon,b, \beta_3,\theta)}= \sqrt{ \frac{b +\epsilon \beta_3}{2b}} \hat{a}_1 - \epsilon e^{-i \theta} \sqrt{ \frac{b - \epsilon \beta_3}{2b}} \hat{a}_2
\end{equation}
and
\begin{equation}
\hat{T}^\dagger_{(\epsilon,b, \beta_3,\theta)} \hat{a}_2 \hat{T}_{(\epsilon,b, \beta_3,\theta)}= \sqrt{ \frac{b +\epsilon \beta_3}{2b}} \hat{a}_2 + \epsilon e^{i \theta} \sqrt{ \frac{b - \epsilon \beta_3}{2b}} \hat{a}_1.
\end{equation}

With the help of this unitary operator we can perform a unitary similarity transformation that brings the Hamiltonian  (\ref{H-2D-beta-0-vec-beta}) and  the lowering operator (\ref{A-2D-beta-0-vec-beta}) to a simpler form, doing this we get

\begin{equation}
\hat{H}^T_{2-D} = \hat{T}^\dagger_{(1,b, \beta_3,\theta)} \hat{a}_1 \hat{T}_{(1,b, \beta_3,\theta)} =    \beta_0 \hat{N} + (2 - \beta_0) \hat{J}_3=
\hat{a}_1^\dagger \hat{a}_1 + (\beta_0  - 1) \hat{a}_2^\dagger \hat{a}_2,  \label{H-T-plus-1-b-beta-0-<-+2} 
\end{equation}
and

\begin{equation}
 \hat{A}^T_{(\vec{u} , \vec{0} )} =   \hat{T}^\dagger_{(1,b, \beta_3,\theta)}           \hat{A}_{(\vec{u} , \vec{0} )}             \hat{T}_{(1,b, \beta_3,\theta)} = \sqrt{\frac{2b}{b+\beta_3}}\mu_1 \hat{a}_1, \label{A-T-plus-1-b-beta-0-<-+2} 
\end{equation} 
when $\epsilon=1$ and $2-b=\beta_0 < 2,$

\begin{equation}
\hat{H}^T_{2-D} = \hat{T}^\dagger_{(1,b, \beta_3,\theta)} \hat{a}_1 \hat{T}_{(1,b, \beta_3,\theta)} =    \beta_0 \hat{N} + (\beta_0 -2) \hat{J}_3=
(\beta_0 -1) \hat{a}_1^\dagger \hat{a}_1 + \hat{a}_2^\dagger \hat{a}_2,  \label{H-T-plus-1-b-beta-0->-+2} 
\end{equation}

and

\begin{equation}
 \hat{A}^T_{(\vec{u} , \vec{0} )} =   \hat{T}^\dagger_{(1,b, \beta_3,\theta)}           \hat{A}_{(\vec{u} , \vec{0} )}             \hat{T}_{(1,b, \beta_3,\theta)} = - e^{-i \theta} \sqrt{\frac{2b}{b -\beta_3}} \mu_2 \hat{a}_2,\label{A-T-plus-1-b-beta-0->-+2}
\end{equation} 
when $\epsilon=1$ and $2+ b = \beta_0 > 2,$

\begin{equation}
\hat{H}^T_{2-D} = \hat{T}^\dagger_{(-1,b, \beta_3,\theta)} \hat{a}_1 \hat{T}_{(-1,b, \beta_3,\theta)} =    \beta_0 \hat{N} -  (2 -\beta_0) \hat{J}_3=
(\beta_0 -1) \hat{a}_1^\dagger \hat{a}_1 + \hat{a}_2^\dagger \hat{a}_2 , \label{H-T-minus-1-b-beta-0-<-+2}
\end{equation}
and

\begin{equation}
 \hat{A}^T_{(\vec{u} , \vec{0} )} =   \hat{T}^\dagger_{(-1,b, \beta_3,\theta)}           \hat{A}_{(\vec{u} , \vec{0} )}             \hat{T}_{(-1,b, \beta_3,\theta)} = e^{-i \theta} \sqrt{\frac{2b}{b +\beta_3}} \mu_2 \hat{a}_2,\label{A-T-minus-1-b-beta-0-<-+2}
\end{equation} 
when $\epsilon=-1$ and $2-b= \beta_0 < 2$ and

\begin{equation}
\hat{H}^T_{2-D} = \hat{T}^\dagger_{(-1,b, \beta_3,\theta)} \hat{a}_1 \hat{T}_{(-1,b, \beta_3,\theta)} =    \beta_0 \hat{N} -  (\beta_0-2) \hat{J}_3=
\hat{a}_1^\dagger \hat{a}_1 + ( \beta_0 -1 ) \hat{a}_2^\dagger \hat{a}_2 ,\label{H-T-minus-1-b-beta-0->-+2}
\end{equation}

and

\begin{equation}
 \hat{A}^T_{(\vec{u} , \vec{0} )} =   \hat{T}^\dagger_{(-1,b, \beta_3,\theta)} \hat{A}_{(\vec{u} , \vec{0} )} \hat{T}_{(-1,b, \beta_3,\theta)} = \sqrt{\frac{2b}{b-\beta_3}} \mu_1 \hat{a}_1, \label{A-T-minus-1-b-beta-0->-+2}
\end{equation}
when $\epsilon=-1$ and $2 +b= \beta_0 > 2.$

We observe that the similarity transformation brings the Hamiltonian (\ref{H-2D-beta-0-vec-beta}) to the basic form  (\ref{H-2D-beta-3-minus}) when $\epsilon =1$ and $ \beta_0 <2$  and when $\epsilon =-1$ and $ \beta_0  >2$ and to the basic form   (\ref{H-2D-beta-3-plus}) when $\epsilon =1$ and $ \beta_0 >2$  and when $\epsilon =-1$ and $ \beta_0  <2.$ The same occurs with the transformed lowering operator (\ref{A-2D-beta-0-vec-beta}) which under the same circumstances we just described above becomes    (\ref{A-2D-beta-3-minus}) and  (\ref{A-2D-beta-3-plus}) respectively.

\subsubsection{Energy spectrum and eigenstates and associated coherent states for a generalized  fractional 2-D anisotropic oscillator}
In this subsection we will compute the energy spectrum and the eigenstates of the Hamintonian  (\ref{H-2D-beta-0-vec-beta}). To accomplish our task, we will first calculate the eigenstates of the associated lowering operator  (\ref{H-2D-beta-0-vec-beta}) and  then the fundamental state of the system. Next, we will apply the  corresponding raising operator to this last states to finally obtain the desired energy eigenstates. According to the general relations (\ref{general-commutator-H-2D-A}) and (\ref{general-commutator-H-2D-A-dagger}), the energy spectrum will be automatically determined once we compute the energy of the fundamental state.

The eigenstates of the lowering operator (\ref{H-2D-beta-0-vec-beta}) verify the eigenvalue equation 
\begin{equation} 
  \mu_1  \left(  \hat{a}_1  + \frac{2 \beta_-}{ 2 - \beta_0 + \beta_3} \hat{a}_2\right) \mid  \psi \rangle = \lambda \mid  \psi \rangle, \label{A-psi-lambda-psi-beta-0}
\end{equation}
where $\lambda \in \mathbb{C},$ and $| \psi \rangle = \sum_{k_1=0}^{\infty}  \sum_{k_2=0}^{\infty} C_{k_1,k_2} |k_1 \rangle \otimes \mid k_2 \rangle,$ where $C_{k_1,k_2}, \; k_1,k_2 =0, \cdots, \infty$ are complex coefficients to be determined. Inserting this state into  
both sides of equation (\ref{A-psi-lambda-psi-beta-0}) and equating to zero each  one of the linear combination of coefficients that multiply  the same linearly independent basis state, we should obtain an algebraic  system of  equations from where we could obtain suitable recurrence relations serving to compute the unknown coefficients and then the eigenstate $\mid \psi \rangle.$  Another way to proceed   would be to use  the Fock-Bargmann \cite{Bargmann} space  of complex  holomorphic  functions where the creation and annihilation operators are represented by
\begin{equation}
\hat{a}_i = \frac{\partial}{ \partial \xi_i}, \quad \hat{a}_i^\dagger = \xi_i, \quad i=1,2,
\end{equation}
and the projection of the state  $\mid  \psi \rangle $ on the  the basis state $\mid \xi_1, \xi_2 \rangle$ is represented by the complex holomorphic function $ \psi(\xi_1 , \xi_2) = \langle \xi_1, \xi_2 \mid \psi\rangle.$ In such a space, the eigenvalue equation (\ref{A-psi-lambda-psi-beta-0}) assumes the form  

\begin{equation} 
  \mu_1  \left(  \frac{\partial}{\partial \xi_1}  + \frac{2 \beta_-}{ 2 - \beta_0 + \beta_3}         \frac{\partial}{\partial \xi_2}     \right)    \psi  (\xi_1, \xi_2)  = \lambda  \psi  (\xi_1,\xi_2), \label{A-psi-lambda-psi-beta-0-differential}
\end{equation}
where we should add a compatible initial condition, for example, $\psi (0, \xi_2) = \varphi_0 (\xi_2),$ where  $\varphi_0 (\xi_2)$ is a one variable complex holomorphic function. In principle $\varphi_0 (\xi_2)$  is an arbitrary function, but it must be compatible with the requirement that the fundamental state of the system, i.e., the eigenstate of the lowering operator  (\ref{A-2D-beta-0-vec-beta})   associated to the eigenvalue $\lambda=0$, is also an eigenstate of the Hamiltonian  (\ref{H-2D-beta-0-vec-beta}). 

A solution of  (\ref{A-psi-lambda-psi-beta-0-differential}) compatible with the given initial condition is 

\begin{equation}
\psi_\lambda  (\xi_1 , \xi_2) = \exp\left[ \frac{\lambda \; \xi_1}{\mu_1}   \right] \varphi_0  \left( \xi_2 - \frac{2 \beta_- \; \xi_1}{2 - \beta_0 +\beta_3}    \right),
 \end{equation}
then the fundamental state of the system is given by
\begin{equation}
\psi_0  (\xi_1 , \xi_2) =  \varphi_0  \left( \xi_2 - \frac{2 \beta_- \; \xi_1}{2 - \beta_0 +\beta_3}    \right). \label{fundamental-0-holomorphic}
 \end{equation}

Now,  as  the representation of the  Hamiltonian  (\ref{H-2D-beta-0-vec-beta}) in the Fock-Bargmann representation reads
\begin{equation}
\hat{H}_{2-D}  = \frac{\beta_0 + \beta_3}{2}  \xi_1 \frac{\partial}{\partial \xi_1}   + \frac{\beta_0 - \beta_3}{2}  \xi_2 \frac{\partial}{\partial \xi_2} + \beta_- \xi_1   \frac{\partial}{\partial \xi_2} + \beta_+ \xi_2   \frac{\partial}{\partial \xi_1} ,
\label{H-2D-beta-0-vec-beta-differential} 
\end{equation} 
then is is direct to show that  its action on the fundamental state (\ref{fundamental-0-holomorphic}) is given by

\begin{equation}
\hat{H}_{2-D} \;  \varphi_0  (x) = (\beta_0 -1)   x  \frac{d \varphi_0 }{dx} (x)
\end{equation}
where
$ x=\left( \xi_2 - \frac{2 \beta_- \; \xi_1}{2 - \beta_0 +\beta_3}    \right). $ Consequently, for the ground state to be also an eigenstate  of $\hat{H}_{2-D}$, the function $\varphi_0 (x)$ must satisfy

\begin{equation}
   x  \frac{d \varphi_0}{dx} (x)  = {\bf \kappa}  \varphi_0 (x),
\end{equation}
then
\begin{equation}
 \varphi_0 (x) =  \mathcal{C} \; x^{\bf \kappa},
\end{equation}
where $\mathcal{C}$ is a normalization constant.  Furthermore, as  $\varphi_0 (x)$ is holomorphic then ${\bf \kappa}$ is a positive integer.

In summary, according to what we have just deduced above, the eigenstates of the lowering operator (\ref{A-2D-beta-0-vec-beta})  are given by

\begin{equation}
\psi_{\lambda} (\xi_1,\xi_2) = \mathcal{C} \; \exp\left[ \frac{\lambda \; \xi_1}{\mu_1}   \right] \left( \xi_2 - \frac{2 \beta_- \; \xi_1}{2 - \beta_0 +\beta_3}    \right)^{\bf \kappa}, \quad {\bf \kappa}  \in \mathbb{N}_+, \label{lambda-states-holomorphic-form}
\end{equation}
the fundamental state is given by 

\begin{equation}
\psi_{0} (\xi_1,\xi_2) = \mathcal{C} \;  \left( \xi_2 - \frac{2 \beta_- \; \xi_1}{2 - \beta_0 +\beta_3}    \right)^{\bf \kappa}, \quad {\bf \kappa}  \in \mathbb{N}_+,  \label{lambda-0-state-holomorphic-form} 
\end{equation}
and energy of the fundamental state is given by ${\bf \kappa} \; (\beta_0 -1), \; \beta_0 \neq 2.$ On the other hand,  according to the commutation relation (\ref{general-commutator-H-2D-A-dagger}), we deduce that the energy spectrum is given by 
\begin{equation}
E^{({\bf \kappa})}_n  = {\bf \kappa}  (\beta_0 - 1) + n, \quad n=0,1,2,\ldots,
\end{equation}
which when $\beta_0 = k_o  \in \mathbb{Z},$  $k_0 \notin \{1,2,3\} ,$ becomes $ E^{(k_0,{\bf \kappa})}_n= {\bf \kappa}  (k_0  - 1) +  n,$  the energy spectrum of a family   of generalized $(k_0 -1):1$ or $1: (k_0-1)$ commensurate  2-D anisotropic quantum oscillator. Moreover, by choosing $\beta_0 =\frac{p}{q},$ where $p$ and $q$ are integers whose decomposition into multiplicative factors has no common factors,  we get fractional $(p-q):q$ or $q : (p-q)$ 2-D commensurate anisotropic oscillator systems.  

\vspace{1.0cm} Let us now return to the number space representation. In this representation the states (\ref{lambda-states-holomorphic-form}) have the form

\begin{equation}
\mid \lambda \rangle = \mathcal{C}_\lambda  \exp\left[ \frac{\lambda \; \hat{a}_1^\dagger}{\mu_1}   \right]  \left( \hat{a}_2^\dagger - \frac{2 \beta_- \; \hat{a}_1^\dagger}{2 - \beta_0 +\beta_3}    \right)^{\bf \kappa} \mid 0, 0 \rangle,  \quad {\bf \kappa}  \in \mathbb{N}_+, \label{lambda-eigenstates-beta-0-beta-3}
\end{equation}
  where $\mid 0, 0 \rangle$ is the two-mode quantum oscillator ground state. Thus, the  normalized ground state of the generalized system takes the form
  \begin{equation}
\mid 0 \rangle =  \frac{\sum_{k=0}^{\kappa}  \sqrt{\binom{\kappa}{k}}   (-1)^k \left(\frac{2 \beta_-}{2-\beta_0 + \beta_3} \right)^k  \; \mid k , \kappa - k \rangle          }{\left(1 + \left\|\frac{2 \beta_-}{2-\beta_0 + \beta_3}\right\|^2\right)^{\frac{\kappa}{2}} },  \quad {\bf \kappa}  \in \mathbb{N}_+. 
\end{equation}

 As for the normalization of the states (\ref{lambda-eigenstates-beta-0-beta-3}), since we are also interested in knowing their global structure, instead of calculating the normalization constants directly from that  expression, we will use the aforementioned similarity transformations for this purpose. Thus, starting with the transformed system (\ref{H-T-plus-1-b-beta-0-<-+2}--\ref{H-T-plus-1-b-beta-0-<-+2}), when $\beta_0 <2 $ and $\beta_0 \neq 1,$ and proceeding with calculations exactly as we have done in the previous paragraphs, we get the following eigenstates of the transformed lowering operator  (\ref{H-T-plus-1-b-beta-0-<-+2}):
 
 \begin{equation}
 \mid \tilde{\lambda} \rangle=  \mathcal{C}_{\tilde{\lambda}} \; \exp\left[ \sqrt{\frac{ b +\beta_3}{2b}} \frac{\lambda}{\mu} \hat{a}_1^\dagger \right]  (\hat{a}_2^\dagger)^{{\bf \kappa}} \mid 0, 0 \rangle, \quad {\bf \kappa}=1,2, \ldots,
 \end{equation}
 where $b= 2 - \beta_0,$ or their normalized version
 
 \begin{equation}
 \mid \tilde{\lambda} \rangle =  \hat{D}_1 \left(  \sqrt{\frac{ b +\beta_3}{2b}} \frac{\lambda}{\mu}  \right) \mid 0 , {\bf \kappa} \rangle, \quad {\bf \kappa}=1,2,\ldots,
 \end{equation}
  where $b= 2 - \beta_0,$ and $\hat{D}_1 (\alpha)$ is the canonical  one-mode quantum oscillator displacement operator.  Finally,  acting on these states  with the unitary operator  $\hat{T}_{(1, b, \beta_3,\theta)}$ we get the normalized eigentates of the lowering operator (\ref{A-2D-beta-0-vec-beta}), i.e.,
  
  \begin{equation}
\mid \lambda \rangle = \hat{T}_{(1, b, \beta_3,\theta)} \hat{D}_1 \left(  \sqrt{\frac{ b +\beta_3}{2b}} \frac{\lambda}{\mu}  \right) \mid 0 , {\bf \kappa} \rangle, \quad {\bf \kappa}=1,2,\ldots,
\end{equation}
 where $b= 2- \beta_0,$ $ \beta_0 <2 $ and $\beta_0  \neq 1.$
 
 We observe that these states have the structure of   two-mode non-separable squeezed states represented by the action of a   two-mode mixing unitary operator  acting on  a  two mode-state which has the structure  of a direct product of  a one-mode coherent state with a pure ${\bf \kappa}$ particle state. 

When $b= \beta_0 -2,$ $  \beta_0 >2$ and $\beta_0 \neq 3,$ we can use the transformed  lowering operator (\ref{A-T-minus-1-b-beta-0->-+2}) to compute the transformed eigenstates $\mid \tilde{\lambda} \rangle$ and then act on it with  unitary operator  $\hat{T}_{(-1, b, \beta_3,\theta)}$ to obtain normalized eigentates of the lowering operator (\ref{A-2D-beta-0-vec-beta}), i.e.,
 \begin{equation}
\mid \lambda \rangle = \hat{T}_{(-1, b, \beta_3,\theta)} \hat{D}_1 \left(  \sqrt{\frac{ b -\beta_3}{2b}} \frac{\lambda}{\mu}  \right) \mid 0 , {\bf \kappa} \rangle, \quad {\bf \kappa}=1,2,\ldots.
\end{equation}

\vspace{1.0cm} There are still other solutions for the differential equation (\ref{A-psi-lambda-psi-beta-0-differential}). For example, if we use the separation of variables method, i.e., if we write $\psi (\xi_1,\xi_2) =\varphi_1 (\xi_1) \varphi_2 (\xi_2),$ we find that a solution, which is compatible with the requirement that the fundamental state be also an eigenstate of the Hamiltonian, is given by

\begin{equation}
\psi_\lambda (\xi_1,\xi_2) = \varphi_1 (0) \varphi_2 (0) \exp\left[\frac{c_1 \lambda}{\mu_1} \xi_1 \right]   \exp\left[\frac{\lambda}{\mu_1} \left(c_1 -1\right) \frac{2 -\beta_0 +\beta_3}{2 \beta_-} \xi_2 \right],
\end{equation} 
  which in the number  representation has the form of a two-mode separable coherent state: 
  
  \begin{equation}
  \mid \lambda \rangle =   \hat{D}_1 \left(\frac{c_1 \lambda}{\mu_1}  \right)   \hat{D}_2 \left( \frac{\lambda}{\mu_1} \left(c_1 -1\right) \frac{2 -\beta_0 +\beta_3}{2 \beta_-}  \right) \; \mid 0,0\rangle,
\end{equation} 
where $c_1$ is an arbitrary constant and $\hat{D}_i (\alpha), i=1,2$ are the standard quantum oscillator displacement operators in the mode one and two, respectively. 

We observe that in this case the fundamental state $\mid 0 \rangle$ coincides with the two mode vacuum $\mid 0, 0 \rangle.$ Therefore, according   \ref{H-2D-beta-0-vec-beta},  $\hat{H}_{2-D} \mid 0,0 \rangle =0, $ then   from the generalized commutation relation (\ref{general-commutator-H-2D-A-dagger}) we deduce that, in this  particular case,  the energy spectrum of $\hat{H}_{2-D} $ is identical to  to that of the canonical one-dimensional  quantum harmonic oscillator. The energy eigenstates are given by

\begin{equation}
\mid E_n \rangle = \mathcal{N}_n^{-\frac{1}{2}}
\; ( \hat{A}^\dagger_{(\vec{u} , \vec{0} )} )^n \mid  0, 0 \rangle =    \mathcal{N}_n^{-\frac{1}{2}} \; (\mu_1^\ast)^n  \left(  \hat{a}_1^\dagger  + \frac{2 \beta_+}{ 2 - \beta_0 + \beta_3} \hat{a}_2^\dagger \right)^n \mid 0,0 \rangle. \label{A-dagger-n-2D-beta-0-vec-beta} 
 \end{equation}
By developing the binomial operator, we get
\begin{equation}
\mid E_n \rangle =  \mathcal{N}_n^{-\frac{1}{2}} \sum_{k=0}^{n}   \sqrt{ \frac{n!}{(n-k)! k!}} \left(\frac{2 \beta_+}{ 2 - \beta_0 + \beta_3}\right)^{n-k} \; \mid k, n-k \rangle, \quad n=0,1,2, \ldots,
\end{equation}
where some terms have been absorbed in the normalization constant, which finally is given by
\begin{equation}
\mathcal{N}_n = \left[  1 +    \left\|   \frac{2 \beta_+}{ 2 - \beta_0 + \beta_3}  \right\|^2 \right]^n.
 \end{equation}
\subsubsection{Case $b=0,$ Energy spectrum and eigenstates and associated coherent states for the isotropic 2D oscillator}
 Let us here treat briefly the case when $\vec{\beta}=0.$ In this case, as we have already computed, the generalized Hamiltonian   
 (\ref{H-2D}) becomes the 2D isotropic Hamiltonian  (\ref{H-2D-b-zero}), and the generalized associated annihilation operator (\ref{A-lowering-basic})  becomes  (\ref{A-2D-b-zero}). The eigenvalue equation of this last operator in the Fock-Bargmann analytic representation is written as follows
 
 \begin{equation}
 \mu_1 \frac{\partial \psi}{\partial \xi_1} (\xi_1,\xi_2) + \mu_2 \frac{\partial \psi}{\partial \xi_2} (\xi_1,\xi_2) = \lambda \psi (\xi_1,\xi_2),
 \end{equation}
 $\lambda \in \mathbb{C}.$
 
The general solution of this partial differential equation is given by

\begin{equation}
 \psi_{\lambda} (\xi_1,\xi_2) = \exp[ \frac{\lambda}{\mu_1} \xi_1 ] \varphi_2   \left( \frac{\mu_1 \xi_2 - \mu_2 \xi_1}{\mu_1} \right) ,
\end{equation}
where $\varphi_2 (z)$ is an arbitrary analytic  function of the  complex variable  $z.$ The condition that the fundamental state $  \psi_{0} (\xi_1,\xi_2)$  is  also an eigenstate of the   isotropic 2D Hamiltonian  (\ref{H-2D-b-zero}) leads us to choose 

\begin{equation}
 \varphi_2 \left(\frac{\mu_1 \xi_2 - \mu_2 \xi_1}{\mu_1} \right)  = \varphi_2 (0)
 \exp[ \lambda c_2 \;  \left(\frac{\mu_1 \xi_2 - \mu_2 \xi_1}{\mu_1} \right)],
\end{equation}
where $c_2 $ is an arbitrary complex constant,  or
\begin{equation}
 \varphi_2   \left( \frac{\mu_1 \xi_2 - \mu_2 \xi_1}{\mu_1} \right) = \varphi_2 (0)       \varphi_2   \left( \frac{\mu_1 \xi_2 - \mu_2 \xi_1}{\mu_1} \right)^{{\bf \kappa}}, \quad {\bf \kappa} \in \mathbb{N}_+.
\end{equation}
The first of these solutions leads to the separable coherent states

\begin{equation}
\psi_{\lambda} (\xi_1, \xi_2) = \varphi_{2} (0) \exp\left[  \frac{\lambda}{\mu_1} ( 1 - c_2 \mu_2) \xi_1\right] \exp\left[  \lambda  c_2 \xi_2  \right],
\end{equation}
while   the second leads to  the non-separable coherent states

\begin{equation}
\psi_{\lambda} (\xi_1, \xi_2) = \varphi_{2} (0) \exp\left[  \frac{\lambda}{\mu_1} \xi_1 \right]  \left( \frac{\mu_1 \xi_2 - \mu_2 \xi_1}{\mu_1} \right)^{{\bf \kappa}}, \quad {\bf \kappa} \in \mathbb{N}_+.
\end{equation}
The normalized version of these states in the number representation are given by, 

\begin{equation}
\mid \lambda \rangle_1 = \hat{D}_1 \left(   \frac{\lambda}{\mu_1} (1 - c_2 \mu_2)  \right) \hat{D}_2   \left( \lambda  c_2 \right) \mid 0,0 \rangle \label{Iso-CS-1}
\end{equation}
and

\begin{equation}
\mid \lambda \rangle_2 = \hat{D}_1\left(   \frac{\lambda}{\mu_1} \right)  \frac{\sum_{k=0}^{{\bf \kappa}}  (-1)^k  \sqrt{\binom{{\bf \kappa}}{k}} \;  \mu_1^{\kappa - k} \;   \mu_2^k \; \frac{1}{\sqrt{k!}} \left( \hat{a_1}^\dagger  + \frac{\lambda^\ast}{\mu_1^\ast} \right)^k \mid 0, {\bf \kappa}-k\rangle }{ \sqrt{  \sum_{k=0}^{{\bf \kappa}} \binom{{\bf \kappa}}{k}  \|\mu_1\|^{2(\kappa-k)}  \|\mu_2\|^{2k}  \sum_{\ell=0}^{k}   \binom{k}{\ell}    \;  \frac{1}{ \ell !} \; \|\frac{\lambda}{\mu_1}\|^{(2 \ell)} }}, \quad \kappa=1,2, \cdots.\label{iso-su2-states}
\end{equation}
 Hence, the fundamental state in each case is given by
 \begin{equation}
 \mid 0  \rangle_1 = \mid 0,0 \rangle, \quad \text{and} \quad  \mid 0  \rangle_2 =  \frac{\sum_{k=0}^{{\bf \kappa}}      (-1)^k  \sqrt{\binom{{\bf \kappa}}{k}}   \;    \mu_1^{\kappa -k} \;  \mu_2^k
   \mid  k, {\bf \kappa}-k\rangle }{ \left( \|\mu_1\|^2 + \|\mu_2\|^2 \right)^{\frac{\kappa}{2}}}, \quad \kappa=1,2, \cdots,  \label{ground-0-1}
 \end{equation}
respectively.

From equation (\ref{Iso-CS-1}), we observe that  when $c_2 = \mu_2^\ast,$ and $\|\mu_1\|^2 + \|\mu_2\|^2 =1,$ we regain the Schr\"odinger coherent states associated to the  isotropic 2D quantum oscillator\cite{JM-VH},\cite{JM-VH-0}. Also,  from ,(\ref{ground-0-1}) we observe that union of the two sets of states  shown there represents the $su(2)$ type coherent states.  They are also eigenstates  of the isotropic 2D quantum Hamiltonian with energy spectrum $E_{\kappa}=\kappa, \; \kappa=0,1,2,\ldots.$ 

\vspace{1.0cm}
On the other hand, the  eigenvalue of the isotropic Hamiltonian  (\ref{H-2D-b-zero}) in these fundamental states is given by $E^{(1)}_0=0$ and $E^{(2)}_0 = \kappa, \; k=1,2, \cdots.$ Thus, according to the general  commutation  relations  (\ref{general-commutator-H-2D-A}  and (\ref{general-commutator-H-2D-A-dagger}), which certainty  hold for the particular case we are dealing with,  the energy spectrum in each of these chains is  given by $E^{(1)}_n= n,  \; n=0, 1,2, \cdots$ and  $E^{(2)}_n= \kappa + n, \; n=0,1, \cdots, \; \kappa=1,2, \cdots, $ respectively.  

\vspace{1.0cm}

Finally, by using  (\ref{general-commutator-H-2D-A-dagger})  adapted to the particular expression of the Hamiltonian (\ref{H-2D-b-zero})  and of the adjoint of the operator   (\ref{A-2D-b-zero}), we get set of energy eigenstates  
\begin{equation}
\mid E_n \rangle_1 = \mathcal{N}_n^{-\frac{1}{2}}
\; ( \hat{A}^\dagger_{(\vec{u} , \vec{0} )} )^n \mid  0, 0 \rangle =\mathcal{N}_n^{-\frac{1}{2}}  \left( \mu_1^\ast  \hat{a}_1^\dagger +  \mu_2^\ast \hat{a}_2^\dagger \right)^n \mid 0,0 \rangle, \quad n=0,1,\ldots,
 \end{equation}
 where $\mathcal{N}_n, \; n=0,1, \ldots,$ is a normalization constant. By developing the binomial operator, we get
\begin{equation}
\mid E_n \rangle_1 =  \mathcal{N}_n^{-\frac{1}{2}} \sum_{k=0}^{n}   \sqrt{ \frac{n!}{(n-k)! k!}}  {\mu_1^\ast}^k  {\mu_2^\ast}^{n-k} \; \mid k, n-k \rangle,
\end{equation}
where  
\begin{equation}
\mathcal{N}_n = \left[  \| \mu_1 \|^2  +    \left\| \mu_2 \right\|^2 \right]^n.
 \end{equation}
In the same way,  from (\ref{ground-0-1}), we get 

\begin{equation}
\mid E_n  \rangle_2 =  \mathcal{N}_n^{-\frac{1}{2}} 
 \sum_{k=0}^{{\bf \kappa}} \sum_{\ell=0}^{n}      (-1)^k  \sqrt{\binom{{\bf \kappa}}{k}} \;   \binom{n}{\ell} \;  \sqrt{ \frac{ (n+k-\ell)! (\kappa - k+\ell)!  }{(k)! (\kappa-k)!} }   \;  \mu_1^{\kappa -k} \;  \mu_2^k \;
  {\mu_1^\ast}^{n-\ell} \; {\mu_2^\ast}^{\ell}    \mid n+ k -\ell, {\bf \kappa}-k + \ell\rangle,
 \end{equation}
where $\kappa=1,2, \cdots$ and  the normalization constant $ \mathcal{N}_n$  should  be computed in the usual way.
\subsection{The case $b^2 =1$ and $\beta_0 \notin \{1,3\}:$ $su(2)$ Ladder operators }
\label{ladder-operators}
In this case, from (\ref{eq-parameters}) we have  $\mu_1=\mu_2 =0,$  and then, in general, 

\begin{equation}
\hat{H}_{2D} = \beta_0 \hat{N} + \vec{\beta} \cdot \vec{\hat{J}} + h_0 \hat{I} \label{H-2D-mu-0-beta-0-not-1-3}
\end{equation}
and
\begin{equation}
\hat{A}_{(\vec{0} , \vec{\alpha} )}
= \vec{\alpha} \cdot \vec{\hat{J}},
\end{equation}
which becomes
\begin{equation}
\hat{A}_{(\vec{0} , \vec{\alpha} )}
= \begin{cases}    \alpha_+  \hat{J}_-, \quad \text{when} \quad     \beta_\pm=0 \quad \text{and} \quad  \beta_3 =1 \\    
 \alpha_-  \hat{J}_+,  \quad \text{when} \quad     \beta_\pm=0 \quad \text{and} \quad  \beta_3 = - 1 \\
\alpha_3  \left(  \hat{J}_3  +\frac{\beta_-}{1 +\beta_3}  \hat{J}_+ - \frac{\beta_+}{1 -\beta_3}  \hat{J}_- \right), \quad \text{when} \quad \beta_\pm \neq 0. \label{A-mu-0-su2} 
 \end{cases}.
\end{equation}
 
We notice that when $\beta_\pm=0,$ we can write  $\hat{H}_{2D} = \beta_0 \hat{N} +  \hat{\mathbb{J}}_3 +  h_0 \hat{I},$ 
$ \hat{A}_{(\vec{0} , \alpha_\pm )} =  \| \alpha_\pm \| \;  \hat{\mathbb{J}}_- ,$   $ \hat{A}^\dagger_{(\vec{0} , \alpha_{\pm} )} = \| \alpha_\pm \| \;   \hat{\mathbb{J}}_+ , $ where $ \hat{\mathbb{J}}_3 = \pm \; \hat{J}_3 $ and $  \hat{\mathbb{J}}_{-} =  \frac{\alpha_{\pm}}{\| \alpha_\pm\|} \;  \hat{J}_{\mp},$ $  \hat{\mathbb{J}}_{+} =  \frac{\alpha^\ast_{\pm}}{\| \alpha_\pm\|}  \; \hat{J}_{\mp}$ are standard $su(2)$ generators. Moreover, in the general case $\beta_\pm \neq 0,$ the same is true for the generalized Hamiltonian but now with $\hat{\mathbb{J}}_3 =  \vec{\beta} \cdot \vec{\hat{J}},$ whereas  the ladder operators are given by  $ \hat{A}_{(\vec{0} , \vec{\alpha} )} = \frac{\alpha_3}{ \sqrt{1-\beta_3^2}} \hat{\mathbb{J}}_-  $ and $ \hat{A}^\dagger_{(\vec{0} , \vec{\alpha} )} = \frac{\alpha_3^\ast}{ \sqrt{1-\beta_3^2}} \hat{\mathbb{J}}_+,  $
where

\begin{equation}
\hat{\mathbb{J}}_- =   \frac{\alpha_3}{ \| \alpha_3\|} \sqrt{1 -\beta_3^2}  \left(  \hat{J}_3  +\frac{\beta_-}{1 +\beta_3}  \hat{J}_+ - \frac{\beta_+}{1 -\beta_3}  \hat{J}_- \right) 
\end{equation}
 and
 \begin{equation}
\hat{\mathbb{J}}_+ =   \frac{\alpha_3^\ast}{ \| \alpha_3\|} \sqrt{1 -\beta_3^2}  \left(  \hat{J}_3  +\frac{\beta_+}{1 +\beta_3}  \hat{J}_- \frac{\beta_-}{1 -\beta_3}  \hat{J}_+ \right). 
\end{equation}
 
\subsubsection{Energy eigenstates and truncate spectrum: $su(2)$ Hamiltonian system}

Let us now  consider the general case $\beta_\pm \neq 0.$ In this case, the eigenvalue equation associated to the lowering operator
$\hat{A}_{(\vec{0} , \vec{\alpha} )}$ in (\ref{A-mu-0-su2}) is given by 

\begin{equation}
\alpha_3  \left(  \hat{J}_3  +\frac{\beta_-}{1 +\beta_3}  \hat{J}_+ - \frac{\beta_+}{1 -\beta_3}  \hat{J}_- \right) \mid \psi \rangle= 
\lambda \mid \psi\rangle.
\end{equation}

By making $\beta_{\pm} = R e^{\pm i \theta}$ and  performing  the unitary transformation on the states  $ \mid \psi \rangle = \hat{T}_{(\epsilon, 1, \beta_3,\theta)} \mid {\tilde \psi} \rangle $ where $\hat{T}_{(\epsilon, 1, \beta_3,\theta)}$ is given by (\ref{T-unitary-epsilon-b})  with $b=1,$ and then  acting on the left  of  both sides of this equation  with the operator $\hat{T}^\dagger_{(\epsilon, 1, \beta_3,\theta)},$ we get 
\begin{equation}
 \alpha_3 \left[
    R e^{- i \theta}  \left[ -  \epsilon  +   \frac{1}{2} \frac{1+\epsilon \beta_3}{1 + \beta_3}    +         \frac{1}{2} \frac{1 -\epsilon \beta_3}{1 - \beta_3}   \right] \hat{J}_+  \nonumber  -   R e^{i \theta} \left[  \epsilon + \frac{1}{2} \frac{1-\epsilon \beta_3}{1 + \beta_3} +             \frac{1}{2} \frac{1+\epsilon \beta_3}{1 - \beta_3}   \right] \hat{J}_-   \right] \mid \tilde{\psi} \rangle
= \lambda \mid \tilde{\psi} \rangle, \quad \epsilon = \pm 1.  
\end{equation}
On the other hand,  by performing a similarity transformation of  the generalized Hamiltonian (\ref{H-2D-mu-0-beta-0-not-1-3}), using the same unitary operator,  we get 

\begin{equation}
{\hat{H}}^{T_{(\epsilon)}}_{2D}  = \hat{T}^\dagger_{(\epsilon, 1, \beta_3,\theta)}  \hat{H}_{2D} \hat{T}_{(\epsilon, 1,  \beta_3,\theta)} = \beta_0 \hat{N} + \epsilon \hat{J}_3+ h_0 \hat{I},  \quad \epsilon= \pm 1.
\end{equation}
  
  Thus, when $\epsilon =1,$ the generalized Hamiltonian reduce to 
  
  \begin{equation}
  {\hat{H}}^{T_{(1)}}_{2D}  =  \beta_0 \hat{N} +  \hat{J}_3 + h_0 \hat{I} \label{H-2D-e=1}
\end{equation}
and the eigenvalue equation for the lowering operator to
\begin{equation}
  - \frac{\alpha_3}{\sqrt{1-\beta_3^2}} e^{i \theta} \hat{J}_- \mid \tilde{\psi} \rangle = \lambda \mid \tilde{\psi} \rangle, \label{A-lambda-e=1}
 \end{equation} and when
${\epsilon} =-1$ these same equations read

\begin{equation}
{\hat{H}}^{T_{(-1)}}_{2D}  = \beta_0 \hat{N} -  \hat{J}_3+ h_0 \hat{I}
\end{equation}
  and
  \begin{equation}
   \frac{\alpha_3}{\sqrt{1-\beta_3^2}} e^{- i \theta} \hat{J}_+ \mid \tilde{\psi} \rangle = \lambda \mid \tilde{\psi} \rangle,
 \end{equation}
  respectively.

 Let us take, for example, the reduced case with $\epsilon =1.$ The eigenvalue equation (\ref{A-lambda-e=1}), written in the 2-D  Fock-Bargmann representation of analytic functions take the form
 
 \begin{equation}
  - \frac{\alpha_3}{\sqrt{1-\beta_3^2}} e^{i \theta}  \xi_2 \frac{\partial \tilde{\psi}}{\partial \xi_1} (\xi_1,\xi_2)  = \lambda  \tilde{\psi} (\xi_1, \xi_2),
\label{H-2D-e=1-analytic}
\end{equation}
and has analytic solutions  only when $\lambda=0,$ in which case the solution are given by

\begin{equation}
\tilde{\psi}_{0} (\xi_1, \xi_2)= \tilde{\varphi}_2 (\xi_2) \label{psi-0-xi-1-xi-2} 
\end{equation} 
where $\tilde{\varphi}_2 (\xi_2)$ is an  analytic function of the variable $\xi_2,$ which must be specified using the requirement that the 
eigenstate represented $ \tilde{\psi}_{0} (\xi_1, \xi_2)$ must also be an eigenstate of   (\ref{H-2D-e=1}). Then, expressing   (\ref{H-2D-e=1}) in the Fock-Barmann representation of analytic  functions as

\begin{equation}
 {\hat{H}}^{T_{(1)}}_{2D}  =  \frac{\beta_0 +1}{2}  \xi_1 \frac{\partial}{\partial \xi_1}  +\frac{\beta_0 -1}{2}  \xi_2 \frac{\partial}{\partial \xi_2}  + h_0,
 \end{equation}
the mentioned requirement implies that $ \xi_2 \frac{\partial \tilde{\varphi}}{\partial \xi_2} (\xi_2) ={\bf \kappa} \tilde{\varphi} (\xi_2) ,  $
from where $\tilde{\varphi} (\xi_2) = \tilde{\varphi}_0  \xi_2^{\bf \kappa},$ where ${\bf \kappa}$ is a positive integer and $ \tilde{\varphi}_0$ an arbitrary complex constant. Thus, by inserting this last result into (\ref{psi-0-xi-1-xi-2}),  we get the normalized fundamental state $\tilde{\psi}_{0} (\xi_1, \xi_2)=  \tilde{\varphi}_0  \xi_2^{\bf \kappa}.$

Coming back to the number representation, we find that  the transformed  fundamental state is given by

\begin{equation}
\mid \tilde{0} \rangle = \mid 0, {\bf \kappa} \rangle,  
\end{equation}
from where we deduce that the ground state associated to the generalized Hamiltonian  (\ref{H-2D-mu-0-beta-0-not-1-3}) is given by
two-mode the non-separable state

\begin{equation}
\mid 0 \rangle = \hat{T}_{(1, 1,  \beta_3,\theta)} \mid 0, {\bf \kappa} \rangle,
\end{equation}
or more explicitly by
 \begin{equation}
\mid 0 \rangle = \left(\frac{1 + \beta_3}{2} \right)^{\frac{{\bf \kappa}}{2}} \sum_{k=0}^{{\bf \kappa}}  (-1)^k \sqrt{\binom{{\bf \kappa}}{k}}  \; \left(\frac{ 2 \beta_-}{1 + \beta_3} \right)^k \;  \mid k , {\bf \kappa}-k \rangle. 
\end{equation}
The energy associated to this fundamental state being  $ E_0 =  {\bf \kappa} \frac{\beta_0 -1}{2} + h_0, $ we  deduce that the energy spectrum is given by $E_n= E_0 + n, \; n=0,1, {\bf \kappa},$ i.e., a truncate or finite  spectrum.  In fact, the set of energy eigentates of (\ref{H-2D-mu-0-beta-0-not-1-3})  is finite, it could be computed by first obtaining the transformed normalized energy eigenstates in the form 

\begin{equation}
\mid \tilde{E}_n \rangle = \begin{cases}  \mathcal{\tilde{N}}_n^{- \frac{1}{2}} {(\hat{A}^{T_{1}}_{(\vec{0} , \vec{\alpha} )})^{\dagger}}^n \mid \tilde{0} \rangle = \mathcal{\tilde{N}}_n^{- \frac{1}{2}} \hat{J}_{+}^{n} \mid 0, {\bf \kappa} \rangle = \mid n, {\bf \kappa}-n \rangle, \quad n=0,1,\ldots,{\bf \kappa},  \\
0, \quad n > {\bf \kappa}. \end{cases}
\end{equation}
where  some parameters have been systematically absorbed in the normalization constant. Then, the energy eigenstates of the original Hamiltonian   (\ref{H-2D-mu-0-beta-0-not-1-3}) can be obtained from these last states by acting on them  with the unitary operator
$\hat{T}_{(1, 1,  \beta_3,\theta)},$ that is,

\begin{equation}
\mid  E_n  \rangle = \hat{T}_{(1, 1,  \beta_3,\theta)} \mid n , {\bf \kappa} -n \rangle, \quad n=0,1, \ldots, {\bf \kappa},
\end{equation}
or more explicitly

\begin{eqnarray}
\mid  E_n  \rangle &=& \left(\frac{2}{1+\beta_3}\right)^{(n - \frac{ {\bf \kappa}}{2})}  \; \sum_{k=0}^{n} \beta_+^k  \sqrt{\binom{n}{k} \binom{{\bf \kappa}-n + k}{k}}   \nonumber \\ &\times&  \sum_{\ell=0}^{{\bf \kappa}-n+k}  \; (-1)^\ell  \left( \frac{ 2 \beta_-}{1 + \beta_3} \right)^\ell \;  \sqrt{\binom{n-k+ \ell}{\ell} \binom{{\bf \kappa} -n +k}{\ell}}  \; \mid n-k + \ell, {\bf \kappa}-n+k-\ell \rangle.
\end{eqnarray}

We observe that when $h_0=0$ and  $\beta_0 < 1,$ the energy of the ground state is less th
an $0$ and when $\beta_0 > 1$ but $\beta_0 \neq 3,$ it is bigger than $0.$ Moreover, when $h_0 =0$ and  $\beta_0 =0,$ the energy eigenvalues are  given by $ E_n =   - \frac{{\bf \kappa}}{2} + n, \; n=0,1, {\bf \kappa},$ which corresponds to the spectrum of a system  whose  Hamiltonian and associated ladder operators generate the $su(2)$ Lie algebra. In such a case, we could say that the integer ${\bf \kappa} \geq 1,$ determines the ${\bf \kappa} +1$ irreducible representation space of the bososnic realization of these generators.

\subsection{The case $b^2=1$ and  $\beta_0 \in \{1,3\}:$  2:1 and 1:2 basic, extended and generalized commensurate  anisotropic quantum oscillator} 

In the case $b^2=1,$  the system of equations (\ref{eq-parameters} ) provides us with different solutions depending on the value of the different parameters involved:

\begin{itemize}

\item When $\beta_+ = \beta_- =0$ and $\beta_3\ne 0,$ we have four possible choices for the couple of parameters $(\beta_0, \beta_3),$ that is,  $(1,1);(1,-1) ;  (3,1)$ and $(3,-1),$ which lead to the basic  relations  

\begin{equation}
\hat{H}_{2D} = \hat{N} + \hat{J}_3 = \hat{a}_{1}^\dagger \hat{a}_1, \quad   \hat{A}_{(\vec{u} , \vec{\alpha} )} = \mu_1 \hat{a}_1 +\alpha_+ \hat{J}_- \label{H-2D-A-1} 
\end{equation} 
whose ladder operators verify
\begin{equation}
 \left[\hat{A}_{(\vec{u} , \vec{\alpha} )},  \hat{A}^\dagger_{(\vec{u} , \vec{\alpha} )} \right]=  {\| \mu_1 \|}^2 \hat{I}  + \mu_1 \alpha_{+}^\ast \hat{a}_2 + \mu_{1}^\ast \alpha_+ \hat{a}^\dagger_2  - 2 {\|\alpha_+  \|}^2 \hat{J}_3,
\end{equation} 
\begin{equation}
\hat{H}_{2D} = \hat{N} - \hat{J}_3=   \hat{a}_{2}^\dagger \hat{a}_2 , \quad   \hat{A}_{(\vec{u} , \vec{\alpha} )} = \mu_2 \hat{a}_2  + \alpha_-  \hat{J}_+  \label{H-2D-A-2}  \end{equation} 
whose ladder operators verify
\begin{equation}\left[\hat{A}_{(\vec{u} , \vec{\alpha} )},  \hat{A}^\dagger_{(\vec{u} , \vec{\alpha} )} \right]=  {\| \mu_2 \|}^2 \hat{I}  + \mu_2 \alpha_{-}^\ast \hat{a}_1 + \mu_{2}^\ast \alpha_- \hat{a}_{1}^\dagger  + 2 {\|\alpha_-  \|}^2 \hat{J}_3,
\end{equation} 

  \begin{equation}
\hat{H}_{2D} = 3 \hat{N} +  \hat{J}_3=2 \hat{a}_{1}^\dagger \hat{a}_1 +  \hat{a}_{2}^\dagger \hat{a}_2, \quad  \hat{A}_{(\vec{u} , \vec{\alpha} )} = \mu_2 \hat{a}_2  + \alpha_+  \hat{J}_-, \label{H-2D-basic-1}
\end{equation}
whose ladder operators verify  
\begin{equation}
\left[\hat{A}_{(\vec{u} , \vec{\alpha} )},  \hat{A}^\dagger_{(\vec{u} , \vec{\alpha} )} \right]=  {\| \mu_2 \|}^2 \hat{I}  -  2 {\|\alpha_+ \|}^2 \hat{J}_3,\label{A-2D-basic-1}
\end{equation}
and
 \begin{equation}
\hat{H}_{2D} = 3 \hat{N} - \hat{J}_3= \hat{a}_{1}^\dagger \hat{a}_1 + 2   \hat{a}_{2}^\dagger \hat{a}_2, \quad  \hat{A}_{(\vec{u} , \vec{\alpha} )} = \mu_1 \hat{a}_1  + \alpha_-  \hat{J}_+,\label{H-2D-basic-2}
\end{equation}
whose ladder operators verify
\begin{equation}
\left[\hat{A}_{(\vec{u} , \vec{\alpha} )},  \hat{A}^\dagger_{(\vec{u} , \vec{\alpha} )} \right]=  {\| \mu_1 \|}^2 \hat{I}  +  2 {\|\alpha_-  \|}^2 \hat{J}_3,\label{A-2D-basic-2}
\end{equation}
respectively. Then, the  above  two former Hamiltonians represent  a one dimensional  canonical quantum harmonic oscillator, and the last two correspond  to the $2:1$ and $1:2$  commensurate  anisotropic quantum harmonic oscillator, respectively.  

\item On the other hand, when $\beta_+ = \beta^{\ast}_{-} \neq 0$ and $\beta_3=0,$ by defining $\beta_{\pm} = R e^{\pm i\theta},$ from (\ref{conditions}),  we get 
$R = \frac{1}{2}$ and the extended relations

\begin{eqnarray} 
\hat{H}_{2D} &=& \beta_{0}  \hat{N} +  \frac{1}{2}  \left( e^{-i \theta} \hat{J}_+ + e^{i \theta} \hat{J}_-  \right),  \label{H-2D-semi-gen}
\\ 
 \hat{A}_{(\vec{u} , \vec{\alpha} )}& =& \mu_1 \hat{a}_1 +  \mu_1 \frac{e^{-i \theta}}{(2 -\beta_0)}  \hat{a}_2  +
  \alpha_3 \left(  \hat{J}_3  +\frac{1}{2} e^{-i \theta } \hat{J}_+ - \frac{1}{2} e^{i \theta} \hat{J}_- \right), \quad \beta_0=1,3. \label{A-2D-semi-gen}
\end{eqnarray}
 
\item Finally, when $\beta_\pm \neq 0 $ and $\beta_3 \neq 0,$ all them verifying $4 \beta_+ \beta_- + \beta_3^2=1,$ the 2-D generalized Hamiltonian,  lowering operator and  ladder operators commutation relation assume the form 

\begin{equation}
\hat{H}_{2D} =\beta_{0}  \hat{N} + \beta_- \hat{J}_+  +   \beta_+ \hat{J}_-  + \beta_3 \hat{J}_3, \quad \beta_0 =1,3, 
 \label{H-2D-generalized}
\end{equation}

\begin{equation}
 \hat{A}_{(\vec{u} , \vec{\alpha} )} = \mu_1 \hat{a}_1 +  \mu_1  \frac{2 \beta_-}{(2 + \beta_3 -\beta_0)}  \hat{a}_2  +
  \alpha_3 \left(  \hat{J}_3  +\frac{\beta_-}{1 +\beta_3}  \hat{J}_+ - \frac{\beta_+}{1 -\beta_3}  \hat{J}_- \right), \quad \beta_0=1,3, \label{A-2D-generalized}
\end{equation}

and
\begin{equation}
\left[\hat{A}_{(\vec{u} , \vec{\alpha} )},  \hat{A}^\dagger_{(\vec{u} , \vec{\alpha} )} \right]=  \frac{2 {\| \mu_1 \|}^2}{1 \pm \beta_3} \hat{I}   \pm \frac{1}{2} (\mu_1 \hat{a}_1 + \mu_1^\ast \hat{a}_{1}^\dagger) +
\frac{\mu_1  \beta_-}{ 1 \pm \beta_3} \hat{a}_2 + \frac{\mu_{1}^\ast  \beta^{\ast}_{-}}{ 1 \pm \beta_3} \hat{a}^\dagger_{2} -  \frac{2\| \alpha_3 \|^2}{1-\beta_3^2} \; \vec{\beta} \cdot \vec{\hat{J}}  ,
\end{equation}
respectively. In this last  expression the sign $+$ represents $\beta_0 =1$ and the sign  $-$  represents  $\beta_0=3.$ 

\end{itemize}

 An important fact of all the above explicit expressions for the Hamiltonian and  for the ladder operators is that, under the  conditions    (\ref{eq-parameters}), the $su (2)$ characteristic   parameter, in all cases, is given by  $b^2= 4 \beta_+ \beta_- + \beta_3^2 =1 $ and $\tilde{b} =\sqrt{4 \alpha_- \alpha_+ + \alpha_3^2} =0,$ respectively. This fact allows us to perform  suitably unitary transformations  on the states or unitary similarity transformations on the operators to reduce the problem of solving the generalized $2:1$ or $1:2$ anisotropic quantum oscillator systems (\ref{H-2D-semi-gen}--\ref{A-2D-semi-gen}) and (\ref{H-2D-generalized}--\ref{A-2D-generalized})  to the problem of solving the basic system  (\ref{H-2D-basic-1}--\ref{A-2D-basic-1}) or (\ref{H-2D-basic-2}--\ref{A-2D-basic-2}).

\subsubsection{The case $ \hat{A}_{(\vec{u} , \vec{\alpha} )} = \mu_2 \hat{a}_2  + \alpha_+  \hat{J}_- : $  2:1  basic  commensurate anisotropic quantum oscillator}
\label{basic-2-1-commensurate}

Let us first  find the algebra eigenstates of the lowering operator defined in (\ref{H-2D-basic-1}). Indeed, these states have already been obtained in the literature\cite{JM-VH-IM}. Here, we will recreate these states following an more direct technique by expressing  the creation and annihilation operators, appearing in the 2-D quantum anisotropic oscillator, as differential operators acting on the  two-dimensional complex analytic function representation space of the algebra. Then, if $\psi (\xi_1, \xi_2)$ is complex analytic function, of two complex variables $\xi_{i}, \; i=1,2,$  representing a stationary state of the system, then the set of creation  and annihilation operators of this system are represented  by

\begin{equation}
\hat{a}_i = \frac{d}{d\xi_i}, \quad \hat{a}^\dagger_{i} = \xi_i, \quad i=1,2.
\end{equation}      
Thus, the problem of finding the algebra eigenstates of the lowering operator  in (\ref{H-2D-basic-1}) reduces to the problem of solving  
the partial differential equation

 \begin{equation}
  \left[\mu_2 \frac{d}{d\xi_2}  +   \alpha_+   \xi_2 \frac{d}{d\xi_1} \right] \psi(\xi_1, \xi_2) = \lambda \psi(\xi_1, \xi_2). \label{Dif-equation-1}
  \end{equation} 
  The initial conditions of this equation have to be fixed according to the requirement of the square-integrable property that a quantum system state must verify. 
  
  For example, if we choose $\psi(0, 0)=1,$ the differential equation (\ref{Dif-equation-1}) can be integrated using the  separation of variables method, doing that we get 
   
   \begin{equation}
   \psi_{\lambda} (\xi_1, \xi_2 ) = \exp\left[ c_1  \lambda  \xi_1 \right]   \exp\left[ \frac{\lambda}{\mu_2} \left( \xi_2 - \frac{1}{2}  c_1 \alpha_+   \xi_{2}^2\right) \right], \label{Eigen-states-A-analytic-base}
  \end{equation} 
  where $c_1$ is a complex constant. Then,  coming back to the number state representation, the states   (\ref{Eigen-states-A-analytic-base}) are given by
  
  \begin{equation}
\mid  \lambda \rangle_1 = \mathcal{N}^{\frac{-1}{2}}  \exp\left[ c_1  \lambda  \hat{a}^{\dagger}_1 \right]   \exp\left[ \frac{\lambda}{\mu_2} \left( \hat{a}^{\dagger}_{2} - \frac{1}{2}  c_1 \alpha_+   (\hat{a}^\dagger_{2})^2\right) \right] \mid 0, 0 \rangle,  \label{A-cano-1-cano-squ-2}
  \end{equation}
  where $\mathcal{N}$ is a normalization constant. We observe that this last state, which is an eigenstate of  the  operator  $  \hat{A}_{(\vec{u} , \vec{\alpha} )} = \mu_2 \hat{a}_2  + \alpha_+  \hat{J}_-$ with associated eigenvalue $\lambda,$  has the structure of a coherent state in the mode one and a squeezed coherent state in the mode two. Moreover, using the general techniques  of the  operator algebra, we can write this last state in the normalized  form

\begin{equation}
\mid \lambda \rangle_1 = \hat{D}_1 \left( c_1 \lambda \right) \hat{S}_2 \left( \chi \right)  \hat{D}_2 \left( \frac{\lambda}{\mu_2}  \cosh (\| \chi \|) \right) \mid  0 , 0 \rangle, \label{separable-coherent-states-2-1}
\end{equation}  
  where $\hat{D}_i (z) = e^{z \hat{a}^\dagger_i - z^\ast \hat{a}_i} , \; i=1,2, $ is the canonical displacement operator, and $\hat{S}_2 (\chi) = e^{\frac{-1}{2} ( \chi  (\hat{a}_{2}^\dagger )^2  - \chi^\ast \hat{a}_2^2)},$ where $$\chi= \frac{\frac{ \lambda c_1 \alpha_+ }{\mu_2} } { \| \frac{ \lambda c_1 \alpha_+ }{\mu_2} \| }  \tanh^{-1} \left(   \| \frac{ \lambda c_1 \alpha_+ }{\mu_2} \|  \right), $$ 
   which must verify   $\| \frac{ \lambda c_1 \alpha_+ }{\mu_2} \| < 1.$  

  \vspace{0.5cm}
 
  On the other hand, when we choose   $\psi(0, 0)=0,$ the partial differential equation (\ref{Dif-equation-1}) can be integrated, but its solutions   does not allow a  separation of variables, the analytic solution is given by
  
  \begin{equation}
   \psi_{\lambda} (\xi_1, \xi_2 ) = \exp\left(  \frac{\lambda \xi_2}{\mu_2}   \right)   
\left( \frac{\xi_{2}^2}{2}  -  \frac{\mu_2}{\alpha_+} \xi_1 \right),   
   \end{equation}
   whose  normalized version in the   number space representation is
   
 \begin{equation}
  \mid  \lambda \rangle_2 = \frac{ \hat{D}_2 \left( \frac{\lambda}{\mu_2} \right)}{\sqrt{ \frac{1}{2}  + \frac{\| \mu_2 \|^2}{ \| \alpha_+ \|^2}   +  \frac{\| \lambda \|^2}{ \| \mu_2 \|^2}  +  \frac{\| \lambda \|^4 }{4  \| \mu_2\|^4}    }} \left[     \frac{ {\left( \hat{a}^{\dagger}_{2}  + \frac{\lambda^\ast}{ \mu_2^\ast}  \right)}^2 }{2} | 0, 0 \rangle  - \frac{\mu_2}{\alpha_+}  
   | 1 , 0  \rangle  \right]. \label{two-mode-non-separable-coherent-states}  \end{equation}

 We notice that when $\lambda=0,$ the states $\mid \lambda \rangle_1$  and $\mid \lambda \rangle_2$ become
  \begin{equation} \mid 0  \rangle_1 =\mid 0 ,0 \rangle \quad \text{ and}  \quad \mid 0  \rangle_2 = 
  \frac{  \left[     \frac{1}{\sqrt{2}} | 0, 2 \rangle  - \frac{\mu_2}{\alpha_+}  
   | 1 , 0  \rangle  \right]}{\sqrt{ \frac{1}{2}  + \frac{\| \mu_2 \|^2}{ \| \alpha_+ \|^2} }} \label{ground-states-commensurate-1-2}
   \end{equation}
   respectively.  These states are eigenstates of  $ \hat{A}_{(\vec{u} , \vec{\alpha} )} = \mu_2 \hat{a}_2  + \alpha_+  \hat{J}_-$ with eigenvalue equal to $ 0$, and also they are eigenstates of the 2:1 anisotropic oscillator Hamiltonian described in  (\ref{H-2D-basic-1}) with associated eigenvalue  $0$ and $2,$ respectively. Moreover, according to the relation  (\ref{general-commutator-H-2D-A-dagger}), 
the non-normalized states    
  \begin{equation} \mid E_n \rangle_1 =  \left(\hat{A}^\dagger_{(\vec{u} , \vec{\alpha} )} \right)^n  \mid  0 \rangle_1, \quad n=0,1,\ldots
   \end{equation} and
   \begin{equation} \mid E_n \rangle_2 =  \left(\hat{A}^\dagger_{(\vec{u} , \vec{\alpha} )} \right)^n  \mid  0 \rangle_2, \quad n=0,1, \ldots,
   \end{equation}
   are eigenstates of the Hamiltonian shown in  (\ref{H-2D-basic-1}) with associated eigenvalues equal to $n$ and $n+2,$ respectively.
   
\vspace{1.0cm}
We notice that there are also other solutions to equation (\ref{Dif-equation-1}) which are consistent with the requirement that the eigenstate with $\lambda=0$ be also an eigenstate of the 2:1 commensurate anisotropic Hamiltonian, that is,  
  
  \begin{equation}
   \psi_{\lambda} (\xi_1, \xi_2 ) = \exp\left(  \frac{\lambda \xi_2}{\mu_2}   \right)   
\left( \frac{\xi_{2}^2}{2}  -  \frac{\mu_2}{\alpha_+} \xi_1 \right)^\kappa, \quad \kappa=1,2, \dots.  
   \end{equation}
   The fundamental state in this representation is given by 
   \begin{equation}
   \psi_{0} (\xi_1, \xi_2 ) =   
\left( \frac{\xi_{2}^2}{2}  -  \frac{\mu_2}{\alpha_+} \xi_1 \right)^\kappa, \quad \kappa=1,2, \dots.  \label{2-1-analytic-ground-state}
   \end{equation}
  Then, if we express the 2:1 as
  
  \begin{equation}
  H_{2D} = 2 \xi_1 \frac{\partial}{\partial \xi_1} +   \xi_2 \frac{\partial}{\partial \xi_2},
  \end{equation}
   we can directly prove that
   
   \begin{equation}
   H_{2D} \psi_{0} (\xi_1, \xi_2 ) = 2 \kappa    \psi_{0} (\xi_1, \xi_2 ), \quad \kappa =1,2, \ldots,
   \end{equation}
 i.e.,  energy of the ground state is given by $E_0^{(\kappa)} = 2 \kappa, \; \kappa=1,2, \ldots.$ Hence, according to the general commutation relation  (\ref{general-commutator-H-2D-A-dagger}) we deduce that in this case the energy spectrum is given by  $E_{n}^{(\kappa)} = 2 \kappa +n, \; n=0,1, \ldots,$ for each fixed $\kappa=1,2, \ldots.$

   \subsubsection{ 2:1 Basic commensurate anisotropic Hamiltonian energy eigenstates}
  \label{anisotropic-2:1-normal-ordering}
   Let us now compute the energy eigenstates $ \mid E_n \rangle_1$  and $\mid E_n \rangle_2$ for the 2:1 anisotropic quantum oscillator. To do this, we need first to write the operator $\hat{A}^\dagger_{(\vec{u} , \vec{\alpha} )} = \mu_2^\ast  \hat{a}^\dagger_2 + \alpha^\ast_+  \hat{a}^\dagger_1 \hat{a}_2 $ 
 raised to the power of $n$ in the normal ordering, that is, with an annihilation operator $\hat{a}_2$ standing always to the right of any creation operator $\hat{a}_2^\dagger.$ Thus, if we define the normal ordering binomial power of  $n$ operator   
 \begin{equation}
 \left( \hat{A}^\dagger_{(\vec{u} , \vec{\alpha} )}\right)^n_{\lozenge} = \sum_{k=0}^{n} \binom{n}{k}  (\mu^\ast_2 \hat{a}^\dagger_2)^{k}  (\alpha^\ast_+  \hat{a}^\dagger_1 \hat{a}_2)^{n-k},  \label{A-n-normal-order}
 \end{equation} 
we get the following sequence for the power of  $\hat{A}^\dagger_{(\vec{u} , \vec{\alpha} )}:$
 
\begin{eqnarray*}
\left( \hat{A}^\dagger_{(\vec{u} , \vec{\alpha} )}\right)^0&=&\left( \hat{A}^\dagger_{(\vec{u} , \vec{\alpha} )} \right)^{0}_{\lozenge} =1, \\
\left( \hat{A}^\dagger_{(\vec{u} , \vec{\alpha} )}\right)^1 &=&\left( \hat{A}^\dagger_{(\vec{u} , \vec{\alpha} )}\right)^{1}_{\lozenge} =  \mu_2^\ast  \hat{a}^\dagger_2 + \alpha^\ast_+  \hat{a}^\dagger_1 \hat{a}_2 
\\
\left( \hat{A}^\dagger_{(\vec{u} , \vec{\alpha} )} \right)^2& =&  \left( \hat{A}^\dagger_{(\vec{u} , \vec{\alpha} )} \right)^{2}_{\lozenge}  +     (\mu_2^\ast \alpha^\ast_+  \hat{a}^\dagger_1) \left( \hat{A}^\dagger_{(\vec{u} , \vec{\alpha} )} \right)^{0}_{\lozenge} \\ \left( \hat{A}^\dagger_{(\vec{u} , \vec{\alpha} )} \right)^3& =&  \left( \hat{A}^\dagger_{(\vec{u} , \vec{\alpha} )} \right)^{3}_{\lozenge}  +  3   (\mu_2^\ast \alpha^\ast_+  \hat{a}^\dagger_1) \left( \hat{A}^\dagger_{(\vec{u} , \vec{\alpha} )} \right)^{1}_{\lozenge} \\ \left( \hat{A}^\dagger_{(\vec{u} , \vec{\alpha} )} \right)^4& =&  \left( \hat{A}^\dagger_{(\vec{u} , \vec{\alpha} )} \right)^{4}_{\lozenge}  +  6   
(\mu_2^\ast \alpha^\ast_+  \hat{a}^\dagger_1) \left( \hat{A}^\dagger_{(\vec{u} , \vec{\alpha} )} \right)^{2}_{\lozenge} + 3  (\mu_2^\ast \alpha^\ast_+  \hat{a}^\dagger_1)^2  \left( \hat{A}^\dagger_{(\vec{u} , \vec{\alpha} )} \right)^{0}_{\lozenge} \\
 \left( \hat{A}^\dagger_{(\vec{u} , \vec{\alpha} )} \right)^5& =&  \left( \hat{A}^\dagger_{(\vec{u} , \vec{\alpha} )} \right)^{5}_{\lozenge}  +  10   
(\mu_2^\ast \alpha^\ast_+  \hat{a}^\dagger_1) \left( \hat{A}^\dagger_{(\vec{u} , \vec{\alpha} )} \right)^{3}_{\lozenge} + 15  (\mu_2^\ast \alpha^\ast_+  \hat{a}^\dagger_1)^2  \left( \hat{A}^\dagger_{(\vec{u} , \vec{\alpha} )} \right)^{1}_{\lozenge}, 
\end{eqnarray*}
 and so on until reach the nth term
 \begin{equation}
 \left( \hat{A}^\dagger_{(\vec{u} , \vec{\alpha} )}\right)^n = \sum_{k=0}^{[\frac{n}{2}]} C^{(n)}_{n-2k}  \left( \hat{A}^\dagger_{(\vec{u} , \vec{\alpha} )} \right)^{n-2k}_{\lozenge}, \label{power-n-A-dagger}
 \end{equation}
 where $[\frac{n}{2}]$ represents the integer part of $\frac{n}{2}.$ 
To compute the coefficients $C^{(n)}_{n-2k},$ we can use (\ref{power-n-A-dagger})
in \begin{equation}
 \left( \hat{A}^\dagger_{(\vec{u} , \vec{\alpha} )}\right)^{(n+1)} = \left(\mu_2^\ast  \hat{a}^\dagger_2 + \alpha^\ast_+  \hat{a}^\dagger_1 \right) \left( \hat{A}^\dagger_{(\vec{u} , \vec{\alpha} )}\right)^n, 
\end{equation} 
and  compare the coefficients having the same power of  $\left( \hat{A}^\dagger_{(\vec{u} , \vec{\alpha} )}\right)^{(n-2k-1)}_{\lozenge}$ in both  sides of the equation, doing that we  find that the coefficients $ C^{(n)}_{n-2k}, \quad k=0,1,\ldots, [\frac{n}{2}], n=0,1,\ldots,$ verify the recurrence relation
 
\begin{equation}
C^{(n)}_{n-2k} = (n+1 -2k)  (\mu_2^\ast \alpha^\ast_+  \hat{a}^\dagger_1) C^{(n-1)}_{n-1-2(k-1)} + C^{(n-1)}_{n-1 -2k}, \quad k=1,\ldots, [\frac{n}{2}] , \label{eq-recurrence-n}
\end{equation} 
with the initial conditions $C^{(n)}_{n}=1, \; n=1,2, \ldots.$ 
 The solution of the recurrence series equation (\ref{eq-recurrence-n}) is given by\cite{WW}
\begin{equation}
C^{(n)}_{n-2k} = \frac{n!}{(n-2k)!} \frac{(\mu_2^\ast \alpha^\ast_+  \hat{a}^\dagger_1)^k}{2^k \, k!}, \quad k=0,1, \ldots, [\frac{n}{2}]. \label{C-coefficients-n-k}
\end{equation}  
   
 Then, by  inserting (\ref{C-coefficients-n-k}) into (\ref{power-n-A-dagger}), we get 
   
   \begin{equation}
 \left( \hat{A}^\dagger_{(\vec{u} , \vec{\alpha} )}\right)^n = \sum_{k=0}^{[\frac{n}{2}]}    \frac{n!}{(n-2k)!} \frac{(\mu_2^\ast \alpha^\ast_+  \hat{a}^\dagger_1)^k}{2^k \, k!}   \left( \hat{A}^\dagger_{(\vec{u} , \vec{\alpha} )} \right)^{n-2k}_{\lozenge}. \label{power-n-A-dagger-ordering}
 \end{equation}
  
  Now we are ready to compute the energy eigenstates. From   (\ref{A-n-normal-order}) we deduce that
  
  \begin{equation} {\left( \hat{A}^\dagger_{(\vec{u} , \vec{\alpha} )}\right)}^{n-2k}_{\lozenge} \mid 0 \rangle_1 = {\mu_{2}^\ast}^{n-2k} \sqrt{(n-2k)!} \mid 0, n-2k\rangle,  \label{A-normal-n-2k-onto-state-0-0}\end{equation} 
  then using  (\ref{power-n-A-dagger-ordering}) we have
  \begin{equation} \mid E_n \rangle_1 =  \mathcal{N}_{1,n}^{-\frac{1}{2}} \sum_{k=0}^{[\frac{n}{2}]}  \frac{1}{\sqrt{(n-2k)! k!} }       \left(\frac{\alpha^\ast_+  }{2 \mu^\ast_2 }\right)^k   \mid k, n-2 k\rangle, \label{lambda-1-base}
  \end{equation} where $\mathcal{N}_{1,n},$ is a normalization constant, in which we have absorbed a priory a factor $ n! {\mu^\ast_2}^n, $ which is given by
  
\begin{equation}
\mathcal{N}_{1,n} =  \sum_{k=0}^{[\frac{n}{2}]}   \frac{1}{(n-2k)! k!}       \left(\frac{ \|\alpha_+ \|  }{2 \|\mu_2\| }\right)^{2k}. \label{N-1-n-normalization-constant}
\end{equation}  
     
  On the other hand, 
   
  \begin{eqnarray} {\left( \hat{A}^\dagger_{(\vec{u} , \vec{\alpha} )}\right)}^{n-2k}_{\lozenge} \mid 0 \rangle_2 &=& \frac{1}{\sqrt{2}}
  {\mu_2^\ast}^{n-2k} \sqrt{(n-2k)!} \left[ \frac{\sqrt{(n-2k+1)!(n-2k+2)!}}{2} \mid 0, n-2k+2 \rangle   \right. \nonumber \\ &+&
\left.  \left(\frac{\alpha^\ast_+}{\mu_2^\ast}\right)^2 \sqrt{(n-2k-1)!(n-2k)!} \mid 2, n-2k-2 \rangle + \sqrt{2} \left(   
\frac{\alpha^\ast_+}{\mu_2^\ast} (n-2k) - \frac{\mu_2}{\alpha_+}  \right) \mid 1, n-2k\rangle,
  \right] \nonumber \\ 
   \end{eqnarray} 
the energy eigenstate

   \begin{eqnarray}
 \mid E_n \rangle_2 &=& \mathcal{N}_{2,n}^{-\frac{1}{2}} \sum_{k=0}^{[\frac{n}{2}]}  \frac{1}{\sqrt{(n-2k)! k!} }       \left(\frac{\alpha^\ast_+  }{2 \mu^\ast_2 }\right)^k  \left[ \frac{\sqrt{(n-2k+1)!(n-2k+2)!}}{2} \mid k, n-2k+2 \rangle   \right. \nonumber \\ &+&
 \left(\frac{\alpha^\ast_+}{\mu_2^\ast}\right)^2  \sqrt{\frac{(k+1)(k+2)}{2}}  \sqrt{(n-2k-1)!(n-2k)!} \mid k+2, n-2k-2 \rangle \nonumber \\ &+&  \left. \sqrt{2 (k+1)} \left(   
\frac{\alpha^\ast_+}{\mu_2^\ast} (n-2k) - \frac{\mu_2}{\alpha_+}  \right) \mid k+1, n-2k\rangle
  \right], \nonumber \\ \label{lambda-2-base}
   \end{eqnarray}
   where, as before, $\mathcal{N}_{2,n}$ is a normalization constant given by $\mathcal{N}_{2,n} = {}_2\langle E_n \mid E_n \rangle_2.$

\subsubsection{The case $\beta_{\pm} \neq 0$ and $\beta_3=0$ 2:1 Extended commensurate anisotropic quantum oscillator }

As we have seen in the above sections, when   $\beta_\pm \neq 0$ and $\beta_3=0,$ the lowering operator $ \hat{A}_{(\vec{u} , \vec{\alpha} )}$ takes the form
\begin{equation}
\hat{A}_{(\vec{u} , \vec{\alpha} )}  =  \mu_1 \hat{a}_1 +  \mu_1 \frac{e^{-i \theta}}{(2 -\beta_0)}  \hat{a}_2  +
  \alpha_3 \left(  \hat{J}_3  +\frac{1}{2} e^{-i \theta } \hat{J}_+ - \frac{1}{2} e^{i \theta} \hat{J}_- \right), \quad \beta_0=1,3. \end{equation}
and the extended 2:1 commensurate anisotropic Hamiltonian is written this way
\begin{equation}
\hat{H}_{2D} = \beta_{0}  \hat{N} +  \frac{1}{2}  \left( e^{-i \theta} \hat{J}_+ + e^{i \theta} \hat{J}_-  \right).  \label{H-2:1-2D-beta-3=0}
\end{equation}

We observe that, in the case when $\beta_0=3,$ by performing a unitary similarity transformation to 
the Hamiltonianl (\ref{H-2:1-2D-beta-3=0}), it can be put in the form of a $2:1$ or $1:2$  anisotropic  two-dimensional  quantum oscillator Hamiltonian.
Indeed,
\begin{equation}
\hat{T}^\dagger_{(\epsilon,\theta)} \hat{H}_{2D} \hat{T}_{(\epsilon,\theta)} = \beta_{0}  \hat{N} + \epsilon \hat{J}_3, \label{T-similarity-H-2-D}
\end{equation}
 where
 \begin{equation}
 \hat{T}_{(\epsilon,\theta)} = \exp\left[  - \epsilon \frac{\pi}{4}\left(  e^{-i \theta} \hat{J}_+ - e^{i \theta} \hat{J}_-   \right)                \right], \quad \text{ and} \quad \epsilon = \pm 1.
 \end{equation}
The explicit and disentangled form of this  last operator in terms of the original   creation and annihilation bosonic operators is

 \begin{eqnarray}
 \hat{T}_{(\epsilon, \theta)} &=& \exp\left[  - \epsilon \frac{\pi}{4} \left(  e^{-i \theta} \hat{a}_1^\dagger \hat{a}_2 - e^{i \theta}   \hat{a}_1  \hat{a}_2^\dagger \right) \right], \nonumber \\ &=& \exp\left( - \epsilon   e^{-i \theta} \hat{a}_1^\dagger \hat{a}_2\right) \exp\left(\ln{\sqrt{2}} \; (\hat{a}_1^\dagger \hat{a}_1 - \hat{a}_2^\dagger \hat{a}_2  ) \right)  \exp\left( \epsilon   e^{i \theta} \hat{a}_1 \hat{a}_2^\dagger\right)
  \quad  \quad \epsilon = \pm 1,
\end{eqnarray} 
 which has the structure of a mixing operator.
 
The same transformation on the ladder operator produces
\begin{equation}
\hat{T}^\dagger_{(\epsilon,\theta)} \hat{A}_{(\vec{u} , \vec{\alpha} )} \hat{T}_{(\epsilon,\theta)} = \frac{\mu_1}{\sqrt{2}} \left( 1 + \frac{\epsilon}{2 -\beta_0}  \right) \hat{a}_1 +  \frac{\mu_1 e^{-i\theta}}{\sqrt{2}} \left( \frac{1}{2 -\beta_0}  - \epsilon \right) \hat{a}_2 +  \alpha_3 \left( \frac{1-\epsilon}{2} e^{-i \theta } \hat{J}_+ - \frac{1+\epsilon}{2} e^{i \theta} \hat{J}_- \right). \label{T-similarity-A}
\end{equation}

Then in the case $\beta_0=3$ and $\epsilon=1,$ we have
\begin{equation}
\hat{T}^\dagger_{(1,\theta)} \hat{H}_{2D} \hat{T}_{(1,\theta)} = \beta_{0}  \hat{N} + \hat{J}_3, \quad \text{and} \quad  \hat{T}^\dagger_1  \hat{A}_{(\vec{u} , \vec{\alpha} )} \hat{T}_{(1,\theta)}= - \sqrt{2} \mu_1 e^{-i \theta} \hat{a}_2  - \alpha_3 e^{i \theta} \hat J_{-},
\end{equation}
which corresponds to the 2:1  anisotropic  quantum oscillator system.

\vspace{0.5cm}

On the other hand, when  $\beta_0=3$ and $\epsilon=-1,$ we have
\begin{equation}
\hat{T}^\dagger_{(1,\theta)} \hat{H}_{2D} \hat{T}_{(1,\theta)} = 3 \hat{N} -  \hat{J}_3, \quad \text{and} \quad  \hat{T}^\dagger_{(1,\theta)} \hat{A}_{(\vec{u} , \vec{\alpha} )} \hat{T}_{(1,\theta)}= \sqrt{2} \mu_1  \hat{a}_1 +  \alpha_3 e^{-i \theta} \hat J_{+},
\end{equation}
which corresponds to the 1:2  anisotropic  quantum oscillator system.

\vspace{0.5cm}

When $\beta_0=\epsilon=1$ or $\beta_0 = - \epsilon=1,$ equations (\ref{T-similarity-H-2-D}) and (\ref{T-similarity-A}) lead the generalized  system to the particular cases  (\ref{H-2D-A-1}) and   (\ref{H-2D-A-2}), respectively. As we have mention above, in both cases, the algebra eigenstates of the ladder are non-normalizable.

\vspace{1.00cm}

 To find the algebra eigenstates of $ \hat{A}_{(\vec{u} , \vec{\alpha} )},$ as before, we must solve the eigenvalue equation
 \begin{equation} 
  \hat{A}_{(\vec{u} , \vec{\alpha} )} \mid \lambda \rangle = \lambda \mid \lambda \rangle. \label{A-psi-lambda-psi}
 \end{equation}

Let us illustrate the resolution technique for the case $\beta_0=3,$ and $\epsilon=1.$ In this case, the lowering operator is given by
\begin{equation}
\hat{A}_{(\vec{u} , \vec{\alpha} )}  =  \mu_1 \left( \hat{a}_1 -  e^{-i \theta} \hat{a}_2 \right)  +
  \alpha_3 \left(  \hat{J}_3  +\frac{1}{2} e^{-i \theta } \hat{J}_+ - \frac{1}{2} e^{i \theta} \hat{J}_- \right), \quad \beta_0=1,3. 
  \label{G-A-b0-3-epsilon-1} \end{equation}

By  expressing 
the state $\mid \lambda \rangle$ in (\ref{A-psi-lambda-psi} ) in the form
 \begin{equation}\mid \lambda \rangle = \hat{T}_{(1,\theta)} \mid \tilde{\lambda} \rangle, \label{T-1-on-tilde-lambda} \end{equation}
 
  and then acting on both sides of that equation with the operator $\hat{T}^\dagger_{(1,\theta)},$ we get

     \begin{equation} \left[ \tilde{\mu}_2 \hat{a}_2  +   \tilde{\alpha}_+  \hat J_{-} \right] \mid \tilde{\lambda} \rangle= \lambda  \mid \tilde{\lambda} \rangle \end{equation}
where
\begin{equation}\tilde{\mu}_2 = - \sqrt{2} \mu_1 e^{-i \theta} \quad \text{and } \quad  \tilde{\alpha}_+ =  - \alpha_3 e^{i \theta}. \end{equation}
Thus, using the results of previous section, we get the first set of solutions

\begin{equation}
\mid \tilde{\lambda} \rangle_1 = \hat{D}_1 \left( \tilde{c}_1 \lambda \right) \hat{S}_2 \left(\tilde{\chi}  \right)  \hat{D}_2 \left( \frac{\lambda}{\tilde{\mu}_2}  \cosh (\|\tilde{\chi} \|\right) \mid  0 , 0 \rangle,
\end{equation} 
where $$\tilde{\chi}= \frac{\frac{ \lambda \tilde{c}_1 \tilde{\alpha}_+ }{\tilde{\mu}_2} } { \| \frac{ \lambda c_1 \tilde{\alpha}_+ }{\tilde{\mu}_2} \| }  \tanh^{-1} \left(   \| \frac{ \lambda \tilde{c}_1 \tilde{\alpha}_+ }{\tilde{\mu}_2} \|  \right), $$ 
   which must verify   $$\| \frac{ \lambda \tilde{c}_1 \tilde{\alpha}_+ }{\tilde{\mu_2}} \| < 1.$$ Finally, inserting this result in (\ref{T-1-on-tilde-lambda}) we obtain the normalized eigenstates associated to the $\lambda$ eigenvalue  of the ladder operator  (\ref{G-A-b0-3-epsilon-1}), that is,
   
   \begin{equation}
   \mid  \lambda \rangle_1 = \hat{T}_{(1,\theta)} \;   \hat{D}_1 \left( \tilde{c}_1 \lambda \right) \hat{S}_2 \left(\tilde{\chi}  \right)  \hat{D}_2 \left( \frac{\lambda}{\tilde{\mu}_2}  \cosh (\|\tilde{\chi} \|\right) \mid  0 , 0 \rangle. 
\end{equation} 
 We notice that, the eigenstate of $(\ref{G-A-b0-3-epsilon-1})$   corresponding to the  $\lambda =0$ eigenvalue  is equal to $ \mid 0 \rangle_1  = \hat{T}_{(1,\theta)}  \mid \tilde{0} \rangle_1=  \hat{T}_{(1,\theta)}  \mid 0,0 \rangle= \mid 0,0 \rangle ,$  because $\hat{T}_{(1,\theta)} \mid 0,0 \rangle= \mid 0,0 \rangle.$ 
 
\subsubsection{ 2:1 Extended commensurate anisotropic quantum oscillator energy eigenstates }

 Similarly, the normalized energy eigenstates of $\hat{H}_{2D}$ are given by
 
 \begin{equation}
 \mid E_n \rangle_1  = \hat{T}_{(1,\theta)} \mid  \tilde{E}_n \rangle_1, \end{equation}
 where
 \begin{equation}
 \tilde{E}_n \rangle_1,   =     \tilde{\mathcal{N}}_{1,n}^{-\frac{1}{2}} \sum_{k=0}^{[\frac{n}{2}]}  \frac{1}{\sqrt{(n-2k)! k!} }       \left(\frac{\tilde{\alpha}^\ast_+  }{2 \tilde{\mu}^\ast_2 }\right)^k   \mid k, n-2 k\rangle,  \label{E-n-1-tilde}
  \end{equation} where the normalization constant $\tilde{\mathcal{N}}_{1,n}$  is given by
  
\begin{equation}
\tilde{\mathcal{N}}_{1,n} =  \sum_{k=0}^{[\frac{n}{2}]}   \frac{1}{(n-2k)! k!}       \left(\frac{ \|\tilde{\alpha}_+ \|  }{2 \|\tilde{\mu}_2\| }\right)^{2k}. 
\end{equation}  
 
 \vspace{0.5cm}
The second set of solution of the eigenvalue equation (\ref{A-psi-lambda-psi}) is given by

 \begin{equation}
  \mid  \tilde{\lambda} \rangle_2 = \frac{ \hat{D}_2 \left( \frac{\lambda}{\tilde{\mu}_2}  \right)}{\sqrt{ \frac{1}{2}  + \frac{\| \tilde{\mu}_2 \|^2}{ \| \tilde{\alpha}_+ \|^2}   +  \frac{\| \lambda \|^2}{ \| \tilde{\mu}_2 \|^2}  +  \frac{\| \lambda \|^4 }{4  \| \tilde{\mu}_2\|^4}    }} \left[     \frac{ {\left( \hat{a}^{\dagger}_{2}  + \frac{\lambda^\ast}{ \tilde{\mu}_2^\ast}  \right)}^2 }{2} | 0, 0 \rangle  - \frac{\tilde{\mu}_2}{\tilde{\alpha}_+}  
   | 1 , 0  \rangle  \right],   \end{equation}
from where
 \begin{equation} \mid \tilde{0}  \rangle_2 = 
  \frac{  \left[     \frac{1}{\sqrt{2}} | 0, 2 \rangle  - \frac{\tilde{\mu}_2}{\tilde{\alpha}_+}  
   | 1 , 0  \rangle  \right]}{\sqrt{ \frac{1}{2}  + \frac{\| \tilde{\mu}_2 \|^2}{ \| \tilde{\alpha}_+ \|^2} }}.
   \end{equation}  Thus, following the same procedure used previously, we obtain the second set of  eigenstates of the ladder operator (\ref{G-A-b0-3-epsilon-1}),
   
   \begin{equation}
     \mid  \lambda \rangle_2 = \hat{T}_{(1,\theta)}  \mid  \tilde{\lambda} \rangle_2,
   \end{equation}
   and the second set of energy eigenstates of the generalized anisotropic quantum oscillator Hamiltonian (\ref{H-2:1-2D-beta-3=0}), that is, 
   \begin{equation}\mid  E_n  \rangle_2 = \hat{T}_{(1,\theta)}  \mid  \tilde{E}_n\rangle_2,
 \end{equation}  
 where
 
   \begin{eqnarray}
 \mid\tilde{E}_n \rangle_2 &=& \tilde{\mathcal{N}}_{2,n}^{-\frac{1}{2}} \sum_{k=0}^{[\frac{n}{2}]}  \frac{1}{\sqrt{(n-2k)! k!} }       \left(\frac{\tilde{\alpha}^\ast_+  }{2 \tilde{\mu}^\ast_2 }\right)^k  \left[ \frac{\sqrt{(n-2k+1)!(n-2k+2)!}}{2} \mid k, n-2k+2 \rangle   \right. \nonumber \\ &+&
 \left(\frac{\tilde{\alpha}^\ast_+}{\tilde{\mu}_2^\ast}\right)^2  \sqrt{\frac{(k+1)(k+2)}{2}}  \sqrt{(n-2k-1)!(n-2k)!} \mid k+2, n-2k-2 \rangle \nonumber \\ &+&  \left. \sqrt{2 (k+1)} \left(   
\frac{\tilde{\alpha}^\ast_+}{\tilde{\mu}_2^\ast} (n-2k) - \frac{\tilde{\mu}_2}{\tilde{\alpha}_+}  \right) \mid k+1, n-2k\rangle
  \right],\nonumber \\ \label{E-n-2-tilde}
   \end{eqnarray}
  where, $\tilde{\mathcal{N}}_{2,n}$ is a normalization constant given by $\tilde{\mathcal{N}}_{2,n} = {}_2\langle \tilde{E}_n \mid  
  \tilde{ E}_n \rangle_2.$  
 \subsubsection{The case $\beta_{\pm} \neq 0$ and $\beta_3 \neq 0$ Generalized  2:1 commensurate anisotropic quantum oscillator }
As we have seen in a previous section,   when $\beta_\pm \neq 0 $ and $\beta_3 \neq 0,$ the generalized quantum Hamiltonian $\hat{H}_{2D} $ and lowering operator  $\hat{A}_{(\vec{u} , \vec{\alpha} )} $  are given by equations (\ref{H-2D-generalized}) and 
(\ref{A-2D-generalized}), respectively. In this case, the parameters $\beta_\pm$ and $\beta_3$ are   constrained by the equation $4 \beta_+ \beta_- + \beta_3^2=1.$ Then,  if we write $\beta_\pm = R e^{i \theta},$ this last requirement implies  $4R^2 + \beta_3^2 =1,$ where $ 0 < R < \frac{1}{2}$ and $0< \beta_3 <1.$  Taking into account the new  reparametrization   we have

\begin{equation}
\hat{H}_{2D} =\beta_{0}  \hat{N} +  R  e^{-i \theta} \hat{J}_+  +  R  e^{i \theta}   \hat{J}_-  + \beta_3 \hat{J}_3, \quad \beta_0 =1,3,  \label{H-2D-gen-R}
\end{equation}
and
\begin{equation}
 \hat{A}_{(\vec{u} , \vec{\alpha} )} = \mu_1 \hat{a}_1 +  \mu_1  \frac{2 R e^{- i \theta}} {(2 + \beta_3 -\beta_0)}  \hat{a}_2  +
  \alpha_3 \left(  \hat{J}_3  +\frac{R e^{- i \theta}}{1 +\beta_3}  \hat{J}_+ - \frac{R e^{i \theta}}{1 -\beta_3}  \hat{J}_- \right), \quad \beta_0=1,3,
  \label{A-2-D-gen-R}
\end{equation}

Thus, by defining the unitary operator

\begin{equation}
\hat{T}_{(\epsilon, 1, \beta_3,\theta)} = \exp\left[ - \arctan \left(  \epsilon \sqrt{\frac{1 -\epsilon \beta_3}{1+ \epsilon \beta_3}} \right)  
\left( e^{-i \theta} \hat{J}_+ - e^{i \theta}  \hat{J}_- \right)
\right], \quad \epsilon=\pm 1, \label{T-unitary-epsilon} 
\end{equation}
which disentangled form, among others,  is
\begin{equation}
\hat{T}_{(\epsilon, 1, \beta_3,\theta)} = \exp\left[ -  \epsilon \sqrt{\frac{1 -\epsilon \beta_3}{1+ \epsilon \beta_3}}  e^{-i \theta} \hat{J}_+ \right]
\exp \left[ \ln\left(   \frac{2}{1 + \epsilon \beta_3}     \right) \hat{J}_{3}  \right]
\exp\left[   \epsilon \sqrt{\frac{1 -\epsilon \beta_3}{1+ \epsilon \beta_3}}  e^{i \theta} \hat{J}_- \right], \quad \epsilon=\pm 1,
\end{equation}
we obtain

\begin{equation}
 \widetilde{\hat{H}}_{2D}  = \hat{T}^\dagger_{(\epsilon,1, \beta_3,\theta)}  \hat{H}_{2D} \hat{T}_{(\epsilon,1, \beta_3,\theta)} = \beta_0 \hat{N} + \epsilon \hat{J}_3, \quad \epsilon= \pm 1,
\end{equation}
 and
\begin{eqnarray}  
  \widetilde{\hat{A}}_{(\vec{u} , \vec{\alpha} )}  = \hat{T}^\dagger_{(\epsilon, 1,\beta_3,\theta)}  \hat{A}_{(\vec{u} , \vec{\alpha} )}   \hat{T}_{(\epsilon, 1,\beta_3,\theta)} &= & \mu_1 \left[  \frac{1}{\sqrt{2 (1 + \epsilon \beta_3) }} \left( 1  +  \epsilon \beta_3  
    +   \frac{4 \epsilon R^2}{2 + \beta_3 - \beta_0}  \right)  \right] \hat{a}_1 \nonumber \\  &+& \mu_1 e^{-i \theta} \left[ \frac{1}{\sqrt{2}} \left( 2 R \frac{\sqrt{1 + \epsilon \beta_3}}{2 + \beta_3 - \beta_0} - \epsilon \sqrt{1 - \epsilon \beta_3} \right)\right] \hat{a}_2  \nonumber \\  &+& \alpha_3
    R e^{- i \theta}  \left[-  \epsilon  +   \frac{1}{2} \frac{1+\epsilon \beta_3}{1 + \beta_3}    +         \frac{1}{2} \frac{1 -\epsilon \beta_3}{1 - \beta_3}                 \right] \hat{J}_+  \nonumber \\ &-& \alpha_3 R e^{i \theta} \left[  \epsilon + \frac{1}{2} \frac{1-\epsilon \beta_3}{1 + \beta_3} +             \frac{1}{2} \frac{1+\epsilon \beta_3}{1 - \beta_3}   \right] \hat{J}_-, \quad \epsilon = \pm 1.
 \end{eqnarray}
Then when $\beta_0 =3$ and $\epsilon=1,$ we get

\begin{equation}
\widetilde{\hat{H}}_{2D} = 3 \hat{N} + \hat{J}_3 = 2 \hat{a}_1^\dagger \hat{a}_1  +  \hat{a}_2^\dagger \hat{a}_2
\end{equation}
and
\begin{equation}
\widetilde{\hat{A}}_{(\vec{u} , \vec{\alpha} )}  = - \sqrt{\frac{2}{1-\beta_3}} \mu_1 e^{- i \theta} \hat{a}_2 - \frac{\alpha_3}{2R} e^{i \theta} \hat{J}_-, 
\end{equation}
i.e., the 2:1 commensurate quantum oscillator system,  and when  $\beta_0 =3$ and $\epsilon=-1,$ we get
\begin{equation}
\widetilde{\hat{H}}_{2D} = 3 \hat{N} - \hat{J}_3 =  \hat{a}_1^\dagger \hat{a}_1  + 2   \hat{a}_2^\dagger \hat{a}_2
\end{equation}
and
\begin{equation}
\widetilde{\hat{A}}_{(\vec{u} , \vec{\alpha} )}  =  \sqrt{\frac{2}{1-\beta_3}} \mu_1 \hat{a}_1  +  \frac{\alpha_3}{2R} e^{- i \theta} \hat{J}_+, 
\end{equation}
i.e, the commensurate 1:2 quantum oscillator system.

\vspace{1cm}
On the other hand, when $\beta_0 = \epsilon=1,$ we get

\begin{equation}
\widetilde{\hat{H}}_{2D} =  \hat{N} +  \hat{J}_3 = \hat{a}_1^\dagger \hat{a}_1  
\end{equation}
and
\begin{equation}
\widetilde{\hat{A}}_{(\vec{u} , \vec{\alpha} )}  = \sqrt{\frac{2}{1+\beta_3}} \mu_1 \hat{a}_1 - \frac{\alpha_3}{2R} e^{i \theta} \hat{J}_-, 
\end{equation}
and  when $\beta_0 = - \epsilon=1,$ we get

\begin{equation}
\widetilde{\hat{H}}_{2D} =  \hat{N} -  \hat{J}_3 = \hat{a}_2^\dagger \hat{a}_2  
\end{equation}
and
\begin{equation}
\widetilde{\hat{A}}_{(\vec{u} , \vec{\alpha} )}  = \sqrt{\frac{2}{1+\beta_3}} \mu_1 e^{-i \theta} \hat{a}_2 + \frac{\alpha_3}{2R} e^{-i \theta} \hat{J}_+, 
\end{equation}

We notice that the   1:2 anisotropic system  is connected to the 2:1 anisotropic system by a unitary similarity transformation. Indeed, the    $\widetilde{\hat{H}}_{2D}|_{1:2} $ Hamiltonian can be obtained from   $\widetilde{\hat{H}}_{2D}|_{2:1} $ in the form:

\begin{equation}
\widetilde{\hat{H}}_{2D}|_{1:2} = \hat{T}^\dagger_{(-1, 1,\beta_3,\theta)} \hat{T}_{(1, 1,\beta_3,\theta)} \widetilde{\hat{H}}_{2D}|_{2:1} \hat{T}^\dagger_{(1,1, \beta_3,\theta)}  \hat{T}_{(-1, 1,\beta_3,\theta)},
\end{equation} 
and the same is true for the transformed ladder operators. It means that the generalized $su(2) \roplus \left\{h(1) \oplus h(1)\right\}$ 2-D Hamiltonian  with its associated ladder operators generates a family of isospectral 2:1 and 1:2 commensurate quantum oscillator systems, all them connected by a unitary transformation. The unitary operator $     \hat{T}^\dagger_{(1, 1,\beta_3,\theta)}  \hat{T}_{(-1,1, \beta_3,\theta)},$  charged of this transformation is a mixing type operator that interchanges the roles of two modes.

\subsubsection{Generalized 2:1 commensurate anisotropic quantum oscillator energy eigenstates}

The computation of the  generalized lowering operator and  of its corresponding  $su(2) \roplus \left\{h(1) \oplus h(1)\right\}$ Hamiltonian eigenstates turns out to be simpler by using the unitary transformation on the states, as we have shown in the previous subsection. If again, we choose $\beta_0=3$ and $\epsilon=1,$   the two  chains of eigenstates of (\ref{A-2-D-gen-R})  can be obtained by acting with the unitary  operator  $\hat{T}_{(1, 1,\beta_3,\theta)}$ on the states   (\ref{lambda-1-base}) and (\ref{lambda-2-base}), respectively,  where the parameter $\mu_2$ and $\alpha_+$ must be replaced by  $\tilde{u}_2 =  - \sqrt{\frac{2}{1-\beta_3}} \mu_1 e^{- i \theta} $
and $\tilde{\alpha}_+ =  - \frac{\alpha_3}{2R} e^{i \theta},$ respectively, that is,

 \begin{equation}
   \mid  \lambda \rangle_1 = \hat{T}_{(1,1, \beta_3 ,\theta)}  \;  \hat{D}_1 \left( \tilde{c}_1 \lambda \right) \hat{S}_2 \left(\tilde{\chi}  \right)  \hat{D}_2 \left( \frac{\lambda}{\tilde{\mu}_2}  \cosh (\|\tilde{\chi} \|\right) \mid  0 , 0 \rangle. 
\end{equation} 
where $$\tilde{\chi}= \frac{\frac{ \lambda \tilde{c}_1 \tilde{\alpha}_+ }{\tilde{\mu}_2} } { \| \frac{ \lambda c_1 \tilde{\alpha}_+ }{\tilde{\mu}_2} \| }  \tanh^{-1} \left(   \| \frac{ \lambda \tilde{c}_1 \tilde{\alpha}_+ }{\tilde{\mu}_2} \|  \right), $$ 
   which must verify   $$\| \frac{ \lambda \tilde{c}_1 \tilde{\alpha}_+ }{\tilde{\mu_2}} \| < 1,$$ 
   and
   \begin{equation}
  \mid  \lambda \rangle_2 =  \hat{T}_{(1, 1,\beta_3,\theta)}  \;    \frac{ \hat{D}_2 \left( \frac{\lambda}{\tilde{\mu}_2}  \right)}{\sqrt{ \frac{1}{2}  + \frac{\| \tilde{\mu}_2 \|^2}{ \| \tilde{\alpha}_+ \|^2}   +  \frac{\| \lambda \|^2}{ \| \tilde{\mu}_2 \|^2}  +  \frac{\| \lambda \|^4 }{4  \| \tilde{\mu}_2\|^4}    }} \left[     \frac{ {\left( \hat{a}^{\dagger}_{2}  + \frac{\lambda^\ast}{ \tilde{\mu}_2^\ast}  \right)}^2 }{2} | 0, 0 \rangle  - \frac{\tilde{\mu}_2}{\tilde{\alpha}_+}  
   | 1 , 0  \rangle  \right].   \end{equation}
   
   \vspace{1.0cm}
   On the other hand, under the same circumstances,  the corresponding energy eigenstates can be computed, in the same way,  from  
   (\ref{E-n-1-tilde}) and  (\ref{E-n-2-tilde}), we get
   
   \begin{equation}
 \mid E_n \rangle_1  = \hat{T}_{(1, 1,\beta_3,\theta )}  \;  \tilde{\mathcal{N}}_{1,n}^{-\frac{1}{2}} \sum_{k=0}^{[\frac{n}{2}]}  \frac{1}{\sqrt{(n-2k)! k!} }       \left(\frac{\tilde{\alpha}^\ast_+  }{2 \tilde{\mu}^\ast_2 }\right)^k   \mid k, n-2 k\rangle,
  \end{equation} where the normalization constant $\tilde{\mathcal{N}}_{1,n}$  is given by
  
\begin{equation}
\tilde{\mathcal{N}}_{1,n} =  \sum_{k=0}^{[\frac{n}{2}]}   \frac{1}{(n-2k)! k!}       \left(\frac{ \|\tilde{\alpha}_+ \|  }{2 \|\tilde{\mu}_2\| }\right)^{2k},
\end{equation}  
  and
 
 \begin{eqnarray}
 \mid E_n \rangle_2 &=&    \hat{T}_{(1, 1,\beta_3,\theta )}    \;   \tilde{\mathcal{N}}_{2,n}^{-\frac{1}{2}} \sum_{k=0}^{[\frac{n}{2}]}  \frac{1}{\sqrt{(n-2k)! k!} }       \left(\frac{\tilde{\alpha}^\ast_+  }{2 \tilde{\mu}^\ast_2 }\right)^k  \left[ \frac{\sqrt{(n-2k+1)!(n-2k+2)!}}{2} \mid k, n-2k+2 \rangle   \right. \nonumber \\ &+&
 \left(\frac{\tilde{\alpha}^\ast_+}{\tilde{\mu}_2^\ast}\right)^2  \sqrt{\frac{(k+1)(k+2)}{2}}  \sqrt{(n-2k-1)!(n-2k)!} \mid k+2, n-2k-2 \rangle \nonumber \\ &+&  \left. \sqrt{2 (k+1)} \left(   
\frac{\tilde{\alpha}^\ast_+}{\tilde{\mu}_2^\ast} (n-2k) - \frac{\tilde{\mu}_2}{\tilde{\alpha}_+}  \right) \mid k+1, n-2k\rangle
  \right],\nonumber \\
   \end{eqnarray}
  where, $\tilde{\mathcal{N}}_{2,n}$ is a normalization constant given by $\tilde{\mathcal{N}}_{2,n} = {}_2\langle E_n \mid  
  E_n \rangle_2.$

\section{External linear coupling terms added }
\label{sec-three}
    \setcounter{equation}{0}
    Let us now  consider a general  Hermitian  Hamiltonian formed from the linear combination of all  generators of the
algebra   $ \left\{ h (1) \oplus h(1) \right\} \roplus u(2)  :$    
   
   \begin{equation} 
   \hat{H}_{(\beta_0,\vec{\beta},\vec{\gamma}_1, \vec{\gamma}_2)} = \beta_{0} \hat{N} + \vec{\beta} \cdot \vec{\hat{J}} +  \gamma_1  \hat{a}_1^\dagger +  \gamma_1^\ast \hat{a}_1 + \gamma_2 \hat{a}_2^\dagger  + \gamma_2^\ast \hat{a}_2  +  h_0  \hat{I}, \label{H-the-most-general}
    \end{equation}
    where $h_0$ is a real constant and $\vec{\gamma}_i, \; i=1,2$ represent the pair $(\gamma_i, \gamma_i^\ast),  \; i=1,2, $ respectively. 
The most general lowering operator $\hat{A}_{(\vec{\mu},\vec{v},\vec{\alpha},a_0)}$ that can be formed from a linear combination of the generators of the above mentioned algebra, which is compatible with the relation

\begin{equation}
\left[ \hat{H}_{(\beta_0,\vec{\beta},\vec{\gamma}_1, \vec{\gamma}_2)} , \hat{A}_{(\vec{\mu},\vec{\nu},\vec{\alpha},a_0)} \right] = - \hat{A}_{(\vec{\mu},\vec{\nu},\vec{\alpha},a_0)}, \label{H-2-D-gen-A-gen}
\end{equation}      
    is
    \begin{equation}
     \hat{A}_{(\vec{\mu},\vec{\nu},\vec{\alpha},a_0)}  = \mu_{1} \hat{a}_1 + \mu_{2} \hat{a}_2 + \nu_{1} \hat{a}_1^\dagger + \nu_{2} \hat{a}_2^\dagger  +  \vec{\alpha} \cdot \vec{\hat{J}} +  a_0  \hat{I}. \label{A-the-most-general}
    \end{equation}
    where $a_0$ is a complex constant. 
    
    \vspace{1.0cm}
    
    Taking into account the semi-direct structure of  Lie algebra we are dealing with, from  (\ref{H-2-D-gen-A-gen}) we deduce that
    
\begin{equation}
\left[\vec{\beta} \cdot \vec{\hat{J}},  \vec{\alpha} \cdot \vec{\hat{J}}\right] = -  \vec{\alpha} \cdot \vec{\hat{J}}, \label{su2-part-relation}
\end{equation}     
 which implies that, as we have seen in the above sections, the underlying algebraic equation system
     \begin{equation}     
     \begin{pmatrix}
 2 \beta_-  & - 2 \beta_+ &  1 \\
1- \beta_3 & 0 &   \beta_+ \\
0& 1 + \beta_3 & - \beta_-  
 \end{pmatrix}
\begin{pmatrix}
\alpha_+ \\ \alpha_-  \\ \alpha_3 \end{pmatrix}= \begin{pmatrix}
0 \\ 0  \\0 
\end{pmatrix} , \label{alpha-equation-system}
\end{equation}    
associated to (\ref{su2-part-relation})  has non-null solutions if and only if    $b^2=  4 \beta_+ \beta_- + \beta_3^2 =1.$

\vspace{1.0cm}

  On the other hand,  from (\ref{H-2-D-gen-A-gen}), equating  the coefficients of the one-mode creation and annihilation operators we get the algebraic equation system
  
\begin{eqnarray}
\left( 1 - \frac{\beta_0 + \beta_3}{2}\right) \mu_1 - \beta_+ \mu_2  &=& -  \frac{\gamma_1^\ast}{2} \alpha_3 - \gamma_2^\ast \alpha_+ \label{eq-1-linear} \\  
 - \beta_- \mu_1  + \left( 1 - \frac{\beta_0 - \beta_3}{2}\right) \mu_2 &=&   \frac{\gamma_2^\ast}{2} \alpha_3 - \gamma_1^\ast \alpha_- \label{eq-2-linear} \\
 \left( 1 + \frac{\beta_0 + \beta_3}{2}\right) \nu_1 +  \beta_- \nu_2  &=&   \frac{\gamma_1}{2} \alpha_3  + \gamma_2 \alpha_-  \label{eq-3-linear} \\  
 \beta_+ \nu_1  + \left( 1 + \frac{\beta_0 - \beta_3}{2}\right) \nu_2 &=&   - \frac{\gamma_2}{2} \alpha_3 + \gamma_1  \alpha_+  \label{eq-4-linear} 
 \end{eqnarray}  
 and equating the identity coefficients we obtain
 
 \begin{equation}
a_0 ( \vec{\gamma} \cdot \vec{\mu}, \vec{\gamma^\ast} \cdot \vec{\nu} )  = \gamma_1 \mu_1 + \gamma_2 \mu_2- \gamma_1^\ast \nu_1 - \gamma_2^\ast \nu_2. \label{a-0-constant}
\end{equation} 
  
  \subsection{The case $ b^2 \neq 1$} 
  When $b^2 \neq 1 $, all the $\alpha$ coefficients are null, i.e.,  the lowering operator becomes $\hat{A}_{(\vec{\mu},\vec{\nu},\vec{0},a_0)},$ with  non   $su(2)$ components. Also, equations (\ref{eq-1-linear}) to (\ref{eq-4-linear}) become uncoupled, that is,
  
\begin{equation}
  \begin{pmatrix}
        \left( 1 - \frac{\beta_0 + \beta_3}{2}\right)   &  - \beta_+             \\
         - \beta_-     &  \left( 1 - \frac{\beta_0 - \beta_3}{2}\right) 
        \end{pmatrix}
\begin{pmatrix}
\mu_1 \\ \mu_2
\end{pmatrix}  =
        \begin{pmatrix}
        0 \\ 0 
        \end{pmatrix} \quad \text{and} \quad
           \begin{pmatrix}
        \left( 1 + \frac{\beta_0 + \beta_3}{2}\right)   &   \beta_-             \\
         \beta_+     &  \left( 1 +  \frac{\beta_0 - \beta_3}{2}\right) 
        \end{pmatrix}
\begin{pmatrix}
\nu_1 \\ \nu_2
\end{pmatrix}  =
        \begin{pmatrix}
        0 \\0 
        \end{pmatrix} . \label{mu-nu-equation-system}
\end{equation}  
   We notice that former  of  these two system of algebraic equations has non-null solutions if and only if 
   
   \begin{equation}
   (2 - \beta_0)^2  = b^2
   \end{equation}
   and the latter if and only if
    
   \begin{equation}
   (2 + \beta_0)^2  = b^2.
   \end{equation}

 \subsubsection{The case $b=0$ two non-interacting quantum oscillators with linear coupling}
 When $b=0,$ we have $\vec{\beta}= \vec{0}$ and $\vec{\alpha}= \vec{0},$ and the only possibility of having non-trivial solutions for  the algebraic  system of equations (\ref{mu-nu-equation-system}) is to make $\beta_0=2$ or $\beta_0 =-2.$  When $\beta_0 =-2,$ $\mu_=\mu_2=0$ and $\nu_1$ and $\nu_2$ can take any complex value, but   the eigenstates of the lowering operator are non-normalizable.   When $\beta_0 =2,$ $\nu_=\nu_2=0$ and $\mu_1$ and $\mu_2$ can take any complex value, and the Hamiltonian and lowering operator are given by
 
 \begin{equation}
 \hat{H}_{(2,\vec{0},\vec{\gamma}_1, \vec{\gamma}_2)} = \hat{a}_1^\dagger \hat{a}_1 +  \hat{a}_2^\dagger \hat{a}_2  +  \gamma_1  \hat{a}_1^\dagger +  \gamma_1^\ast \hat{a}_1 + \gamma_2 \hat{a}_2^\dagger  + \gamma_2^\ast \hat{a}_2  +  h_0  \hat{I}, 
 \end{equation}
 
\begin{equation}
 \hat{A}_{(\vec{\mu},\vec{0},\vec{0},a_0)}  =  \mu_{1} \hat{a}_1 +  \mu_2 \hat{a}_2 + \gamma_1 \mu_1 + \gamma_2 \mu_2,  \label{A-iso-linear}
\end{equation} 
 respectively.
 
 Thus, by performing a unitary similarity transformation of the Hamiltonian with the help of the two-mode separable displacement  unitary operator $\hat{D}_1 (- \gamma_1) \hat{D}_2 (- \gamma_2) $  and choosing $h_0 = - ( \|\gamma_1\|^2 + \| \gamma_2|\|^2),$ we get the transformed Hamiltonian
 
 \begin{equation}
 \hat{H}^{D_{(1,2)}} =  \hat{a}_1^\dagger \hat{a}_1 +  \hat{a}_2^\dagger \hat{a}_2  .\label{H-similarity-d1-d2}
 \end{equation}
 The same unitary similarity transformation  for the lowering operator produces
 
\begin{equation}
\hat{A}^{D_{(1,2)}} = \mu_1 \hat a_1 + \mu_2  \hat{a}_2. \label{a-similarity-d1-d2}
\end{equation} 
 
 Hence, the normalized separable and non-separable eigenstates of  (\ref{a-similarity-d1-d2}) are given by
 
 \begin{equation}
 \widetilde{\mid \lambda \rangle}_1 = \hat{D}_1 \left(\frac{\lambda}{\mu_1} c_1  \right)   \hat{D}_2 \left(\frac{\lambda}{\mu_2} (1-c_1)  \right) \mid 0 , 0 \rangle, 
 \end{equation}
where $c_1$ is arbitrary complex constant, and

 \begin{eqnarray}
 \widetilde{\mid \lambda \rangle}_2 &=& \mathcal{C}_{\lambda}^{\frac{1}{2}} \; \exp\left[ \frac{\lambda}{\mu_1}  \hat{a}_1^\dagger  \right] \; \left( \hat{a}_2^\dagger - \frac{\mu_2}{\mu_1} \hat{a}_1^\dagger \right)^\kappa \; \mid 0 ,0\rangle \nonumber \\  &=&
 \hat{D}_1 \left(  \frac{\lambda}{\mu_1}\right)  \frac{ \sum_{k=0}^{\kappa} \sqrt{\binom{\kappa}{k}} (-1)^k \left( \frac{\mu_2}{\mu_1}\right)^{k}  \frac{ \left(  \hat{a}_1^\dagger + \frac{\lambda^\ast}{\mu_1^\ast}  \right)^k}{\sqrt{k!}}  \mid 0, \kappa-k \rangle                                          }{\left[\sum_{k=0}^{\kappa}    \binom{\kappa}{k}   \left\| \frac{\mu_2}{\mu_1} \right\|^{2k}  \sum_{\ell=0}^{k} \binom{k}{\ell} 
 \frac{1}{\ell !}  \left\| \frac{\lambda}{\mu_1} \right\|^{2 \ell} \right]^{\frac{1}{2}}   } , \quad \kappa \in \mathbb{N}_+,
 \end{eqnarray}
where $\mathcal{C}_\lambda$ is a normalization constant,   respectively. Thus, when $\lambda=0$ the normalized  transformed fundamental states in each case are given by

\begin{equation}
\widetilde{\mid 0 \rangle}_1 = \mid 0 ,0 ,\rangle, \quad \text{and} \quad \widetilde{\mid 0 \rangle}_2 =   \hat{D}_1 \left(  \frac{\lambda}{\mu_1}\right)  \frac{ \sum_{k=0}^{\kappa} \sqrt{\binom{\kappa}{k}} (-1)^k \left( \frac{\mu_2}{\mu_1}\right)^{k}  \mid k, \kappa-k \rangle                                          }{\left[  1 +    \left\| \frac{\mu_2}{\mu_1} \right\|^2\right]^{\frac{\kappa}{2}}   } , \quad \kappa \in \mathbb{N}_+,
\end{equation}
 from where, using (\ref{a-similarity-d1-d2}), we can directly deduce that $E^{(1)}_0 = 0$ and $E^{(2)}_0 = \kappa, \; \kappa=1,2, \ldots.$ Hence, by the fact that $ [ \hat{H}^{D_{(1,2)}},  (\hat{A}^{D_{(1,2)}})^\dagger] = (\hat{A}^{D_{(1,2)}})^\dagger,$ we find that the  eigenvalues of $ \hat{H}^{D_{(1,2)}}$  in the energy  eigenstates    $ \mid E^{(\delta)}_n \rangle =( (\hat{A}^{D_{(1,2)}})^\dagger)^n \mid E^{(\delta)}_{0} \rangle, \; \delta=1,2, $ are given by   $E^{(1)}_{n} =n, \; n=0,1,2,\ldots, $ and $E^{(2)}_{n} = \kappa + n,  \; n=0,1,2, \ldots,  \; \kappa \in \mathbb{N}_+ .$

\vspace{1.0cm}
Now, coming back to the non-transformed system,   we  get the following expressions  for the eigenstates of the displaced lowering operator  (\ref{A-iso-linear}):

\begin{equation}
\mid \lambda \rangle_1 = \hat{D}_1 (- \gamma_1) \hat{D}_2 (- \gamma_2) \hat{D}_1 \left(\frac{\lambda}{\mu_1} c_1  \right)   \hat{D}_2 \left(\frac{\lambda}{\mu_2} (1-c_1)  \right) \mid 0 , 0 \rangle
\end{equation} 
 and
 \begin{equation}
\mid \lambda \rangle_2 = \hat{D}_1 (- \gamma_1) \hat{D}_2 (- \gamma_2)  \hat{D}_1 \left(  \frac{\lambda}{\mu_1}\right)  \frac{ \sum_{k=0}^{\kappa} \sqrt{\binom{\kappa}{k}} (-1)^k \left( \frac{\mu_2}{\mu_1}\right)^{k}  \mid k, \kappa-k \rangle                                          }{\left[  1 +    \left\| \frac{\mu_2}{\mu_1} \right\|^2\right]^{\frac{\kappa}{2}}   } , \quad \kappa \in \mathbb{N}_+.
\end{equation}

   \subsubsection{ The case $b=2,$  1 : (-1)  quantum oscillator isospectral to the one-mode canonical quantum oscillator}
These last conditions are compatibles only when $\beta_0 =0,$ or equivalently when   $b=2.$  In such a case, when $\beta_\pm \neq 0,$  we can write

\begin{equation}
\mu_2 = \frac{2 \beta_-}{2 + \beta_3} \mu_1 \quad \text{and } \quad \nu_2=   \frac{- 2 \beta_+}{2 - \beta_3} \nu_1,
\end{equation}
 and when  $\beta_\pm=0,$
  
  \begin{equation}
 \mu_2=\nu_1 =0,  \quad \text{and} \quad \mu_1 , \nu_2  \quad \text{ arbitrary  when}  \quad \beta_3= 2
  \end{equation}
   \begin{equation}
 \mu_1=\nu_2 =0,  \quad\text{ and } \quad \mu_2 , \nu_1  \quad  \text{ arbitrary  when} \quad \beta_3= - 2.
  \end{equation}

 \vspace{1.0cm}
 The form of the generalized Hamiltonian and lowering operator in the above cases is
 
 \begin{equation}
\hat{H}_{(0,\vec{\beta},\vec{\gamma}_1, \vec{\gamma}_2)} =  \vec{\beta} \cdot \vec{\hat{J}}  
+  \gamma_1  \hat{a}_1^\dagger +  \gamma_1^\ast \hat{a}_1 + \gamma_2 \hat{a}_2^\dagger  + \hat{\gamma}_2^\ast \hat{a}_2  +  h_0  \hat{I}, \label{H-iso-spectral-h-oscillator}
 \end{equation}
and 
 \begin{equation}
     \hat{A}_{(\vec{\mu},\vec{\nu},\vec{0},a_0)}  =  \mu_{1} \left( \hat{a}_1 +  \frac{2 \beta_-}{2 +\beta_3}  \hat{a}_2  \right) +  \nu_{1} \left(\hat{a}_1^\dagger -   \frac{2 \beta_+}{2- \beta_3} \hat{a}_2^\dagger \right)  +  a_0 ( \vec{\gamma} \cdot \vec{\mu}, \vec{\gamma^\ast} \cdot \vec{\nu} )  \hat{I},\label{A-squeezing-1}
    \end{equation}
when $\beta_\pm \neq 0,$ and

\begin{equation}\hat{H}_{(0, \beta_3 \hat{k},\vec{\gamma}_1, \vec{\gamma}_2)} =
\begin{cases}
2 \hat{J}_3 + +  \gamma_1  \hat{a}_1^\dagger +  \gamma_1^\ast \hat{a}_1 + \gamma_2 \hat{a}_2^\dagger  + \hat{\gamma}_2^\ast \hat{a}_2  +  h_0 \hat{I} , \quad \text{if} \quad \beta_3 =2 \\ - 2 \hat{J}_3 + +  \gamma_1  \hat{a}_1^\dagger +  \gamma_1^\ast \hat{a}_1 + \gamma_2 \hat{a}_2^\dagger  + \hat{\gamma}_2^\ast \hat{a}_2  +  h_0 \hat{I}  \quad \text{if} \quad \beta_3 =-2 
\end{cases}, \label{2J-3-hamiltonian-beta-pm=0}
\end{equation}
 and

\begin{equation}
\hat{A}_{(\vec{\mu},\vec{\nu},\vec{0},a_0)} =
\begin{cases}
\mu_1 \hat{a}_1 + \nu_2 \hat{a}_{2}^\dagger +  a_0 ( \vec{\gamma} \cdot \vec{\mu}, \vec{\gamma^\ast} \cdot \vec{\nu} )  \hat{I}
\quad \text{if} \quad \beta_3=2 \\
\mu_2 \hat{a}_2 + \nu_1 \hat{a}_{1}^\dagger +  a_0 ( \vec{\gamma} \cdot \vec{\mu}, \vec{\gamma^\ast} \cdot \vec{\nu} )  \hat{I}
\quad \text{if} \quad \beta_3=-2
\end{cases}, \label{A-lowering-beta-pm=0} 
\end{equation}
 when $\beta_\pm=0.$ 
\vspace{1.0cm}

We notice that when $\beta_\pm \neq 0,$ the system governed by the Hamiltonian (\ref{H-iso-spectral-h-oscillator}), together with
its associated ladder operators, (\ref{A-squeezing-1}) and its corresponding  adjoint, is isospectral to the quatum harmonic oscillator. Indeed, the commutator  

\begin{equation}
\left[ \hat{A}_{(\vec{\mu},\vec{\nu},\vec{0},a_0)} ,        \hat{A}^\dagger_{(\vec{\mu},\vec{\nu},\vec{0},a_0)} \right] = 4 \left( \frac{\|\mu_1 \|^2}{2+\beta_3} - \frac{\|\nu_1 \|^2}{2-\beta_3} \right) \hat{I}.
\end{equation}
Then, by redefining  \begin{equation}  {\hat{\mathcal{A}}_{(\vec{\mu},\vec{\nu},\vec{0},a_0)}} =  \frac{\hat{A}_{(\vec{\mu},\vec{\nu},\vec{0},a_0)}}{\sqrt{ 4 \left( \frac{\|\mu_1 \|^2}{2+\beta_3} - \frac{\|\nu_1 \|^2}{2-\beta_3} \right)   }  },  \end{equation}
we get the oscillator algebra

\begin{equation}
\left[ \hat{H}_{(0,\vec{\beta},\vec{\gamma}_1, \vec{\gamma}_2)}  , {\hat{\mathcal{A}}_{(\vec{\mu},\vec{\nu},\vec{0},a_0)}} \right] = -  {\hat{\mathcal{A}}_{(\vec{\mu},\vec{\nu},\vec{0},a_0)}}, \quad \left[ \hat{H}_{(0,\vec{\beta},\vec{\gamma}_1, \vec{\gamma}_2)}  , {\hat{\mathcal{A}}^\dagger_{(\vec{\mu},\vec{\nu},\vec{0},a_0)}} \right] =  {\hat{\mathcal{A}}^\dagger_{(\vec{\mu},\vec{\nu},\vec{0},a_0)}} 
\end{equation}
 and
  \begin{equation}
\left[ {\hat{\mathcal{A}}_{(\vec{\mu},\vec{\nu},\vec{0},a_0)}} ,   {\hat{\mathcal{A}}^\dagger_{(\vec{\mu},\vec{\nu},\vec{0},a_0)}} \right] =\hat{I}.
\end{equation}  
Thus, the canonical normalized  two-mode eigenstates of $  {\hat{\mathcal{A}}_{(\vec{\mu},\vec{\nu},\vec{0},a_0)}} ,$ associated to  the complex eigenvalue  $\lambda, $ i.e.,  the  coherent states associated to this system, are given by

\begin{equation}
\mid \lambda \rangle = \exp \left[ \lambda \; {\hat{\mathcal{A}}^\dagger_{(\vec{\mu},\vec{\nu},\vec{0},a_0)}} -   \lambda^\ast   \; {\hat{\mathcal{A}}_{(\vec{\mu},\vec{\nu},\vec{0},a_0)}}  \right] \mid \tilde{0} \rangle, 
\end{equation} 
  where $\mid \tilde{0} \rangle $ is the fundamental state of the system, which is an eigenstate of $ {\hat{\mathcal{A}}_{(\vec{\mu},\vec{\nu},\vec{0},a_0)}}$ with $0$ eigenvalue. Furthermore, as in the case of the one-dimensional quantum harmonic oscillator,  the normalized energy eigenstates of  $ \hat{H}_{(0,\vec{\beta},\vec{\gamma}_1, \vec{\gamma}_2)},$ associated to the integer eigenvalue $n,$ are given by
  \begin{equation}
  \mid  E_n \rangle = \mid \tilde{n} \rangle =  \frac{\left({ \hat{\mathcal{A}}^\dagger_{(\vec{\mu},\vec{\nu},\vec{0},a_0)}}\right)^n}{\sqrt{n !}}  \mid \tilde{0} \rangle.
  \end{equation}
 
\vspace{1.0cm} 
  
For example, let us consider  the case where $h_0 $  and all $\gamma's$ in (\ref{H-iso-spectral-h-oscillator}) are equal to $0.$ Then, the lowering operator ${\hat{\mathcal{A}}_{(\vec{\mu},\vec{\nu},\vec{0},a_0)}} $ takes the form

\begin{equation}
{\hat{\mathcal{A}}_{(\vec{\mu},\vec{\nu},\vec{0},0)}} =  \hat{A}_{(\vec{\tilde{\mu}},\vec{\tilde{\nu}},0,0)} =    \tilde{\mu}_{1} \left( \hat{a}_1 +  \frac{2 \beta_-}{2 +\beta_3}  \hat{a}_2  \right) +  \tilde{\nu}_{1} \left(\hat{a}_1^\dagger -   \frac{2 \beta_+}{2- \beta_3} \hat{a}_2^\dagger \right) ,\label{Cal-A-squeezing-1}
\end{equation} 
  where \begin{equation} 
  \tilde{\mu}_{1} = \frac{\mu_1}{2   \sqrt{  \frac{\|\mu_1\|^2}{2+\beta_3} -       \frac{\|\nu_1\|^2}{2-\beta_3} } } \quad  \text{and}
\quad \tilde{\nu}_{1} = \frac{\nu_1}{2   \sqrt{  \frac{\|\mu_1\|^2}{2+\beta_3} -       \frac{\|\nu_1\|^2}{2-\beta_3} } }.
  \end{equation}
 Thus, the eigenvalue equation for this last lowering operator, in the two-dimensional representation of complex analytic  functions (Fock-Bargmann representation) is written as
 
 \begin{equation}
 \left( \tilde{\mu}_1 \frac{\partial}{\partial\xi_1}  + \tilde{\nu}_1  \xi_1 \right) \psi(\xi_1,\xi_2) +
\left(    \frac{2 \beta_-   \tilde{\mu}_1}{2+\beta_3}    \frac{\partial}{\partial\xi_2}  -   \frac{2 \beta_+   \tilde{\nu}_1}{2-\beta_3} \xi_2 \right) \psi(\xi_1,\xi_2) = \lambda  \psi(\xi_1,\xi_2),
 \end{equation} 
 and its separable and non-separable solution are given by
 
\begin{equation}
\psi_1 (\xi_1, \xi_2) =  \psi_1 (0,0) \exp\left[ \frac{(\lambda - \lambda_2)}{\tilde{\mu}_1} \xi_1  - \frac{1}{2} \frac{\tilde{\nu}_1}{\tilde{\mu}_1}  \xi_1^2 \right] \; \exp\left[  \frac{2+\beta_3}{ 2 \beta_- }  \frac{\lambda_2}{\tilde{\mu}_1}  \xi_2  -  \frac{1}{2}  \frac{\beta_3 + 2}{\beta_3 -2} 
\frac{\beta_+ \tilde{\nu_1} }{\beta_- \tilde{\mu}_1}  \xi_2^2 \right]
\end{equation} 
and
 
\begin{equation}
\psi_2 (\xi_1, \xi_2) =   \mathcal{C}  \; \exp\left[ \frac{\lambda}{\tilde{\mu}_1} \xi_1  -  \frac{\tilde{\nu}_1}{\tilde{\mu}_1}  \xi_1^2  - 
\frac{2 \beta_+ }{\beta_3 -2} \frac{\tilde{\nu}_1}{\tilde{\mu}_1}  \xi_1 \xi_2 \right] \left( \xi_2 - \frac{2 \beta_-}{2 + \beta_3} \xi_1 \right),
\end{equation}
respectively, where   $\mathcal{C}$ and  $\lambda_2$ are arbitrary constants\footnote{The choice of the $\lambda_2$ constant will directly influence the structure of the ground state. Depending on this choice, the ground state could turn out to be a separable two-mode squeezed  vacuum or a separable two-mode squeezed coherent state, and vice versa. A good choice for  $\lambda_2$ is to take it  proportional to $\lambda,$ for example, $\lambda_2=\frac{\lambda}{2}. $ }.

Then, coming back to the number representation, these  non-normalized states take the form

\begin{equation}
\mid \lambda \rangle_1 =  \mathcal{N}_1^{- \frac{1}{2}} \exp\left[ \frac{(\lambda - \lambda_2)}{\tilde{\mu}_1} \hat{a}_1^\dagger  - \frac{1}{2} \frac{\tilde{\nu}_1}{\tilde{\mu}_1}  {\hat{a}_1^\dagger}{}^2 \right] \; \exp\left[  \frac{2+\beta_3}{ 2 \beta_- }  \frac{\lambda_2}{\tilde{\mu}_1}  \hat{a}_2^\dagger  -  \frac{1}{2}  \frac{\beta_3 + 2}{\beta_3 -2} 
\frac{\beta_+ \tilde{\nu_1} }{\beta_- \tilde{\mu}_1}  {\hat{a}^\dagger_2}{}^2 \right] \mid 0, 0 \rangle \label{cohe-1-b=2}
\end{equation} 
     and
\begin{equation}
\mid \lambda \rangle_2 =   \mathcal{N}_2^{-\frac{1}{2}} \; \exp\left[ \frac{\lambda}{\tilde{\mu}_1} \hat{a}_1^\dagger  -  \frac{\tilde{\nu}_1}{\tilde{\mu}_1}  {\hat{a}_1^\dagger}{}^2  - 
\frac{2 \beta_+ }{\beta_3 -2} \frac{\tilde{\nu}_1}{\tilde{\mu}_1}  \hat{a}_1^\dagger  \hat{a}_2^\dagger \right] \left( \mid 0,1 \rangle - \frac{2 \beta_-}{2 + \beta_3} \mid 1,0 \rangle\right),  \label{cohe-2-b=2}
\end{equation}
  respectively.
  
  Making  $\lambda_2 = \frac{\lambda}{2}$  in (\ref{cohe-1-b=2} ) and following the standard procedure for constructing unitary displacement and  squeezing operators in this kind of situations, the   normalized coherent state of the type $1,$ are given by 
 \begin{equation}
\mid  \lambda \rangle_1 = \hat{S}_1  \left( \chi_1    \right) \hat{S}_2 \left(  \chi_2  \right) \hat{D}_1 \left(  \frac{\lambda}{2 \tilde{\mu}_1} \cosh \|\chi_1\|   \right) \hat{D}_2 \left(  \frac{\lambda}{2 \tilde{\mu}_1} \frac{2+ \beta_3}{2 \beta_-}  \cosh \|\chi_2\|   \right) \mid 0,0 \rangle, \label{squeezed-iso-gen}
\end{equation}   
   where \begin{equation}  \chi_1 = \frac{ \frac{\nu_1}{\mu_1}}{   \| \frac{\nu_1}{\mu_1}   \|} \arctan \left(  \left\rVert \frac{\nu_1}{\mu_1}   \right\rVert   \right) \quad \text{and }  \quad       \chi_2 = \frac{ \frac{ (\beta_3 +2 )  \beta_+ \nu_1}{(\beta_3-2) \beta_- \mu_1}}{   \left\rVert \frac{ (\beta_3 +2)  \nu_1}{(\beta_3 -2) \mu_1}   \right\rVert} \arctan \left(  \left\rVert \frac{  (\beta_3 +2)   \nu_1}{ (\beta_3 - 2) \mu_1}   \right\rVert   \right), \end{equation}   
   where  $ \left\rVert \frac{\nu_1}{\mu_1}   \right\rVert <1 $ and   $  \left| \frac{\beta_3+2}{\beta_3  - 2} \right|  <1. $ From here, we deduce that the fundamental state of the system is of the type two-mode separable squeezed vacuum:
   \begin{equation}
   \mid \tilde{0} \rangle_1 = \hat{S}_1  \left( \chi_1    \right) \hat{S}_2 \left(  \chi_2  \right)  \mid 0,0 \rangle.
   \end{equation}
   Clearly the states (\ref{squeezed-iso-gen}) represent a realization of the squeezed states defined in  \cite{JM-VH-0}, but now we have that these states are connected to the  Hamiltonian   (\ref{H-iso-spectral-h-oscillator}).

\vspace{1.0cm}  
  
  The normalization constant of the lowering operator eigenstates   (\ref{cohe-2-b=2}) is not easy to compute.  Some simplifications can be done if  first we perform a unitary similarity transformation on the operators of the system. Thus, if we  define the mixing unitary operator
  
  \begin{equation}
  \hat{T}_{(\epsilon, 2, \beta_3, \theta)} = \exp\left[ - \arctan\left( \epsilon \sqrt{\frac{2 - \epsilon \beta_3}{2 + \epsilon \beta_3} } \right)  \left(  e^{- i \theta} \hat{J}_+  - e^{i \theta} \hat{J}_- \right)   \right], \quad \epsilon = \pm 1,    
\end{equation}   
   then we realize that the original Hamiltonian reduces to the difference number operator 
   \begin{equation}
 \hat{H}^{T_{\epsilon}}_ {(0,\vec{\beta},\vec{0}, \vec{0})} =    \hat{T}^\dagger_{(\epsilon, 2, \beta_3, \theta)}\hat{H}_{(0,\vec{\beta},\vec{0}, \vec{0})}  \hat{T}_{(\epsilon, 2, \beta_3, \theta)} = 2 \epsilon \hat{J}_3 \label{H-vec-beta-to-beta-3}
 \end{equation}
   and
   \begin{equation}
  {\hat{\mathcal{A}}^{T_{\epsilon}}_{(\vec{\mu},\vec{\nu},\vec{0},0)}} =   \hat{T}^\dagger_{(\epsilon, 2 ,\beta_3, \theta)} {\hat{\mathcal{A}}_{(\vec{\mu},\vec{\nu},\vec{0},0)}}   \hat{T}_{(\epsilon, 2, \beta_3, \theta)} =\begin{cases}  \frac{2 \tilde{\mu}_1}{\sqrt{2 + \beta_3}} \hat{a}_1 -  \frac{2  e^{i \theta} \tilde{\nu_1}}{\sqrt{2 - \beta_3}} \hat{a}_2^\dagger, \quad \text{if}  \quad \epsilon=1                  \\           \frac{2 \tilde{\mu}_1  e^{-i \theta}}{\sqrt{2 + \beta_3}} \hat{a}_2 +  \frac{2  \tilde{\nu_1}}{\sqrt{2 - \beta_3}} \hat{a}_1^\dagger, \quad \text{if}  \quad \epsilon=-1.         \end{cases} \label{Cal-A-beta-to-beta-3}
 \end{equation}
Then, if we choose $\epsilon =1$ and follow the same procedure that we have just detailed above, we find that the non-separable eigenstates of  lowering operator  $ {\hat{\mathcal{A}}^{T_{1}}_{(\vec{\mu},\vec{\nu},\vec{0},0)}}   $ are given by

\begin{equation}
\mid \lambda \rangle^{T}_2 = \mathcal{N}^{- \frac{1}{2}} \exp\left[    \frac{\sqrt{ 2 + \beta_3}}{2 \tilde{\mu}_1} \lambda \hat{a}_1^\dagger \right] \exp\left[ \frac{\nu_1}{\mu_1}   \sqrt{\frac{2 + \beta_3}{2-\beta_3}} e^{i \theta} \hat{a}_1^\dagger  \hat{a}_2^\dagger \right] \mid 0,1 \rangle.
\end{equation} 

When $\lambda=0,$ we can express the normalized transformed ground state in two ways, as  an infinite sum of states whose number of bosons in the second mode is always one unit greater than the number of bosons in the first mode, or, as an $su(1,1)$ type squeezed operator acting on the state $\mid 0 ,1 \rangle$, that is

\begin{equation}
\mid 0  \rangle^{T}_2 =   \left(  1-  \left\rVert \frac{\nu_1}{\mu_1}   \right\rVert^2   \frac{2 + \beta_3}{2-\beta_3}  \right)  
\sum_{k=0}^{\infty}          \left( \frac{\nu_1}{\mu_1}   \sqrt{\frac{2 + \beta_3}{2-\beta_3}} e^{i \theta} \right)^k           \sqrt{k+1}   \mid k , k+1 \rangle , \label{Ground-state-vec-beta-to-beta-3}
\end{equation} 
where  we suppose  $\left\rVert \frac{\nu_1}{\mu_1}   \right\rVert  <1 $ and  $ \frac{2 + \beta_3}{2-\beta_3} <1,$
or
\begin{equation}
\mid 0  \rangle^{T}_2 =  \hat{S} \left( \tilde{\vartheta}, \tilde{\varphi}    \right)   \mid 0, 1 \rangle
\end{equation} 
where
\begin{equation}
\hat{S} \left( \tilde{\vartheta}, \tilde{\varphi}    \right) = \exp\left[ - \frac{\tilde{\vartheta}}{2}  \left(  e^{-i \tilde{\varphi}}  \hat{a}_1^\dagger \hat{a}_2^\dagger  -  e^{i \tilde{\varphi}}  \hat{a}_1  \hat{a}_2 \right) \right],
\end{equation}
and

\begin{equation}
\tanh \left(\frac{\tilde{\vartheta}}{2} \right) =  \left\rVert \frac{\nu_1}{\mu_1}   \right\rVert  \sqrt{\frac{2 + \beta_3}{2-\beta_3}} <1, \quad \text{and} \quad e^{-i \tilde{\varphi}} =  - \; e^{i \theta} e^{i \arctan \left( i   \frac{\nu_1^\ast \mu_1 -\nu_1 \mu_1^\ast}{\nu_1^\ast \mu_1 +\nu_1 \mu_1^\ast}    \right)}. 
\end{equation}

Let us to use the states  (\ref{Ground-state-vec-beta-to-beta-3}) to generate the transformed energy eigenstates of the original Hamiltonian, i.e., the eigenstates of the transformed  Hamiltonian 
 (\ref{H-vec-beta-to-beta-3}) when $\epsilon=1.$  They are given by
 
\begin{equation}
 \mid E_n  \rangle^T_2 = [\mathcal{N}^T_{(n)}]^{- \frac{1}{2}} \left(\frac{2 \tilde{\mu}_1^\ast}{\sqrt{2 + \beta_3}} \hat{a}_1^\dagger -  \frac{2  e^{- i \theta} \tilde{\nu}_1^\ast }{\sqrt{2 - \beta_3}} \hat{a}_2 \right)^n \mid 0  \rangle^{T}_2.
 \end{equation}

Thus, when $n=0,$   from (\ref{Ground-state-vec-beta-to-beta-3}), we deduce

\begin{equation}
 \hat{H}^{T_{1}}_ {(0,\vec{\beta},\vec{0}, \vec{0})}  \mid 0  \rangle^{T}_2 =  2 \hat{J}_3   \mid 0  \rangle^{T}_2 = -  \mid 0  \rangle^{T}_2,
\end{equation}
and as 
\begin{equation} 
\left[  2 \hat{J}_3 ,  \left( {\hat{\mathcal{A}}^{T_{\epsilon}}_{(\vec{\mu},\vec{\nu},\vec{0},0)}} \right)^n \right] = n \left( {\hat{\mathcal{A}}^{T_{\epsilon}}_{(\vec{\mu},\vec{\nu},\vec{0},0)}} \right)^n,
\end{equation}
it is straightforward to show that
\begin{equation}   2 \hat{J}_3  \mid E_n  \rangle^T_2 = (n-1) \mid E_n  \rangle^T_2, \end{equation}
i.e., the  operator $2 \hat{J}_3,$ which  represent the difference between  the number of bosons in the  first mode  and the number of bosons in the second mode, in this case,    has the same spectrum as the  one-mode canonical quantum harmonic oscillator Hamiltonian, except that the former is shifted in one unity with respect to the latter.       

\vspace{1.0cm}
Let us   finish this section making some comments about the case $\beta_\pm=0,$ and $\beta_3=2.$ By comparing  the system (\ref{2J-3-hamiltonian-beta-pm=0}--\ref{A-lowering-beta-pm=0}) with the system (\ref{H-vec-beta-to-beta-3}--\ref {Cal-A-beta-to-beta-3}) we observe that, for the particular choice of the parameters $h_0=0$ and $\gamma_i=0, \;  i=1,2,$  they have the same structure, then the former leads to the same energy eigestates  and coherent states structures as the latter, i.e,  to two-mode non-separable set of eigenstates. The structure of this set of eigenstates is characterized by the presence of a standard $su(1,1)$ two-mode squeezed operator.

\subsubsection{The case $b^2 \neq 1$ and $\beta_0 \neq 0:$ A family of  generalized fractional commensurate anisotropic 2D  quantum  oscillators}   
   
Let us now study the case where $b^2 \neq 1,$ and $\beta_0 \ne 0.$  In this case, the two system of algebraic equations shown  (\ref{mu-nu-equation-system}) are incompatible,  in the sense that we cannot have non-null solutions for both system  simultaneously. So, if we choose $(2 -\beta_0 )^2 = b^2,$ where $\beta_0 \neq 0,$ then $(2 + \beta_0 )^2 \neq b^2,$ i.e.,    the only solution for these system are  

\begin{equation} 
\mu_2 = \frac{2 \beta_- \;  \mu_1}{2 + \beta_3 - \beta_0}  \quad \text{and } \quad \nu_1 = \nu_2=0    \end{equation}
and when we choose $(2 + \beta_0  )^2 = b^2,$ where $\beta_0 \neq 0,$ then  $(2 - \beta_0)^2 \neq b^2,$ i.e.,    the only solution for these system are  
\begin{equation} \mu_1= \mu_2 =0   \quad \text{and } \quad  \nu_2= -  \frac{2 \beta_+ \;  \nu_1}{2 + \beta_0 - \beta_3}. \end{equation}

As the second of these choices leads to non-normalizable eigenstates for the generalized  lowering operator, we will only treat here the 
first choice. Then, again, if we make $h_0$ and all $\gamma$ parameters equal to $0,$ the generalized Hamiltonian becomes

\begin{equation}
\hat{H}_{(\beta_0, \vec{\beta},\vec{0}, \vec{0})} = \beta_0 \hat{N} + \vec{\beta}  \cdot \vec{\hat{J}}   \label{H-fractional-anisotropic}
 \end{equation}
and 
 \begin{equation}
     \hat{A}_{(\vec{\mu},\vec{0},0,0)}  =  \mu_{1} \left( \hat{a}_1  +  \frac{2 \beta_-}{2 +\beta_3 - \beta_0}  \hat{a}_2  \right) \label{A-fractional-anysotropic},
    \end{equation}
or if  we define $\beta_\pm = R e^{i \pm  \theta},$ they can be written as
\begin{equation}
\hat{H}_{(\beta_0, R, \theta, \beta_3)} = \beta_0 \hat{N} + \frac{\sqrt{b^2 - \beta_3^2}}{2} \left( e^{- i\theta} \hat{J}_+ + e^{i \theta}  \hat{J}_-    \right) + \beta_3 \hat{J}_3   \label{H-fractional-anisotropic-R-theta-beta-3}
 \end{equation}
and 
 \begin{equation}
     \hat{A}_{(\vec{\mu},R,\theta,\beta_3)}  =  \mu_{1} \left( \hat{a}_1  +  \frac{ \sqrt{b^2 - \beta_3^2} \; e^{-i \theta}}{2 +\beta_3 - \beta_0}  \hat{a}_2  \right) \label{A-fractional-anysotropic-R-theta-beta-3},
    \end{equation}
where $b =|2 - \beta_0|. $  Thus, we regain the special case already treated in section \ref{sec-fractional}. Here we will solve the same problem but now we will use a slightly different technique which will give us more  information about the way we can choose the initial conditions at the moment of solving the partial differential equation giving rise to the present lowering operator eigenstates.

\vspace{1,0cm} 

We notice that, according to the imposing condition in this subsection, the  $\beta_0$ parameter can take any real value but must be   different from $0, \pm 1, $ and $ \pm 3.$  

We are interested in computing the algebra eigenstates of  (\ref{A-fractional-anysotropic-R-theta-beta-3}), that is, to solve the eigenvalue equation
    
    \begin{equation}
    \left[ \mu_{1} \left( \hat{a}_1  +  \frac{ \sqrt{b^2 - \beta_3^2} \; e^{-i \theta}}{2 +\beta_3 - \beta_0}  \hat{a}_2  \right) \right] \mid \lambda \rangle =  \lambda \mid \lambda \rangle.
    \end{equation}
   Then, following the same procedure as in the  previous sections and suitably choosing  the initial conditions and/or  the integration constants, we  obtain a set of  separable two-mode normalized states
   
   \begin{equation}
   \mid \lambda \rangle_1 =  \hat{D}_1 \left( \frac{\lambda}{2 \mu_1}\right) \hat{D}_2 \left(  \frac{  \lambda (2 +\beta_3 - \beta_0 )  e^{i \theta} }{2 \mu_1 \; \sqrt{b^2 - \beta_3^2}} \right) \; \mid 0,0 \rangle \label{A-e-states-separable}
   \end{equation}
and a set of non-separable two mode states,

\begin{equation}
\mid \lambda \rangle_2 = \mathcal{N}^{- \frac{1}{2}} e^{ \frac{\lambda}{\mu_1} \hat{a}_1^\dagger} \left[ \mid 0, 1 \rangle - e^{-i \theta} \frac{\sqrt{{b^2 - \beta_3^2}}}{2 +\beta_3 - \beta_0} \mid 1,0\rangle  \right]. \label{A-e-states-non-separable}
\end{equation}
 among others solutions.

Thus, the normalized eigenstates of the lowering operator corresponding to the eigenvalue $0$ are given by

\begin{equation}
\mid \tilde{0} \rangle_1 = \mid 0,0  \rangle \quad \text{and } \quad \mid \tilde{0} \rangle_2= \sqrt{\frac{(2-\beta_0 + \beta_3)^2}{2\left[b^2 - (2-\beta_0) \beta_3\right]}} \left[ \mid 0, 1 \rangle - e^{-i \theta} \frac{\sqrt{{b^2 - \beta_3^2}}}{2 +\beta_3 - \beta_0} \mid 1,0\rangle  \right].
\end{equation} 
Then, as the energy eigenvalue associated to these states are

\begin{equation}
\hat{H}_{(\beta_0, R, \theta, \beta_3)} \mid \tilde{0} \rangle_1 = 0 \mid \tilde{0} \rangle_1
\end{equation}
 and
 \begin{equation}
\hat{H}_{(\beta_0, R, \theta, \beta_3)}  \mid \tilde{0} \rangle_2 = 1  \mid \tilde{0} \rangle_2,
 \end{equation}
respectively, from the constructed  commutation relation $[\hat{H}_{(\beta_0, R, \theta, \beta_3)}, \hat{A}^\dagger_{(\vec{\mu},R,\theta,\beta_3)}   ] =  \hat{A}^\dagger_{(\vec{\mu},R,\theta,\beta_3)}, $ which implies $[\hat{H}_{(\beta_0, R, \theta, \beta_3)}, (\hat{A}^\dagger_{(\vec{\mu},R,\theta,\beta_3)} )^n  ] = n  (\hat{A}^\dagger_{(\vec{\mu},R,\theta,\beta_3)} )^n  $ we deduce that

\begin{eqnarray}
\hat{H}_{(\beta_0, R, \theta, \beta_3)}   (\hat{A}^\dagger_{(\vec{\mu},R,\theta,\beta_3)} )^n   \mid \tilde{0} \rangle_{\ell} &=&
 (\hat{A}^\dagger_{(\vec{\mu},R,\theta,\beta_3)} )^n  \hat{H}_{(\beta_0, R, \theta, \beta_3)}  + n  \;  (\hat{A}^\dagger_{(\vec{\mu},R,\theta,\beta_3)} )^n \mid \tilde{0} \rangle_{\ell}  \\ &=& \left[ (\ell -1) +  n  \right]  (\hat{A}^\dagger_{(\vec{\mu},R,\theta,\beta_3)} )^n \mid \tilde{0} \rangle_{\ell}, \quad \ell=1,2,
\end{eqnarray}
i.e., the  states
\begin{equation}
 \mid E_{n} \rangle_\ell = \left(\mathcal{N}^{(\ell)}_n \right)^{- \frac{1}{2}} (\hat{A}^\dagger_{(\vec{\mu},R,\theta,\beta_3)} )^n   \mid \tilde{0} \rangle_{\ell} , \quad \ell=1,2,
\end{equation}
where $\mathcal{N}^{(\ell)}_n, \; \ell=1,2\; n=1,2,\ldots$  are normalization constants,  are eigenstates of the generalized Hamiltonian (\ref{H-fractional-anisotropic-R-theta-beta-3}) with associated eigenvalues equal to
 $(2-\ell) +n, \; \ell=1,2; n=1,2, \ldots.$ In other words, the Hamiltonian (\ref{H-fractional-anisotropic-R-theta-beta-3}) is isospectral to the canonical one-dimensional quantum oscillator, or almost isospectral in some cases.  
 
\vspace{1.0cm}

Let us now analyze the structure of the generalized Hamiltonian  (\ref{H-fractional-anisotropic-R-theta-beta-3}) by performing a unitary similarity transformation on it. For that, let us define the unitary operator

\begin{equation}
\hat{T}_{(\epsilon,b,\beta_3,\theta)} = \exp\left[ - \arctan\left(\epsilon  \sqrt{\frac{b - \epsilon \beta_3}{b + \epsilon \beta_3} } \left(  e^{- i \theta} \hat{J}_+ - e^{i \theta} \hat{J}_-             \right)       \right)\right] 
\end{equation}
where $b =|2  - \beta_0 |$ and $\epsilon = \pm 1.$ Then,   

\begin{equation}
 \hat{H}^T_{(\beta_0, \beta_3, \epsilon)}  = \hat{T}^\dagger_{(\epsilon,b,\beta_3,\theta)}  \hat{H}_{(\beta_0, \vec{\beta},\vec{0}, \vec{0})}  \hat{T}_{(\epsilon,b,\beta_3,\theta)} = \beta_0 \hat{N} + \epsilon  | 2 - \beta_0| \hat{J}_3,
\end{equation} 
 i.e., the similarity transformation reduces it to a simpler Hamiltonian, which expressed in  the original two-modes creation and annihilation operators look like this
 
 \begin{equation}
  \hat{H}^T_{(\beta_0,\epsilon)} = \frac{1}{2} \left(\beta_0 + \epsilon |2 - \beta_0| \right) \hat{a}_1^\dagger \hat{a}_1 +   \frac{1}{2} \left(\beta_0  - \epsilon |2 - \beta_0|\right) \hat{a}_2^\dagger \hat{a}_2.  \label{H-T-anisotropic-fractional}
 \end{equation}
So, the transformed Hamiltonian represents a two dimensional non-interacting quantum harmonic oscillators of different frequencies $w_1 =  \frac{1}{2} \left(\beta_0 + \epsilon |2 - \beta_0| \right) $ and  $w_2 =  \frac{1}{2} \left(\beta_0 - \epsilon |2 - \beta_0| \right).$
The frequency ratio is given by

\begin{equation}
\frac{w_2}{w_1} =   \frac{\beta_0 - \epsilon |2 - \beta_0|}{\beta_0 + \epsilon |2 - \beta_0|},
\end{equation}
or more explicitly by
\begin{equation}
\frac{w_2}{w_1} =\begin{cases}  \frac{1}{ \beta_0 -1}, \quad \text{if} \quad  \epsilon=1,  \beta_0 \geq 2 \quad \text{or  if} \quad  \epsilon=-1,  \beta_0 < 2 \\ 
 \\ \frac{2 \beta_0 - 2}{2}, \quad \text{if} \quad  \epsilon=1,  \beta_0  <  2, \quad \text{or  if} \quad \epsilon=-1,  \beta_0  \geq   2 
 \end{cases} .
\end{equation}
So, we can say that, in general,  $\hat{H}^T_{(\beta_0, \epsilon)} $ represents a $w_1 : w_2$ 2-D anisotropic quantum harmonic oscillator. Furthermore,   for example, if  $\beta_0 = k,$ where $k$ is an integer such that $k \geq 2,$ and  $k \ne 3,$ and $\epsilon =1,$ the frequency ratio is given by
\begin{equation}
\frac{w_2}{w_1} = \frac{1}{k-1}; \quad k=2,3, \ldots,
\end{equation}
i.e,  $\hat{H}^T_{(\beta_0, \beta_3, 1)} $ represents a $(k-1) : 1$ 2-D commensurate quantum oscillator. Moreover, if $\epsilon=1$ and $\beta_0 = \frac{p}{q},$ where $p$ and $q$ are non-negative integers such that $ p > 2 q,$ with $\frac{p}{q} \neq 3, $ and   $\epsilon =1,$ the frequency ratio becomes

\begin{equation}
\frac{w_2}{w_1} = \frac{q}{p-q},
\end{equation}
i.e.,  $\hat{H}^T_{(\beta_0, \beta_3, 1)} $ represents a $(p-q) : q$ 2-D commensurate quantum oscillator. Also, when $\beta_0 = \frac{p}{q},$ where $p$ and $q$ are non-negative integers such that $  1 < \frac{ p}{  q} < 2,$  and $\epsilon =1,$ the frequency ratio becomes 
\begin{equation}
\frac{w_2}{w_1} = \frac{p-q}{q},
\end{equation}
i.e.,  $\hat{H}^T_{(\beta_0, \beta_3, 1)} $ represents a $q :(p- q)$ 2-D commensurate quantum oscillator. The analysis with $\epsilon =-1$ is similar, but we do not show the details here.

\vspace{1.0cm}

On the other hand, when $\epsilon =1,$ the unitary similarity transformation on the lowering operator  (\ref{A-fractional-anysotropic-R-theta-beta-3}) produces   \begin{equation} 
 \hat{A}^T_{(\vec{\mu},R,\theta,\beta_3)}   =   \hat{T}^\dagger_{(1,b,\beta_3,\theta)}  \hat{A}_{(\vec{\mu},R,\theta,\beta_3)} \hat{T}_{(1,b,\beta_3,\theta)} = \begin{cases}
\sqrt{\frac{2b}{b+ \beta_3}}  \mu_1 \hat{a}_1, \quad \quad  \text{when} \quad \beta_0 < 2 \\ 
-  \sqrt{\frac{2b}{b- \beta_3}} \; e^{-i \theta} \; \mu_1 \hat{a}_2, \quad \text{when} \quad \beta_0  \geq 2.
 \end{cases} \label{A-T-1-splited}    \end{equation}
    Hence, the algebra eigenstates of the transformed lowering operator can  be  computed by using  the Fock-Bargmann representation space of complex  analytic functions to express the operators and states, doing that we obtain the following partial differential equations with suitably initial conditions.
    
\begin{equation}
\sqrt{\frac{2b}{b+ \beta_3}}  \mu_1 \frac{\partial \psi_< }{\partial \xi_1}(\xi_1 , \xi_2) = \lambda   \psi_{<}(\xi_1 , \xi_2),  \quad \psi_{<} (0, \xi_2) = \varphi_<  (\xi_2),  \quad \text{ if} \quad  \beta_0 <2,
\end{equation}    
and
    \begin{equation}
- \sqrt{\frac{2b}{b- \beta_3}}   e^{-i \theta} \mu_1 \frac{\partial  \psi_{\geq} }{\partial \xi_2}(\xi_1 , \xi_2) = \lambda   \psi_{\geq}(\xi_1 , \xi_2), \quad \psi_{\geq}(\xi_1 ,0) = \varphi_{\geq} (\xi_1),  \quad \text{if} \quad  \beta_0 \geq 2. 
\end{equation}
       
The solution of these elementary partial differential equations are given by
       
       \begin{equation}
       \psi_{<} (\xi_1 ,\xi_2) = \exp\left[ \sqrt{\frac{ b +\beta_3}{2b}} \frac{\lambda}{\mu_1} \xi_1 \right] \varphi_<  (\xi_2)
       \end{equation}
and

 \begin{equation}
       \psi_{\geq} (\xi_1 ,\xi_2) = \exp\left[ - \sqrt{\frac{ b - \beta_3}{2b}}  e^{i \theta} \frac{\lambda}{\mu_1} \xi_2 \right] \varphi_\geq  (\xi_1),
       \end{equation}
respectively, whose form in the number representation space is given by

 \begin{equation}
      \mid \lambda \rangle_<  =   \exp\left[ \sqrt{\frac{ b +\beta_3}{2b}} \frac{\lambda}{\mu_1} \hat{a}_1^\dagger \right] \varphi_<  (\hat{a}_2^\dagger) \mid 0,0 \rangle \label{AT-e-states-beta-0-little}
       \end{equation}
and

 \begin{equation}
    \mid \lambda \rangle_\geq     = \exp\left[ - \sqrt{\frac{ b - \beta_3}{2b}}  e^{i \theta} \frac{\lambda}{\mu_1} \hat{a}_2^\dagger \right] \varphi_\geq  (\hat{a}_1^\dagger) \mid 0,0 \rangle, \label{AT-e-states-beta-0-greater}
       \end{equation}
respectively.

\vspace{1.0cm}

We notice that, when $ \beta_0 < 2, $ the two-mode separable solution (\ref{A-e-states-separable}) corresponds to the choice 
 
\begin{equation}
\varphi_<  (\hat{a}_2^\dagger) = \exp\left[ \frac{\beta_3 e^{i \theta}}{\sqrt{2b (b -\beta_3) }} \frac{\lambda}{\mu_1} \hat{a}_2^\dagger \right]
\end{equation} 
in equation (\ref{AT-e-states-beta-0-little}),
and when $\beta_0 \geq 2,$ it corresponds to the choice
\begin{equation}
\varphi_\geq  (\hat{a}_1^\dagger) = \exp\left[ \frac{\beta_3}{\sqrt{2b (b + \beta_3) }} \frac{\lambda}{\mu_1} \hat{a}_1^\dagger \right],
\end{equation}
 in (\ref{AT-e-states-beta-0-greater}),
 once we apply the inverse transformation on the states, of course. 
 
 Also, when $ \beta_0 < 2, $ the two-mode non-separable solution (\ref{A-e-states-non-separable}) corresponds to the choice 
 
 \begin{equation}
\varphi_<  (\hat{a}_2^\dagger) = \sqrt{\frac{2b}{b+\beta_3}} \;  \exp\left[ -   e^{i \theta} \sqrt{\frac{b -\beta_3}{2b}} \frac{\lambda}{\mu_1} \hat{a}_2^\dagger \right] \; \hat{a}_2^\dagger
\end{equation} 
in equation (\ref{AT-e-states-beta-0-little}),
and when $\beta_0 \geq 2,$ it corresponds to the choice
\begin{equation}
\varphi_\geq  (\hat{a}_1^\dagger) =   \sqrt{\frac{2b}{b-\beta_3}}  \; e^{-i \theta}\;  \exp\left[ \sqrt{\frac{b + \beta_3}{2b}} \frac{\lambda}{\mu_1} \hat{a}_1^\dagger \right] \; \hat{a}_1^\dagger
\end{equation}
 in (\ref{AT-e-states-beta-0-greater}),
 once we apply the inverse transformation on the states. 
 
 \vspace{1.0cm}
 Other choices of $\varphi_<  (\hat{a}_2^\dagger) $ and $ \varphi_\geq  (\hat{a}_1^\dagger)$ are also possible, the only limit we have a at this stage  is that the states   (\ref{AT-e-states-beta-0-little}) and (\ref{AT-e-states-beta-0-greater}) be normalizable. In general, as we have shown in the particular cases just seen above,   these function can also depend on $\lambda.$ Thus, if we  write   $\varphi_<  (\lambda, \hat{a}_2^\dagger) $ and  $ \varphi_\geq  (\lambda, \hat{a}_1^\dagger),$ the transformed fundamental states, which are eigenstates of the lowering operator described in (\ref{A-T-1-splited}), corresponding to the $\lambda=0$ eigenvalue, are given by
 
 \begin{equation}
  \mid \tilde{0} \rangle_{T} = \begin{cases} \varphi_<  (0, \hat{a}_2^\dagger) \mid 0,0 \rangle, \quad \text{if} \quad \beta_0 <2, \\
     \varphi_\geq  (0, \hat{a}_1^\dagger) \mid 0, 0 \rangle, \quad \text{if}  \quad \beta_0 \geq 2 \end{cases}.
       \end{equation}
 
 From these last states we can compute the non-normalized energy eigenstates of the transformed  Hamiltonian (\ref{H-T-anisotropic-fractional}). They should be given by 
 
 \begin{equation}
 \mid \tilde{E}_n \rangle^T = \left( \hat{A}^T_{(\vec{\mu},R,\theta,\beta_3)} \right)^n    \mid \tilde{0} \rangle_{T} =  \begin{cases} \varphi_<  (0, \hat{a}_2^\dagger) \mid n, 0 \rangle, \quad \text{if} \quad \beta_0 <2, \\
     \varphi_\geq  (0, \hat{a}_1^\dagger) \mid 0, n \rangle, \quad \text{if}  \quad \beta_0 \geq 2 \end{cases}, \label{En-T}
 \end{equation}
  where some parameters have been absorbed by the  $\varphi$'s functions. Thus, by acting with the Hamiltonian on these states we get
  
  \begin{equation}
   \hat{H}^T_{(\beta_0,1)} \mid \tilde{E}_n \rangle^T =   \begin{cases}   n  \;   \varphi_<  (0, \hat{a}_2^\dagger) \;  \mid n, 0 \rangle + w_2 \;  \hat{a}_2^\dagger \;  \frac{\partial \varphi_< }{\partial \hat{a}_2^\dagger}  (0, \hat{a}_2^\dagger) \; \mid n,0\rangle , \quad \text{if} \quad \beta_0 <2, \\
  w_1 \;  \hat{a}_1^\dagger \;  \frac{\partial \varphi_\geq }{\partial \hat{a}_1^\dagger}  (0, \hat{a}_1^\dagger) \;     \mid  0 , n \rangle +  n  \; \varphi_\geq  (0, \hat{a}_1^\dagger) \mid 0,n \rangle, \quad \text{if}  \quad \beta_0 \geq 2 \end{cases},  \label{H-T-En-E-En}
  \end{equation} 
  which are consistent with the requirement of being eigenstates  of  $\hat{H}^T_{(\beta_0,1)}$  if and only if 
 
 \begin{equation}
 \frac{\partial \varphi_< }{\partial \hat{a}_2^\dagger}  (0, \hat{a}_2^\dagger) =  c_{<} \; \;  \varphi_<  (0, \hat{a}_2^\dagger) \quad \text{ and} \quad   \frac{\partial \varphi_\geq }{\partial \hat{a}_1^\dagger}  (0, \hat{a}_1^\dagger) = c_{\geq} \; \; \varphi_\geq  (0, \hat{a}_1^\dagger) ,
 \end{equation}
 where $ c_{<}$ and $c_{\geq}$ are real constants, i.e,
 
 \begin{equation}
 \varphi_<  (0, \hat{a}_2^\dagger) = (\hat{a}_2^\dagger)^{c_{<}} \quad \text{and }
 \quad  \varphi_\geq  (0, \hat{a}_1^\dagger) = (\hat{a}_1^\dagger)^{c_{\geq}} .   
 \end{equation} 
Finally, inserting this last results into (\ref{En-T}) and  (\ref{H-T-En-E-En}), we get

 \begin{equation}
 \mid \tilde{E}_n \rangle^T =    \begin{cases} (\hat{a}_2^\dagger)^{c_{<}} \;  \mid n, 0 \rangle, \quad \text{if} \quad \beta_0 <2, \\
     ( \hat{a}_1^\dagger)^{c_{\geq}} \; \mid 0, n \rangle, \quad \text{if}  \quad \beta_0 \geq 2 \end{cases} \label{E-n-T-final}
 \end{equation}
 and
 \begin{equation}
   \hat{H}^T_{(\beta_0,1)} \mid \tilde{E}_n \rangle^T =   \begin{cases}   ( n   + w_2 \;  c_{<}  )  \mid \tilde{E}_n \rangle^T, \quad \text{if} \quad \beta_0 <2, \\
 ( w_1 \;  c_{\geq} +  n  )  \mid \tilde{E}_n \rangle^T, \quad \text{if}  \quad \beta_0 \geq 2 \end{cases}. 
  \end{equation} 
We therefore observe that the energy spectrum in both situations, i.e., when $\beta_0 <2$ and when $\beta_0 \geq 2$ is isospectral to the one dimensional quantum oscillator up to a constant. In case that  $w_1$ and $w_2$ be integer or fractional numbers, the energy spectrum represents those of a commensurate 2-D quantum harmonic oscillator. As for the $c_{<}$ and $c_{\geq}$  constants, they should preferably be chosen as non-negative integers.  Finally, the energy eigenstates of the original Hamiltonian (\ref{H-fractional-anisotropic-R-theta-beta-3}) can be computed by acting with the unitary operator $\hat{T}_{(1,b,\beta_3,\theta)} $ on the states (\ref{E-n-T-final}), i.e., 
$\mid E_n \rangle = \hat{T}_{(1,b,\beta_3,\theta)}  \mid \tilde{E}_n \rangle^T.$

\subsection{Generalized 2:1 commensurate quantum oscillator: case $ b^2=1$}
When $ b^2= 4 \beta_+ \beta_- + \beta_3^2 =1,$  the $\alpha$ parameters could be taken different from $0$,  then various   situations can occur. Indeed, from the condition just  mentioned and from  equation (\ref{alpha-equation-system}), we deduce that  when $\beta_3 =0,$ 
\begin{equation} 
\beta_{\pm} = \frac{ e^{\pm i \theta} }{2} \quad \text{  and } \quad \alpha_{\pm} = \mp \; \frac{ e^{\pm  i \theta}}{2} \alpha_{3},
   \end{equation}
when $\beta_{\pm} =0,$   
    
\begin{equation}
    \begin{cases} \alpha_- = \alpha_3 = 0, \quad \alpha_{+} \neq 0, \quad\text{ if} \quad  \beta_3 =1 \\  \alpha_+ = \alpha_3 = 0, \quad \alpha_{-} \neq 0, \quad\text{ if}  \quad \beta_3 =-1\end{cases} \label{beta-pm-conditions}
    \end{equation} 
  and  when
$\beta_\pm  \neq 0$ and $\beta_3 \neq 0,$  
\begin{equation}
\alpha_{+} = - \frac{\beta_+}{1-\beta_3} \alpha_3 \quad\text{ and}  \quad \alpha_{-} =  \frac{\beta_-}{1+\beta_3}  \alpha_3 .  \label{alpha-minus-plus}
\end{equation}   

Let's start with the analysis of the second of these situations and then move on to the last one, because the first one can be obtained from the last one by making $\beta_3=0.$

\subsubsection{Generalized two-dimensional  2:1  commensurate anisotropic quantum oscillator with linear coupling}
When $\beta_{\pm}=0,$ then from the above condition for $b^2$ we have  $\beta_3= \pm 1.$  Then by inserting  the corresponding values of $\alpha$ and $\beta_3,$   which are summarized  in equation
 (\ref{beta-pm-conditions}),  into  the algebraic equation system described in  (\ref{eq-1-linear}--\ref{eq-4-linear}), we get  30  different forms of expressing the set of $\mu$ and  $\nu$ type parameters in terms of the $\alpha$ and $\gamma$ parameters, for given values of $\beta_0$ and $\beta_3,$ 15 for each allowed value of $\beta_3.$ These relations are shown in appendix \ref{appa}. In six of these relations,  three  for each allowed value of $\beta_3,$ the allowed value of $\beta_0$ is arbitrary but different from $\{\pm1,\pm3\}.$ This fact could give raise to different families of generalized  two-dimensional anisotropic  quantum oscillator systems, and  in particular,  of generalized two-dimensional fractional commensurate  anisotropic quantum oscillator systems. Unfortunately, the eigenstates of the corresponding generalized   lowering operator seem to be non-normalizable in the usual way.  We will therefore  focus on the particular case  described in equation (\ref{A-17}), which represents a 2:1 2-D commensurate anisotropic quantum oscillator with linear coupling, that is the case when $\beta_0=3$ and $\beta_3=1,$ with arbitrary nonzero $\gamma$ parameters:

\begin{eqnarray}
 \hat{H}_{3, 1, \gamma_1,  \gamma_2} &=& 3 \hat{N} + \hat{J}_3 + \gamma_1 \hat{a}_1^\dagger + \gamma_1^\ast \hat{a}_1 +  \gamma_2 \hat{a}_2^\dagger + \gamma_2^\ast \hat{a}_2,  \label{H-A-17-2:1} \\
 \hat{A}_{3,1,\gamma_1,\gamma_2, \alpha_+} & =&  \gamma_2^\ast \alpha_+  \hat{a}_1 + \mu_2 \hat{a}_2  + \frac{\gamma_1 \alpha_+}{2} \hat{a}_2^\dagger + \alpha_+ \hat{J}_-  +  
 \left( \gamma_2 \mu_2 +  \frac{\gamma_1 \gamma_2^\ast \alpha_+}{2}  \right)  \hat{I}. \label{A-A-17-2:1}
 \end{eqnarray}
 
 Let us now to compute the eigenstates of ladder operator ( \ref{A-A-17-2:1}), that is,
 
\begin{equation} 
 \hat{A}_{3,1,\gamma_1,\gamma_2, \alpha_+} \mid \lambda \rangle=  \lambda \mid \lambda \rangle, \quad 
 \lambda \in \mathbb{C}.  \label{A-3-1-gamma-1-gamma-2-alpha-plus}
\end{equation}
To solve this equation, let us first write it in the two-dimensional  Fock-Bargmann representation space of complex analytic functions, that is,
\begin{equation}
  \alpha_+ \left( \gamma_2^\ast  +  \xi_2  \right)   \frac{\partial \psi}{\partial \xi_1} (\xi_1 , \xi_2) +  \mu_2  \frac{\partial \psi}{\partial \xi_2} (\xi_1 , \xi_2)   + \left(  \frac{\gamma_1 \alpha_+}{2} \xi_2  +  \gamma_2 \mu_2 +  \frac{\gamma_1 \gamma_2^\ast \alpha_+}{2}  -\lambda \right)  \psi(\xi_1,\xi_2)   =0, \quad \psi (0, \xi_2) = \phi_{2} (\xi_2), \label{A-lambda-psi=lambda-psi}
\end{equation}
where by convenience  we have added the initial condition $ \psi(0, \xi_2) = \phi_{2} (\xi_2),$ where $ \phi_{2} (\xi_2)$ is an arbitrary function of $\xi_2,$ which will influence directly  the form that the resulting fundamental state of the system will take.

\vspace{1.0cm}
The choice $\psi(0, 0) = 1,$   allows us express the analytic  function eigenstate   in the separable form $\psi (\xi_1, \xi_2) = \varphi_1 (\xi_1) \varphi_2 (\xi_2) \varphi_{2}.$ Then, by substituting this last expression into  (\ref{A-lambda-psi=lambda-psi}),  using the standard  techniques of resolution of linear partial differential equations, making  the following substitution during the development $\frac{\frac{d \varphi_1}{d \xi_1} (\xi_1)}{\varphi_{1} (\xi_1)}= c_1,$ where $c_1$ is an arbitrary complex  constant,  and then integrating the resulting first order linear differential equations in the variables $\xi_1$ and  $\xi_2 $, we get the function

\begin{equation}
\psi_{1} (\xi_1, \xi_2) = \exp\left[ c_1 \xi_1 \right] \exp\left[   \frac{1}{\mu_2}  \left( (\lambda-   \alpha_+ \gamma_2^\ast  c_1 )  -  (\gamma_2 \mu_2 + \frac{\gamma_1 \gamma_2^\ast \alpha_+} {2} ) \right) \xi_2  - \frac{\alpha_+}{2 \mu_2}  \left( c_1 \lambda + \frac{\gamma_1}{2}   \right) \xi_2^2   \right],
\end{equation}
whose form in the two-mode number representation is given by

\begin{equation}
\mid \lambda \rangle_1 = \mathcal{N}_1^{- \frac{1}{2}} \;  \exp\left[c_1 \hat{a}_1^\dagger \right] \exp\left[   \frac{1}{\mu_2}  \left( (\lambda -   \alpha_+ \gamma_2^\ast  c_1 )  -  (\gamma_2 \mu_2 + \frac{\gamma_1 \gamma_2^\ast \alpha_+} {2} ) \right)  
\hat{a}_2^\dagger  - \frac{\alpha_+}{2 \mu_2}  \left( c_1 + \frac{\gamma_1}{2}   \right) {\hat{a}_2^\dagger}{}^2   \right] \mid 0,0\rangle,
\end{equation}
 which represent a set of two-mode separable eigenstates of the ladder operator $ \hat{A}_{3,1,\gamma_1,\gamma_2, \alpha_+},$ having the  form of a canonical coherent state in the mode one and of a squeezed coherent state in the mode two. We observe that these states reduce to (\ref{A-cano-1-cano-squ-2}) when all $\gamma$ parameters are equal $0.$
 
 \vspace{1.0cm}
 The fundamental state candidate, which we define as the eigenstate of  (\ref{A-3-1-gamma-1-gamma-2-alpha-plus}) associated to the eigenvalue 
 $\lambda =0,$ is given by
 
\begin{equation}
\mid \tilde{0}  \rangle_1 = \mathcal{N}_1^{- \frac{1}{2}} \;  \exp\left[c_1 \hat{a}_1^\dagger \right] \exp\left[  -  \frac{1}{\mu_2} \left( \gamma_2^\ast  \alpha_+  c_1       +  (\gamma_2 \mu_2 + \frac{\gamma_1 \gamma_2^\ast \alpha_+} {2} ) \right)  
\hat{a}_2^\dagger  - \frac{\alpha_+}{2 \mu_2}  \left( c_1 + \frac{\gamma_1}{2}   \right) {\hat{a}_2^\dagger}{}^2   \right] \mid 0,0\rangle,
\end{equation}
is also an eigenstate of  the Hamiltonian (\ref{H-A-17-2:1}), if and only if,  $c_1 = - \frac{\gamma_1}{2}. $
Thus, the true fundamental state of the system is

\begin{equation}
\mid \tilde{0}  \rangle_1 = \mathcal{N}_1^{- \frac{1}{2}} \;  \exp\left[- \frac{\gamma_1}{2} \hat{a}_1^\dagger \right] \exp\left[  - \gamma_2  
\hat{a}_2^\dagger   \right] \mid 0,0\rangle,
\end{equation}
 which assumes the normalized form
\begin{equation} 
 \mid \tilde{0}  \rangle_1 = \hat{D}_1 \left( - \frac{\gamma_1}{2} \right) \hat{D}_2 \left( - \gamma_2 \right) \mid 0,0\rangle, \label{2:1-linear-commensurate-separable-coherent-states}
 \end{equation}
i.e., a set of  separable two-mode coherent states.
 
 \vspace{1.0cm}

We observe that
\begin{equation}
 \hat{H}_{3, 1, \gamma_1,  \gamma_2}  \mid \tilde{0}  \rangle_1 =  - \left(  \frac{\|\gamma_1\|^2}{2}  + \|\gamma_2\|^2 \right)   \mid \tilde{0}  \rangle_1,
\end{equation}
i.e, according to the freedom we have to choose the parameter $h_0$ in equation  (\ref{H-the-most-general}), which does not modify the general structure of the system in any way,  if we add   $  \left(\frac{\|\gamma_1\|^2}{2}  + \|\gamma_2\|^2  \right) \hat{I}$  to the Hamiltonian (\ref{H-A-17-2:1}), we realize that the quantum system governed by the Hamiltonian  $\hat{H}_{3, 1, \gamma_1,  \gamma_2} +  \left(\frac{\|\gamma_1\|^2}{2}  + \|\gamma_2\|^2 \right) \hat{I}, $ together with the set of ladder operators $\hat{A}_{3,1,\gamma_1,\gamma_2, \alpha_+} $ and $\hat{A}^\dagger_{3,1,\gamma_1,\gamma_2, \alpha_+} ,$  has a discrete energy spectrum which is identical to that of the canonical  quantum harmonic oscillator with associated energy eigenstates given by

\begin{equation}
\mid E_n \rangle_1 = \mathcal{N}_{n,1}^{-\frac{1}{2}} \left( \hat{A}^\dagger_{3,1,\gamma_1,\gamma_2, \alpha_+}\right)^n   \mid \tilde{0}  \rangle_1. \label{En-2:1-commensurate-linear-1}
\end{equation}   

\vspace{1.0cm}
By writing the raisin operator $ \hat{A}^\dagger_{3,1,\gamma_1,\gamma_2, \alpha_+}$ in the form

\begin{equation}
\hat{A}^\dagger_{3,1,\gamma_1,\gamma_2, \alpha_+} = \hat{\Delta}_1 (\hat{a}_1^\dagger, \hat{a}_2^\dagger) +  \hat{\Delta}_2 (\hat{a}_1^\dagger, \hat{a}_2),
\end{equation}
where
\begin{equation}
\hat{\Delta}_1 (\hat{a}_1^\dagger, \hat{a}_2^\dagger) = \left( \gamma_2^\ast \mu_2^\ast + \frac{\gamma_1^\ast \gamma_2 \alpha_+^\ast}{2}  \right) \hat{I} +  \gamma_2 \alpha_{+}^\ast \hat{a}_1^\dagger + \mu_2^\ast \hat{a}_2^\ast, \quad \text{and} \quad
  \hat{\Delta}_2 (\hat{a}_1^\dagger, \hat{a}_2) = \alpha_+^\ast \left(   \hat{a}_1^\dagger + \frac{ \gamma_1^\ast}{2}   \right) \hat{a}_2,
\end{equation}   
satisfy the commutation relation
\begin{equation}
\left[  \hat{\Delta}_2 (\hat{a}_1^\dagger, \hat{a}_2),\hat{\Delta}_1 (\hat{a}_1^\dagger, \hat{a}_2^\dagger)   \right] = \alpha_+^\ast \mu_2^\ast \left(   \hat{a}_1^\dagger + \frac{ \gamma_1^\ast}{2}   \right) \hat{I},
\end{equation}
and adapting suitably the results of section \ref{anisotropic-2:1-normal-ordering}, in particular equation (\ref{power-n-A-dagger-ordering}),  we get the normal ordering operator

   \begin{equation}
 \left( \hat{A}^\dagger_{3,1,\gamma_1,\gamma_2, \alpha_+}\right)^n = \sum_{k=0}^{[\frac{n}{2}]}    \frac{n!}{(n-2k)!} \frac{\left[
 \alpha_+^\ast \mu_2^\ast \left(   \hat{a}_1^\dagger + \frac{ \gamma_1^\ast}{2}   \right) 
 \right]^k}{2^k \, k!}   \left( \hat{A}^\dagger_{3,1,\gamma_1,\gamma_2, \alpha_+}\right)^{n-2k}_{\lozenge}. \label{power-n-A-2:1-commensurate-linear-dagger-normal-ordering}
 \end{equation}
Thus, by inserting this last result into equation  (\ref{En-2:1-commensurate-linear-1}), taking into account the structure of the  
two-mode fundamental state  (\ref{2:1-linear-commensurate-separable-coherent-states}) and using the standard properties of the harmonic oscillator unitary displacement  operators,  we obtain the following expression for the energy eigenstates of this system: 

\begin{equation}
\mid E_n \rangle_1 = \mathcal{N}_{n,1}^{-\frac{1}{2}} =  \mathcal{N}_{n,1}^{-\frac{1}{2}} \hat{D}_1 \left( - \frac{\gamma_1}{2} \right) \hat{D}_2 \left( - \gamma_2 \right)  \sum_{k=0}^{[\frac{n}{2}]}   \frac{n!}{(n-2k)!} \frac{\left(
 \alpha_+^\ast \mu_2^\ast   \hat{a}_1^\dagger\right)^k}{2^k \, k!}   \left(\mu_2^\ast \hat{a}_2^\dagger \right)^{n-2k} \mid 0,0 \rangle. 
\end{equation}
Finally, using the results  (\ref{A-normal-n-2k-onto-state-0-0} -- \ref{N-1-n-normalization-constant}), we get the normalized energy eigenstates 
\begin{equation}
\mid E_n \rangle_1 =  \hat{D}_1 \left( - \frac{\gamma_1}{2} \right) \hat{D}_2 \left( - \gamma_2 \right) \frac{ \sum_{k=0}^{[\frac{n}{2}]}  \frac{1}{\sqrt{(n-2k)! k!} }       \left(\frac{\alpha^\ast_+  }{2 \mu^\ast_2 }\right)^k   \mid k, n-2 k\rangle}{\sqrt{  \sum_{k=0}^{[\frac{n}{2}]}   \frac{1}{(n-2k)! k!}  \left(\frac{ \|\alpha_+ \|  }{2 \|\mu_2\| }\right)^{2k} }}. 
\end{equation}

\vspace{1.0cm}
The above procedure suggest us that there exist a unitary similarity transformation that can bring our Hamiltonian and ladder operators to a simpler structure. Indeed, we can show directly that

 \begin{equation}
\hat{H}_{3,1,0,0} =    \hat{D}^\dagger_1 \left( - \frac{\gamma_1}{2} \right) \hat{D}^\dagger_2 \left( -\gamma_2 \right)  \left[\hat{H}_{3,1,\gamma_1,\gamma_2}  +  \left(\frac{\|\gamma_1\|^2}{2}  + \|\gamma_2\|^2 \right) \hat{I}   \right]  \hat{D}_1 \left( - \frac{\gamma_1}{2} \right) \hat{D}_2 \left( -\gamma_2 \right)  =  2 \hat{a}_1^\dagger \hat{a}_1 + \hat{a}_2^\dagger \hat{a}_2
\end{equation}    and

\begin{equation}
\hat{A}_{3,1,0,0,\alpha_+} =  \hat{D}^\dagger_1 \left( - \frac{\gamma_1}{2} \right) \hat{D}^\dagger_2 \left( -\gamma_2 \right)   \hat{A}_{3,1,\gamma_1,\gamma_2, \alpha_+}   \hat{D}_1 \left( - \frac{\gamma_1}{2} \right) \hat{D}_2\left( -\gamma_2 \right)  =  \mu_2 \hat{a}_2 + \alpha_+ \hat{J}_-  \label{A-two-mode-3-1-0-0-alpha-plus}
\end{equation}    
i.e, the system reduces to  (\ref{H-2D-basic-1}).

\vspace{1.0cm}
There are many other solutions of the partial differential equation (\ref{A-lambda-psi=lambda-psi}) depending on the choice of the initial conditions. We will not try to find all them here, instead,  we will rather use the above fact to simplify the problem by taking advantage of the results of the previous sections. Doing that, we will obtain a particular set of two-mode   non-separable algebra eigenstates of the lowering operator $\hat{A}_{3,1,\gamma_1,\gamma_2, \alpha_+} $.    Thus, using the fact that the states  (\ref{two-mode-non-separable-coherent-states}) are eigenstates of the lowering operator  (\ref{A-two-mode-3-1-0-0-alpha-plus}), then the normalized eigenstates of  $\hat{A}_{3,1,0,0,\alpha_+}$ are given by

\begin{equation}
\mid  \lambda \rangle_{2} =   \frac{\hat{D}_1 \left( - \frac{\gamma_1}{2} \right) \hat{D}_2\left( -\gamma_2 \right)  \hat{D}_2 \left( \frac{\lambda}{\mu_2} \right)}{\sqrt{ \frac{1}{2}  + \frac{\| \mu_2 \|^2}{ \| \alpha_+ \|^2}   +  \frac{\| \lambda \|^2}{ \| \mu_2 \|^2}  +  \frac{\| \lambda \|^4 }{4  \| \mu_2\|^4}    }} \left[     \frac{ {\left( \hat{a}^{\dagger}_{2}  + \frac{\lambda^\ast}{ \mu_2^\ast}  \right)}^2 }{2} | 0, 0 \rangle  - \frac{\mu_2}{\alpha_+}  
   | 1 , 0  \rangle  \right].
\end{equation} 
From this last equation we get the fundamental state of the system

\begin{equation}
\mid \tilde{0} \rangle_2 =   \frac{\hat{D}_1 \left( - \frac{\gamma_1}{2} \right) \hat{D}_2\left( -\gamma_2 \right)}{\sqrt{ \frac{1}{2}  + \frac{\| \mu_2 \|^2}{ \| \alpha_+ \|^2} }    } \left[     \frac{1 }{\sqrt{2}} | 0,  2 \rangle  - \frac{\mu_2}{\alpha_+}  
   | 1 , 0  \rangle  \right],
\end{equation} 
from where we deduce that the action of the shifted  Hamiltonian on it is given by
\begin{equation}\left[\hat{H}_{3,1,\gamma_1,\gamma_2}  +  \left(\frac{\|\gamma_1\|^2}{2}  + \|\gamma_2\|^2 \right) \hat{I}   \right] \mid \tilde{0} \rangle_2 = 2 \mid \tilde{0} \rangle_2,
\end{equation}
i.e., the  energy  spectrum of the system is $E_n= n+2, \; n=0, 1, \ldots.$ Furthermore, using (\ref{lambda-2-base}), we get the energy eigenstates 

  \begin{eqnarray}
 \mid E_n \rangle_2 &=& \mathcal{N}_{2,n}^{-\frac{1}{2}}         \;   \hat{D}_1 \left( - \frac{\gamma_1}{2} \right) \hat{D}_2\left( -\gamma_2 \right)   \;                \sum_{k=0}^{[\frac{n}{2}]}  \frac{1}{\sqrt{(n-2k)! k!} }       \left(\frac{\alpha^\ast_+  }{2 \mu^\ast_2 }\right)^k  \left[ \frac{\sqrt{(n-2k+1)!(n-2k+2)!}}{2} \mid k, n-2k+2 \rangle   \right. \nonumber \\ &+&
 \left(\frac{\alpha^\ast_+}{\mu_2^\ast}\right)^2  \sqrt{\frac{(k+1)(k+2)}{2}}  \sqrt{(n-2k-1)!(n-2k)!} \mid k+2, n-2k-2 \rangle \nonumber \\ &+&  \left. \sqrt{2 (k+1)} \left(   
\frac{\alpha^\ast_+}{\mu_2^\ast} (n-2k) - \frac{\mu_2}{\alpha_+}  \right) \mid k+1, n-2k\rangle
  \right], \nonumber \\ \label{lambda-2-base-gen}
   \end{eqnarray}
   where, as before, $\mathcal{N}_{2,n}$ is a normalization constant given by $\mathcal{N}_{2,n} = {}_2\langle E_n \mid E_n \rangle_2.$

\subsubsection{Interacting two-mode commensurate anisotropic quantum oscillators}
When $b^2= 4 \beta_+ \beta_-  + \beta_3^2 =1 $ and $\beta_+ =\beta_{-}^\ast \neq 0$ and $\beta_3 \neq 0,$ the relations (\ref{alpha-minus-plus}) are satisfied. Then,  inserting them into  the algebraic equation system (\ref{eq-1-linear}--\ref{eq-4-linear}), we obtain

   \begin{equation}
    \begin{pmatrix}
     \left( 1 - \frac{\beta_0 +\beta_{3}}{2} \right)& - \beta_+ \\
    - \beta_-  &    \left( 1 - \frac{\beta_0 - \beta_{3}}{2} \right) \end{pmatrix} \begin{pmatrix}
    \mu_1 \\ \mu_2
    \end{pmatrix}
    = \alpha_3 \; \begin{pmatrix}
    - \left( \frac{\gamma_1^\ast}{2}  - \frac{\gamma_2^\ast \beta_+}{1-\beta_3} \right) \\
    \left( \frac{\gamma_2^\ast}{2}  - \frac{\gamma_1^\ast \beta_-}{1+\beta_3}  \right)
\end{pmatrix}     \label{eq-mu-1-2}
    \end{equation}     
   and
\begin{equation}
    \begin{pmatrix}
     \left( 1 + \frac{\beta_0 +\beta_{3}}{2} \right) &   \beta_- \\
    \beta_+  &   \left( 1 + \frac{\beta_0 -\beta_{3}}{2} \right) \end{pmatrix} \begin{pmatrix}
    \nu_1 \\ \nu_2
    \end{pmatrix}
    = \alpha_3 \; \begin{pmatrix}
    \left( \frac{\gamma_1}{2}  + \frac{\gamma_2  \beta_-}{1 +\beta_3} \right) \\
    - \left( \frac{\gamma_2}{2}  + \frac{\gamma_1  \beta_+}{1-\beta_3}  \right)
\end{pmatrix},     \label{eq-nu-1-2}
    \end{equation}    
where we have taken $\alpha_3$ as an non-zero independent parameter.

Also in this case several situations can occur depending on the allowed values of the parameters. All  these situations are described in Appendices  \ref{appa} and  \ref{appb-0},  where the generalized Hamiltonian with its associated lowering operator is detailed. According to what we have already seen in the previous sections, in the process of searching for normalizable eigenstates for the lowering operator, the use of unitary similarity transformations was very useful. Here we will use this tool  to bring  the generalized Hamiltonian (\ref{H-the-most-general}) and its corresponding associated lowering operator (\ref{A-the-most-general}) to assume a simpler form. In fact, by first performing a convenient unitary similarity transformation of type $su(2)$ and then a second transformation of the same characteristics but of type $h(2)$, that is, using the two-mode separable displacement operator associated with the 2-D quantum harmonic oscillator, it is possible to transform the two interacting quantum harmonic oscillator Hamiltonian with linear external coupling into a two non-interacting quantum oscillator Hamiltonain without external interaction.Moreover, the final structure of the Hamiltonian thus transformed will be identical to one of the basic forms already discussed in section \ref{sec-two} with the only difference lying in the particular value of the new parameters.  In addition to all this, we have that the same successive unitary similarity transformations will lead the associated generalized lowering  operator to one of the  corresponding basic forms  of section   \ref{sec-two}.

\vspace{1.0cm}
In general, a first $su(2)$ unitary similarity transformation of the generalized Hamiltonian (\ref{H-the-most-general}) with the passing operator 

\begin{equation} \hat{T}_{(\epsilon,b, \beta_3, \theta)}= \exp\left[ - \arctan \left( \epsilon \sqrt{\frac{b -\epsilon \beta_3}{b+ \epsilon \beta_3}}  \right)  \left( e^{-i \theta} \hat{J}_+ - e^{i \theta} \hat{J}_-      \right)  \right],
\end{equation}
where $b^2= 4  \beta_+ \beta_- + \beta_3^2,$ and $\beta_\pm = R e^{\pm i \theta},$ will bring it to 

\begin{equation}
\hat{H}^{T}_{(\beta_0, \epsilon,\beta_3, \theta, \vec{\tilde{\gamma}}_1, \vec{\tilde{\gamma}}_2)} = 
\beta_0 \hat{N} + \epsilon b  \hat{J}_3 +  \tilde{\gamma}_{(1; \epsilon, b) }  \hat{a}_1^\dagger +  \tilde{\gamma}^\ast_{(1; \epsilon, b) } \hat{a}_1 +    \tilde{\gamma}_{(2; \epsilon, b, \beta_3) }   \hat{a}_2^\dagger  +   \tilde{\gamma}^\ast_{(2; \epsilon, b, \beta_3)}   \hat{a}_2  +  h_0  \hat{I}, \label{H-the-most-general-T}
\end{equation}      
 where
\begin{eqnarray}
\tilde{\gamma}_{(1; \epsilon, b, \beta_3) }  &=& \sqrt{\frac{b + \epsilon \beta_3}{2b}} \gamma_1 + \epsilon  e^{-i \theta}  \sqrt{\frac{b - \epsilon \beta_3}{2b}} \gamma_2, \\
\tilde{\gamma}_{(2; \epsilon, b, \beta_3) }  &=& \sqrt{\frac{b + \epsilon \beta_3}{2b}} \gamma_2  -  \epsilon  e^{i \theta}  \sqrt{\frac{b - \epsilon \beta_3}{2b}} \gamma_1.
 \end{eqnarray} 
 A second unitary similarity  transformation using the passing operator 
 
\begin{equation}
\hat{D}_{(1,2; \beta_0, \epsilon,b,\tilde{\gamma_1},\tilde{\gamma_2})} = \hat{D}_1 \left(-   \frac{ 2 \tilde{\gamma}_{(1; \epsilon, b,\beta_3) }  }{\beta_0 + \epsilon b} \right)  \hat{D}_2 \left(- \frac{2 \tilde{\gamma}_{(2; \epsilon, b, \beta_3) } }{\beta_0 - \epsilon b} \right),
\end{equation} 
 where $\hat{D}_1$ and $\hat{D}_2$ are the standard one-dimensional  displacement operators in the mode one and two, respectively,
 brings the transformed Hamiltonian  (\ref{H-the-most-general-T}) to

\begin{equation}
\hat{H}^{T,D_{(1,2)}}_{(\beta_0, \epsilon,\beta_3, \theta, \vec{\tilde{\gamma}}_1, \vec{\tilde{\gamma}}_2)} = 
\beta_0 \hat{N} + \epsilon b  \hat{J}_3  - 2 \frac{\|\tilde{\gamma}_{(1; \epsilon, b) }  \|^2}{\beta_0 + \epsilon b}   \hat{I}  -  2 \frac{\|\tilde{\gamma}_{(2; \epsilon, b) }  \|^2}{\beta_0 -  \epsilon b}     \hat{I} + h_0   \hat{I}, \label{H-the-most-general-T-D-1-2}
\end{equation}   
that is, a two mode $(\beta_0 + \epsilon b) : (\beta_0 - \epsilon b)$  anisotropic quantum oscillator. The choice of $h_0 = 2 \frac{\|\tilde{\gamma}_{(1; \epsilon, b) }  \|^2}{\beta_0 + \epsilon b}  +   2 \frac{\|\tilde{\gamma}_{(2; \epsilon, b) }  \|^2}{\beta_0 -  \epsilon b}   $ would lead to a simple energy spectrum. 

\vspace{1.0cm}

 For example,  when  $b^2=1,$    $\frac{\gamma_1}{2} \neq  \frac{\gamma_2 \beta_-}{1-\beta_3},$  $\frac{\gamma_1}{2} \neq - \frac{\gamma_2 \beta_-}{1+\beta_3}$ and   $ \beta_0 \notin \{\pm1,\pm3 \},$ this composed unitary similarity transformation for the lowering (\ref{gen-gen-lowering-beta-0-diff}) operator leads    to
 
 \begin{equation}
\hat{A}^{T, D_{(1,2)}}_{(\vec{0},\vec{0},\vec{\alpha},0)}  = - \epsilon \frac{\alpha_3 e^{i \epsilon \theta}}{2 \sqrt{1 - \beta_3^2} }  \hat{J}_{(-  \epsilon)},
    \end{equation}
    where $\hat{J}_{(-\epsilon)} = \hat{J}_{\mp}, $ when $\epsilon = \pm 1.$ Thus, when $\epsilon=1,$ by 
    using the results of section \ref{ladder-operators},  we get the energy eigenstates
    
    \begin{equation}
    \mid E_n \rangle =   \hat{D}_1 \left(- 2  \frac{ \tilde{\gamma}_{(1; 1, 1, \beta_3) }   }{\beta_0 +  1} \right)  \hat{D}_2 \left(- \frac{2 \tilde{\gamma}_{(2; 1, 1, \beta_3) } }{\beta_0 - 1} \right)  \hat{T}_{(1,1, \beta_3, \theta)} \mid 0, \kappa \rangle, \quad \kappa=1,2, \ldots,
    \end{equation}
and the energy spectrum

\begin{equation}
E_n = (\beta_0 -1) \kappa +n. \quad n=0,1,\ldots, \quad \kappa= 1,2, \ldots.
\end{equation}

\vspace{1.0cm}

On the other hand,   under the same circumstances, when $\beta_0=3,$ under the  similarity transformation the  lowering operator
  (\ref{A-2-D-gen-3-gamma-all-diff})  becomes
    
\begin{equation}    
\hat{A}^{T, D_{(1,2)}}_{(\vec{\mu},\vec{0},\vec{\alpha},0)}  = - \left[\sqrt{\frac{2}{1-\beta_3}}   \mu_1  e^{- i \theta}  + \frac{\alpha_3\tilde{\gamma}_{(2;1,1)}^\ast}{1 - \beta_3}   \right] \hat{a}_2  - \frac{\alpha_3 e^{i \theta}}{2 \sqrt{1 - \beta_3^2} }  \hat{J}_{-}, \label{annihilator-gen-1-2}
\end{equation}
when $\epsilon =1$ and

\begin{equation}    
\hat{A}^{T, D_{(1,2)}}_{(\vec{\mu},\vec{0},\vec{\alpha},0)}  =  \left[\sqrt{\frac{2}{1-\beta_3}}   \mu_1   -  \frac{\alpha_3\tilde{\gamma}_{(1;-1,1)}^\ast}{1 - \beta_3}   \right] \hat{a}_1 + \frac{\alpha_3 e^{-i \theta} }{2 \sqrt{1 - \beta_3^2} }  \hat{J}_{+},
\end{equation}
when $\epsilon =-1.$ 

Thus, when $\epsilon=1,$ to find the eigenstates  of the transformed annihilator (\ref{annihilator-gen-1-2}), we can use the results of section  \ref{basic-2-1-commensurate} with the parameter  $\mu_2$ replaced by ${\tilde \mu_2} =  - \left[\sqrt{\frac{2}{1-\beta_3}}   \mu_1  e^{- i \theta}  + \frac{\alpha_3\tilde{\gamma}_{(2;1,1)}^\ast}{1 - \beta_3}   \right] $ and a the parameter  $\alpha_+$ replaced by
$\tilde{\alpha}_+ =   - \frac{\alpha_3 e^{i \theta}}{2 \sqrt{1 - \beta_3^2} }  .$ Hence, using  (\ref{separable-coherent-states-2-1}) and (\ref{two-mode-non-separable-coherent-states}) we get  

\begin{equation}
\widetilde{\mid  \lambda \rangle}_1 = \hat{D}_1 \left( c_1 \lambda \right) \hat{S}_2 \left( \tilde{\chi} \right)  \hat{D}_2 \left( \frac{\lambda}{\tilde{\mu}_2}  \cosh (\| \tilde{\chi} \|) \right) \mid  0 , 0 \rangle,
\end{equation}  
  where $$\tilde{\chi}= \frac{\frac{ \lambda c_1 \tilde{\alpha}_+ }{\tilde{\mu}_2} } { \| \frac{ \lambda c_1 \tilde{\alpha}_+ }{\tilde{\mu}_2} \| }  \tanh^{-1} \left(   \| \frac{ \lambda c_1 \tilde{\alpha}_+ }{\tilde{\mu}_2} \|  \right), $$ 
   which must verify   $\| \frac{ \lambda c_1 \tilde{\alpha}_+ }{\tilde{\mu}_2} \| < 1,$  
and
 \begin{equation}
  \widetilde{\mid  \lambda \rangle}_2 = \frac{ \hat{D}_2 \left( \frac{\lambda}{\tilde{\mu}_2} \right)}{\sqrt{ \frac{1}{2}  + \frac{\| \tilde{\mu}_2 \|^2}{ \| \tilde{\alpha}_+ \|^2}   +  \frac{\| \lambda \|^2}{ \| \tilde{\mu}_2 \|^2}  +  \frac{\| \lambda \|^4 }{4  \| \tilde{ \mu}_2\|^4}    }} \left[     \frac{ {\left( \hat{a}^{\dagger}_{2}  + \frac{ \lambda^\ast}{ \tilde{\mu}_2^\ast}  \right)}^2 }{2} | 0, 0 \rangle  - \frac{\tilde{\mu}_2}{\tilde{\alpha}_+}  
   | 1 , 0  \rangle  \right].   \end{equation}
 
From these last equations, by making $\lambda=0,$   we  get the fundamental states  
\begin{equation} \widetilde{\mid 0  \rangle}_1 =\mid 0 ,0 \rangle \quad \text{ and}  \quad \widetilde{\mid 0  \rangle}_2 = 
  \frac{  \left[     \frac{1}{\sqrt{2}} | 0, 2 \rangle  - \frac{\tilde{\mu}_2}{\tilde{\alpha}_+}  
   | 1 , 0  \rangle  \right]}{\sqrt{ \frac{1}{2}  + \frac{\| \tilde{\mu}_2 \|^2}{ \| \tilde{\alpha}_+ \|^2} }} \label{ground-states-commensurate-1-2-T-D}
   \end{equation}

Hence, the eigenstates of the lowering operator operator   (\ref{A-2-D-gen-3-gamma-all-diff})  are given by

\begin{equation}
\mid \lambda \rangle_1 =   \hat{D}_1 \left(-  \frac{ \tilde{\gamma}_{(1; 1, 1, \beta_3) }}{2} \right)  \hat{D}_2 \left(-  \tilde{\gamma}_{(2; 1, 1, \beta_3)}   \right)  \hat{T}_{(1,1, \beta_3, \theta)}  \widetilde{ \mid 0 \rangle}_1,
\end{equation}

and

\begin{equation}
\mid \lambda \rangle_2 =   \hat{D}_1 \left(-  \frac{ \tilde{\gamma}_{(1; 1, 1, \beta_3) }   }{2} \right)  \hat{D}_2 \left(-  \tilde{\gamma}_{(2; 1, 1, \beta_3) }  \right)  \hat{T}_{(1,1, \beta_3, \theta)}  \widetilde{ \mid 0 \rangle}_2.
\end{equation}
\section{$SU(2)$ Chen type  $p:q$ commensurate coherent states}
\label{sec-four}
In this section we will compute the energy eigenstates of the commensurate $p:q$ quantum oscillator, 

\begin{equation}
\hat{\mathcal{H}}_{(p,q)} = \frac{\left[ p \; \hat{a}_1^\dagger \hat{a}_1 +  q  \;  \hat{a}_2^\dagger \hat{a}_2 \right]}{p \; q}, \label{H-p-q-commensurate}
\end{equation}
where  $p$ and $q$ are non-negative integer numbers.   To do this, we will use some results already obtained in previous section, in particular, those that relate to the  isotropic and to the  2:1 commensurate anisotropic 2D quantum oscillators. Then, we will generalize the structure of the lowering operators of these systems to include the case of the $p:q$ commensurate anisotropic quantum oscillator.

\vspace{1.0cm}
In previous sections we have seen that when $ p=q=1,$ i.e., when $\hat{H}_{(1,1)}$ represent the 2:D isotropic quantum oscillator, the associated lowering operator was given by $\hat{A}_{(1,1)}= \alpha_+ \hat{a}_1 + \alpha_-  \hat{a}_2, $ where we have renamed  the constants.  The normalized fundamental state in this case, for a given value of $\kappa= 0, 1, \ldots,$ in the number representation, was given by

\begin{equation}
\mid 0 \rangle^{(\kappa)}_{(1,1)} = \mathcal{N}^{- \frac{1}{2}}_{(\kappa)} \;  \left( \alpha_+  \hat{a}_2^\dagger -  \alpha_-  \hat{a}_1^\dagger  \right)^\kappa \mid 0,0 \rangle = \frac{\sum_{k=0}^{\kappa}  \sqrt{\binom{\kappa}{k}}  (-1)^\kappa  \alpha_-^k \alpha_+^{\kappa-k} \mid k, \kappa - k  \rangle }{\left(\| \alpha_+ \|^2 + \| \alpha_- \|^2\right)^{\frac{\kappa}{2}}}, \quad \kappa=0,1, \ldots, \label{kappa-1-1}
\end{equation}
 and the associated chains of energy were computed using the corresponding raisin operator in the following way
 
 \begin{equation}
\mid E_{n} \rangle^{(\kappa)}_{(1,1)} = \mathcal{N}^{- \frac{1}{2}}_{(n, \kappa)}  \; (\hat{A}^\dagger_{(1,1)})^n  \mid 0 \rangle^{(\kappa)}
\end{equation}
where $\mathcal{N}^{- \frac{1}{2}}_{(n, \kappa)}, \; n, \kappa=0,1,2, \ldots,$ are normalization constants. 

 We notice that for a fixed value of $\kappa,$ the energy spectrum was given by $E^{(\kappa)}_{n} = \kappa +n, \; n=0,1,2, \ldots.$ Then when $k=0,$ the  isotropic 2D quantum oscillator was isospectral to the 1-D canonical quantum oscillator. Indeed, in this case we had 
 
 \begin{equation}
[ \hat{\mathcal{H}}_{(1,1)} ,\hat{A}_{(1,1)}] = - \hat{A}_{(1,1)}, \quad  [\hat{\mathcal{H}}_{(1,1)} ,\hat{A}^\dagger_{(1,1)}] =  \hat{A}^\dagger_{(1,1)} \quad \text{and} \quad  [\hat{A}_{(1,1)} , \hat{A}^\dagger_{(1,1)}] = \left(\| \alpha_+ \|^2 + \| \alpha_- \|^2\right) \hat{I}. \label{algebra-standard-oscillator}
 \end{equation}
  
  \vspace{1.0cm}
Let us now introduce a new element to the discussion, that is,  the fact that the   fundamental states  (\ref{kappa-1-1}) form  a complete set of orthonormal states  which are also eigenstates of $\hat{\mathcal{H}}_{(1,1)}.$ Indeed, 

\begin{equation}
\hat{\mathcal{H}}_{(1,1)} \mid 0 \rangle^{(\kappa)} = \kappa  \mid 0 \rangle^{(\kappa)}, \quad \kappa= 0,1, \ldots.
\end{equation}
Furthermore, we observe that the state $ \mid 0 \rangle^{(\kappa)}$ can be considered as the result of the action of the  raising operator
\begin{equation}
\hat{\mathcal{A}}^\dagger_{(1,1)} =   \alpha_+  \hat{a}_2^\dagger -  \alpha_-  \hat{a}_1^\dagger 
\end{equation}
to the power $\kappa$ on the two-mode fundamental  state $\mid 0 , 0 \rangle,$ i.e.,  
 \begin{equation}
 \mid 0 \rangle^{(\kappa)}_{(1,1)} = \left( \hat{\mathcal{A}}^\dagger_{(1,1)} \right)^\kappa \mid 0,0 \rangle.
\end{equation}
Moreover, we observe that the set of operators $\hat{\mathcal{H}}_{(1,1)},$  $\hat{\mathcal{A}}_{(1,1)}$  and   $\hat{\mathcal{A}}^\dagger_{(1,1)}$
 form the algebra
 
  \begin{equation}
 [\hat{\mathcal{H}}_{(1,1)} ,\hat{\mathcal{A}}_{(1,1)}] = - \hat{\mathcal{A}}_{(1,1)}, \quad  [\hat{\mathcal{H}}_{(1,1)} ,\hat{\mathcal{A}}^\dagger_{(1,1)}] =  \hat{\mathcal{A}}^\dagger_{(1,1)} \quad \text{and} \quad  [\hat{\mathcal{A}}_{(1,1)} , \hat{\mathcal{A}}^\dagger_{(1,1)}] = \left(\| \alpha_+ \|^2 + \| \alpha_- \|^2\right) \hat{I},
 \end{equation}
 which is identical to (\ref{algebra-standard-oscillator}). 
 
 \vspace{1.0cm}

When $p=2$ and $q=1,$ we had essentially 

\begin{equation}
\hat{\mathcal{H}}_{(2,1)} =  \frac{\left[ 2 \; \hat{a}_1^\dagger \hat{a}_1 +    \;  \hat{a}_2^\dagger \hat{a}_2 \right]}{2},
\quad  \text{and} \quad \hat{A}_{(2,1)} =  \alpha_-   \; \hat{a}_2  +   \alpha_+  \; \hat{a}_2^\dagger  \hat{a}_1 = \alpha_- \;  \hat{a}_2 + \alpha_+ \; \hat{J}_-  ,
\end{equation} 
and from (\ref{2-1-analytic-ground-state}) we had the ground states, that in the number representation look like
\begin{equation}
\mid 0 \rangle^{(k)}_{(2,1)} =  \mathcal{N}^{- \frac{1}{2}}_{(\kappa)}   \left( \frac{ \alpha_+  \;  (\hat{a}^\dagger_2)^2 - 2 \alpha_-  \; \hat{a}^\dagger_1}{2} \right)^{\kappa} \mid 0,0 \rangle = \frac{\sum_{k=0}^{\kappa}  \sqrt{\binom{\kappa}{k}} \sqrt{\frac{(2 (\kappa-k))! }{(\kappa-k)!}  }   (-1)^k  \alpha_-^k \left( \frac{\alpha_+}{2}\right)^{\kappa-k} \mid k, 2(\kappa - k)  \rangle }{\left( 
\sum_{k=0}^{\kappa} \binom{\kappa}{k} \frac{(2(\kappa-k))!}{(\kappa-k)!}  \left(\frac{\|\alpha\|}{2} \right)^{2 (\kappa-k)} \|\alpha_-\|^{2k}\right)^{\frac{1}{2}}}, 
\end{equation} 
where $ \kappa=0,1, \ldots.$
Again, we observe that the state $ \mid 0 \rangle^{(\kappa)}$ can be considered as the result of the action of the  raising operator
\begin{equation}
\hat{\mathcal{A}}^\dagger_{(2,1)} =    \frac{ \alpha_+  \;  (\hat{a}^\dagger_2)^2 - 2 \alpha_-  \; \hat{a}^\dagger_1}{2} 
\end{equation}
to the power $\kappa$ on the two-mode fundamental  state $\mid 0 , 0 \rangle,$ i.e.,  
 \begin{equation}
 \mid 0 \rangle^{(\kappa)}_{(2,1)} = \left( \hat{\mathcal{A}}^\dagger_{(2,1)} \right)^\kappa \mid 0,0 \rangle.
\end{equation}
Also, in this case, the state $ \mid 0 \rangle^{(\kappa)}_{(2,1)}, \; \kappa=1,2, \ldots,$ is an eigenstate of $ \hat{\mathcal{H}}_{(2,1)}$ with associated eigenvalue equal to $\kappa, \; \kappa=1,2,\ldots,$ then the energy  spectrum is given by $(E_{n}^{(\kappa)})_{(2,1)} = \kappa, \; \kappa=1,2, \ldots.$ Indeed, you can prove it directly or deduce that from  the commutation relations characterizing  the system, that is,

  \begin{equation}
 [\hat{\mathcal{H}}_{(2,1)} ,\hat{\mathcal{A}}_{(2,1)}] = - \hat{\mathcal{A}}_{(2,1)}, \quad  [\hat{\mathcal{H}}_{(2,1)} ,\hat{\mathcal{A}}^\dagger_{(2,1)}] =  \hat{\mathcal{A}}^\dagger_{(2,1)} 
 \end{equation}
 and
 \begin{equation}
 [\hat{\mathcal{A}}_{(2,1)} , \hat{\mathcal{A}}^\dagger_{(2,1)}] = \| \alpha_+ \|^2   \hat{a}_2^\dagger \hat{a}_2 +  \frac{\| \alpha_+ \|^2  + 2   \| \alpha_- \|^2}{2 } \hat{I}.
 \end{equation}

\vspace{1.0cm}
The previous results can be extended as follows:  if $p$ and $q$ are non-negative coprime integers, the special choice of the lowering operator

\begin{equation}
\mathcal{\hat{A}}_{(p,q)} = \frac{\alpha_{+}^\ast\; q \;  \hat{a}_2^p -  \alpha_{-}^\ast \; p \; \hat{a}_1^q}{p \; q}, \label{cal-A-p-q}
\end{equation}  
assures us that the following commutation relation holds
\begin{equation}
[\mathcal{\hat{H}}_{(p,q)},\mathcal{\hat{A}}_{(p,q)}] = -\mathcal{\hat{A}}_{(p,q)} \label{H-cal-lowering-p-q}
\end{equation} 
and consequently
\begin{equation}
[\mathcal{\hat{H}}_{(p,q)},\mathcal{\hat{A}}^\dagger_{(p,q)}] = \mathcal{\hat{A}}^\dagger_{(p,q)}. \label{H-cal-raising-p-q}
\end{equation}

\vspace{1.0cm}

This choice is truly inspired by the natural generalization of the corresponding form of the lowering operator when $p=q=1$ and when $p=2$ and $q=1,$ which for a more general case we estimate it could be 
\begin{equation}
\hat{A}_{(p,q)} = \alpha_- \; ({\hat{a}^\dagger_1})^{(q-1)} \; \hat{a}_2 + \alpha_+ \; \hat{a}_1 \; ({\hat{a}^\dagger_2})^{(p-1)} , \label{lowering-A-generalized}
\end{equation}
and from the fact that its corresponding eigenstates associated  to the eigenvalue $\lambda=0$ are given by
\begin{equation}
\mid 0 \rangle^{(\kappa)}_{(p,q)} =  (\mathcal{\hat{A}}^\dagger_{(p,q)})^\kappa \mid 0,0 \rangle, \quad \kappa=0,1,2,\ldots. \label{kappa-0-p-q}
\end{equation}
Indeed, this last result follows directly from the fact that $[\hat{A}_{(p,q)},\mathcal{\hat{A}}^\dagger_{(p,q)}]=0, \; \kappa=0, 1,2, \ldots,$ and that $\hat{A}_{(p,q)} \mid 0,0 \rangle=0.$

\vspace{1.0cm}

Then, from all the above, we can say that a chain of orthonormalized eigenstates of the  $p:q$ commensurate anisotropic 2D quantum oscillator Hamiltonian  (\ref{H-p-q-commensurate}) is given by the set of states (\ref{kappa-0-p-q}), which have the explicit normalized form 

\begin{equation}
\mid 0 \rangle^{(\kappa)}_{(p,q)} = 
\frac{\sum_{k=0}^{\kappa}  \sqrt{\binom{\kappa}{k}} \sqrt{\frac{(p (\kappa-k))! }{(\kappa-k)!}  }  \sqrt{\frac{(q k)!}{k!}} (-1)^k  \left(\frac{\alpha_-}{q}\right)^k \left( \frac{\alpha_+}{p}\right)^{\kappa-k} \mid q k, p(\kappa - k)  \rangle }{\left( 
\sum_{k=0}^{\kappa} \binom{\kappa}{k} \frac{(p(\kappa-k))!}{(\kappa-k)!} \frac{(q k)!}{k!} 
\left(\frac{\|\alpha_+\|}{p} \right)^{2(\kappa-k)} \left(\frac{\|\alpha_-\|}{q} \right)^{2k} 
\right)^{\frac{1}{2}}}, \quad \kappa=0,1,2,\ldots. \label{p-q-energy-anisotropic-both}
\end{equation}
These states are certainty eigenstates of (\ref{H-p-q-commensurate}) with  associated eigenvalues $E^{(\kappa)}_{(p,q)}= \kappa, \; \kappa=1,2, \ldots. $  Thus, we get suitable Chen $su(2)$ coherent states type\cite{Chen-1}--\cite{Chen-2}, which have been obtained here on a simple way by using classical arguments of the theory of ladder operators.

\vspace{1.0cm}
There is something more, according to equation (\ref{cal-A-p-q}),  the state $\mid 0,0 \rangle$ is not the only  eigenstate of $\hat{\mathcal{A}}$ with associated eigenvalue equal to $0.$ There are others,  depending on the values of $p$ and $q.$ Indeed, for fixed values of $p$ and $q,$ by using equation (\ref{cal-A-p-q}), it is direct to check that

\begin{equation}
\mid 0  \rangle^{(k_1,k_2)}_{(p,q)} = \mid k_1, k_2 \rangle, \quad k_1 =0,\ldots, (q-1), \quad k_2 =0,\ldots, (p-1),
\end{equation}  
are also eigenstates of $\hat{\mathcal{A}}$ with associated eigenvalue equal to $0.$ Then according to (\ref{H-p-q-commensurate}) these states are also eigenstates of $\hat{\mathcal{H}}_{(p,q)}$ with associated eigenvalues equal to $E^{(k_1,k_2)}_{(0,p,q)}= \frac{k_1}{q}+\frac{k_2}{p}.$ Also, according to the commutation relations (\ref{H-cal-lowering-p-q}) and (\ref{H-cal-raising-p-q}) the states defined as

\begin{equation}
\mid E \rangle^{(k_1,k_2)}_{(n, p,q)} = (\mathcal{N}^{(k_1,k_2)}_{(n, p,q)})^{- \frac{1}{2}} 
(\hat{\mathcal{A}}^\dagger_{p,q})^n \mid k_1, k_2 \rangle, \quad k_1 =0,\ldots, (q-1), \quad k_2 =0,\ldots, (p-1),
\end{equation} 
are eigenstates of $\hat{\mathcal{H}}_{(p,q)}$ with associated eigenvalues $ E^{(k_1,k_2)}_{(n, p,q)} = n + \frac{k_1}{q}+\frac{k_2}{p}, \; n=0,1,2,\ldots.$ Thus, again, by using classical arguments of the ladder operators theory, we obtain the right energy spectrum of the $p:q$ commensurate anisotropic quantum oscillator computed by Louck et Al. \cite{Louck-Moshinsky-Wolf}. 

\vspace{1.0cm}

Moreover,  there are yet another set of eigenstates of (\ref{cal-A-p-q}) with eigenvalue equal to $0,$ that is, the two-mode  
non-separable 
state

\begin{equation}
\mid \tilde{0} \rangle_{(p,q)} = \frac{\frac{p}{\alpha^\ast_+ \; \sqrt{p!}} \mid 0 , p \rangle +   \frac{q}{\alpha^\ast_- \; \sqrt{q!}} \mid q , 0 \rangle}{ \left[  \frac{p^2}{|\alpha_+\|^2 \; p!    }        +     \frac{p^2}{|\alpha_-\|^2 \; q!    }         \right]^{\frac{1}{2}}  }          . 
\end{equation} 
It is direct to show that these states are also eigenstates of the $p:q$ commensurate anisotropic Hamiltonian (\ref{H-p-q-commensurate}) with associated eigenvalue equal to $1$. Then, again, with the help of the raising operator $ \mathcal{\hat{A}}^\dagger_{(p,q)}$ we can generate a chain of eigenstates  of  (\ref{H-p-q-commensurate}) in the usual way, i.e.,

\begin{equation}
 \mid \tilde{E}_{n} \rangle_{(p,q)} = {\tilde{\mathcal{N}}}_{(p,q)}^{-\frac{1}{2}} \; (\mathcal{\hat{A}}^\dagger_{(p,q)})^{n} \; \mid \tilde{0} \rangle_{(p,q)}, \quad n=0,1, \ldots.
\end{equation}
 By the way, the  energy spectrum of these states is given by $\tilde{E}_{n} = n+1, \; n=0,1, \ldots. $ The explicit  non-normalized form of these states is  
 
 \begin{eqnarray}
 \mid \tilde{E}_{n} \rangle_{(p,q)} &=& {\tilde{\mathcal{N}}}_{(p,q)}^{-\frac{1}{2}} \;  \sum_{k=0}^{n} \binom{n}{k}  (-1)^k    \left(\frac{\alpha_-}{q}\right)^{k}   \left(\frac{\alpha_+}{p}\right)^{(n-k)} \nonumber \\ &\times& \left[  \frac{\sqrt{ (qk)! (p ((n+1)-k))!}}{\alpha^\ast_+ \; (p-1)!} \mid  qk, p((n+1) -k) \rangle +  \frac{ \sqrt{ (q(k+1))! (p (n-k)!}}{ \alpha^\ast_- \; (q-1)!  }   \mid  q(k+1), p(n -k) \rangle \right]. \nonumber\\
 \end{eqnarray}

 \vspace{1.0cm}

  It might be worthwhile to find the generalized Hamiltonian $\hat{\mathbb{H}}_{(p,q)}$ that verifies $[\hat{\mathbb{H}}_{(p,q)}, \hat{A}_{(p,q)} ],$ where $\hat{A}_{(p,q)} $ is the lowering operator (\ref{lowering-A-generalized}). In general, this Hamiltonian should commute with the commutator $[\hat{A}_{(p,q)},\hat{A}_{(p,q)} ] $ and reduce to the isotropic or to the  $2:1$ commensurate anisotropic Hamiltonain, when $ p=1$ and $q=1$ and $p=2$ and $q=1,$ respectively.  One possibility is
  
  \begin{equation}
   \hat{\mathbb{H}}_{(p,q)} = \frac{p \;  \hat{a}_1^\dagger \hat{a}_1  + q \; \hat{a}_2^\dagger \hat{a}_2}{\left[ 1 - (p-1) (q-1)\right]}. \label{Hamiltonian-p-q-alternative}
  \end{equation}
It is clear that the eigenstates of this Hamiltonian are given by the  states (\ref{p-q-energy-anisotropic-both}). Moreover, its energy spectrum is $\mathbb{E}^{(\kappa)}_{(p,q)}= \frac{{p \; q  \; \kappa }}{1 -(p-1)(q-1)}, \; \kappa=0,1, \ldots. $ For example, when $p=q=1,$ it becomes the same as that of the canonical one-mode quantum oscillator, when $p > q $ and $q=1,$  $\mathbb{E}^{(\kappa)}_{(p,1)} = p \; \kappa, \;  \kappa= 1,2,\ldots,$ and when $p > q >1,$  $ \mathbb{E}^{(\kappa)}_{(p,q)} \leq 0, \forall  \kappa \in \mathbb{N}_{0}. $ In this last case, it could be interesting to define a new $p:q$ anisotropic 2D Hamiltonian as the negative of $\ref{Hamiltonian-p-q-alternative},$ which means that the old lowering operator become a the new raising operator and vice-versa.  Thus, the task   would be now to compute the eigenstates of 

\begin{equation}
\tilde{\hat{A}}_{(p,q)} = \alpha^\ast_- \; ({\hat{a}_1})^{(q-1)} \hat{a}^\dagger_2 + \alpha^\ast_+ \;  \hat{a}^\dagger_1   ({\hat{a}_2})^{(p-1)},\label{lowering-tilde-A-generalized}
\end{equation}
which is harder to do than in the original case.  Indeed, the  corresponding  partial differential equation becomes more difficult to solve according to the increasing in the value of the parameters $p$ and $q.$ 
 
 \vspace{1.0cm}
 
 \section{Conclusion}
\setcounter{equation}{0}
 
 In this article we have  studied the conditions  under which  a quantum system, whose Hamiltonian is an element of the complex 
 $ \left\{ h (1) \oplus h(1) \right\} \roplus u(2)  $  Lie algebra, admits ladder operators which are also elements of this algebra.  As a realization of such a system we have chosen 2 interacting quantum oscillators with linear external coupling. We have proved that there are suitable unitary similarity transformations of the type $su(2)$ and $h(1) \oplus h(1)$ which  bring the Hamiltonian and its associated  ladder operators into a more fundamental  $u(2)$ form. The algebra eigenstates of the lowering operators   constructed  in this way have been computed, and we have found  known as well as new  structures of coherent and squeezed states for the $1:1,$ $2:1,$ and $1:2$ commensurate quantum oscillators. Furthermore, we have shown that the $1:2$ and $2:1$  commensurate anisotropic Hamiltonian are connected by a composed $su(2)$ similarity transformation. We have also shown that for a very special choice of the parameters the Hamiltonian and its associated ladder operators verify the  $su(2)$ Lie algebra. In this case, we have found an interesting truncated bosonic set of energy eigenstates and spectrum.  In the case that the characteristic  $su(2)$ parameter  $b \neq 1,$ we have found that not only the   isotropic  2D quantum oscillator is included in the set of $u(2)$ quantum systems  but also  the  fractional  commensurate and non-commensurate anisotropic  2-D oscillator systems. Finally, based on the experience gained in the treatment of the $1:1$ and $2:1$ systems, more precisely, on the particular form of the lowering operators and their respective eigenstates associated with the eigenvalue equal to $0,$ we have been able to  construct a valid extension of these operators for the   $p:q$ commensurate anisotropic case. We have found that the resulting energy eigenstates states in this case are comparable to the $SU(2)$ type coherent states proposed by Cheng. Something interesting we could do is to construct from these  latter the  corresponding commensurate anisotropic  2D  Schr\"odinger-type coherent states and then compute the  probability density of them  to finally see how the results differ from those obtained in  \cite{JM-VH}.

\section*{Acknowledgments}  The author would like  to thank Dr.  Artorix de la Cruz de O\~na  of Dalhousie University for his valuable support and advice during the preparation of this article. He would also like to thank to his little purple doll Mary and to his beloved children by their encouragement  and great patience.

\appendix

\section{Allowed values of the parameters in the case $b=1$}
\label{appa}
\setcounter{equation}{0}
 When $b^2= 4 \beta_+ \beta_- + \beta_3^2 =1,$ and $\beta_{\pm}=0,$ the only values allowed for $\beta_3$ are $\pm 1 ,$
 which in turn do not allow for non-zero $\alpha$ parameters other than $\alpha_{\pm}$, respectively. These values will split the entire parameter set giving rise to two different classes of generalized two-mode anisotropic Hamiltonian systems that can be obtained from each other by simply exchanging the index of the modes and $\alpha_+$ for $\alpha_-$ and $\gamma_1$ for $\gamma_2.$
 
 \subsection{The case $\beta_3=1$}
 \label{appa-1}

When $\beta_{\pm}=0$ and $\beta_3=1,$ by  inserting the corresponding values of the $\alpha$ parameters  summarized  in equation
 (\ref{beta-pm-conditions})  into  the algebraic equation system (\ref{eq-1-linear}--\ref{eq-4-linear}), we obtain the following values for the vector $(\vec{\mu}, \vec{\nu}) \overset{\operatorname{def}}{=} (\mu_1,\mu_2,\nu_1,\nu_2) $  in terms of the remaining parameters:

 \begin{enumerate} 
 \item  In case that $ \gamma_1=\gamma_2=0,$ 
\begin{equation}
 (\vec{\mu}, \vec{\nu}) =\begin{cases}   (\mu_1,0,0,0), \quad \text{if} \quad \beta_0=1 \\ (0,\mu_2,0,0), \quad \text{if} \quad \beta_0=3   \\  (0,0,0, \nu_2), \quad \text{if} \quad \beta_0=-1   \\
  (0,0,\nu_1,0), \quad \text{if} \quad \beta_0=-3   \\
  (0,0,0,0),\quad \text{if} \quad \beta_0 \notin \{\pm1,\pm3\}  
 \end{cases} 
\end{equation} 
 \item In case that $ \gamma_1 \neq 0,\gamma_2=0  $
 \begin{equation}
 \\ (\vec{\mu}, \vec{\nu}) =\begin{cases}   (\mu_1,0,0,  \gamma_1 \alpha_+), \quad \text{if} \quad \beta_0 = 1  \\ (0,\mu_2,0,\frac{ \gamma_1 \alpha_+}{2}), \quad \text{if} \quad \beta_0=3 \\
 (0,0,\nu_1,- \gamma_1 \alpha_+ ), \quad \text{if} \quad \beta_0=-3  \\
  (0,0,0,\frac{2 \gamma_1 \alpha_+}{1 + \beta_0}),\quad \text{if} \quad \beta_0 \notin \{\pm1,\pm3\}  
 \end{cases} 
 \end{equation}
 \item In case that $  \gamma_1= 0, \gamma_2 \neq 0 $
 \begin{equation}
  (\vec{\mu}, \vec{\nu}) =\begin{cases}   (  \gamma_2^\ast  \alpha_+   ,\mu_2,0,0), \quad \text{if} \quad \beta_0 = 3  \\ ( -  \gamma_2^\ast  \alpha_+ ,0,0,\nu_2), \quad \text{if} \quad \beta_0=-1  \\
 ( - \frac{ \gamma_2^\ast  \alpha_+}{2},0,\nu_1,0 ), \quad \text{if} \quad \beta_0=-3  \\
  ( \frac{2 \gamma_2^\ast  \alpha_+}{\beta_0-1},0,0,0),\quad \text{if} \quad \beta_0 \notin \{\pm1,\pm3\}  
 \end{cases} 
 \end{equation}
\item In case that $  \gamma_1  \neq 0 , \gamma_2 \neq 0  $ 
 \begin{equation}
  (\vec{\mu}, \vec{\nu}) =\begin{cases}   ( \gamma_2^\ast  \alpha_+  ,\mu_2,0,\frac{ \gamma_1 \alpha_+}{2}), \quad \text{if} \quad \beta_0 = 3  \\ 
 ( - \frac{ \gamma_2^\ast  \alpha_+}{2},0,\nu_1, -  \gamma_1 \alpha_+), \quad \text{if} \quad \beta_0=-3  \\
  ( \frac{2 \gamma_2^\ast  \alpha_+}{\beta_0-1},0,0,\frac{2 \gamma_1 \alpha_+}{1 + \beta_0}),\quad \text{if} \quad \beta_0 \notin \{\pm1,\pm3\}  
 \end{cases}
 \end{equation}
 \setcounter{my-counter}{\value{enumi}} 
 \end{enumerate}

\subsection{The case $\beta_3=-1$}
\label{appa-2}
When $\beta_{\pm}=0$ and $\beta_3=-1,$ the same procedure leads to

\begin{enumerate} 
\setcounter{enumi}{\value{my-counter}} 
 \item  In case that $ \gamma_1=\gamma_2=0,$ 
\begin{equation}
 (\vec{\mu}, \vec{\nu}) =\begin{cases}   (0,\mu_2, 0,0), \quad \text{if} \quad \beta_0=1 \\ (\mu_1,0,0,0), \quad \text{if} \quad \beta_0=3   \\  (0,0,\nu_1, 0), \quad \text{if} \quad \beta_0=-1   \\
  (0,0,0,\nu_2), \quad \text{if} \quad \beta_0=-3   \\
  (0,0,0,0),\quad \text{if} \quad \beta_0 \notin \{\pm1,\pm3\}  
 \end{cases} 
\end{equation} 
 \item In case that $ \gamma_1 \neq 0,\gamma_2=0  $
 \begin{equation}
 \\ (\vec{\mu}, \vec{\nu}) =\begin{cases}   (\mu_1, \gamma_1^\ast \alpha_-  ,0,  0), \quad \text{if} \quad \beta_0 = 3  \\ 
 (0,- \gamma_1^\ast \alpha_-, \nu_1,0), \quad \text{if} \quad \beta_0=-1 \\
 (0,  - \frac{\gamma_1^\ast \alpha_-}{2},0,\nu_2), \quad \text{if} \quad \beta_0=-3  \\
  (0, \frac{2 \gamma_1^\ast  \alpha_-}{\beta_0 -1},0,0),\quad \text{if} \quad \beta_0 \notin \{\pm1,\pm3\}  
 \end{cases} 
 \end{equation}
 \item In case that $  \gamma_1= 0, \gamma_2 \neq 0 $
 \begin{equation}
  (\vec{\mu}, \vec{\nu}) =\begin{cases}   (0 ,\mu_2, \gamma_2 \alpha_-,0), \quad \text{if} \quad \beta_0 = 1  
  \\ (\mu_1,0, \frac{\gamma_2 \alpha_-}{2},0), \quad \text{if} \quad \beta_0=3  \\
 (0,0, -\gamma_2 \alpha_-, \nu_2 ), \quad \text{if} \quad \beta_0=-3  \\
  (0,0, \frac{2 \gamma_2  \alpha_-}{1+ \beta_0},0),\quad \text{if} \quad \beta_0 \notin \{\pm1,\pm3\}  
 \end{cases} 
 \end{equation}
\item In case that $  \gamma_1  \neq 0 , \gamma_2 \neq 0  $ 
 \begin{equation}
  (\vec{\mu}, \vec{\nu}) =\begin{cases}   (\mu_1, \gamma_1^\ast \alpha_-, \frac{\gamma_2 \alpha_-}{2},0), \quad \text{if} \quad \beta_0 = 3  \\ 
 (0, - \frac{\gamma_1^\ast \alpha_-}{2},- \gamma_2 \alpha_-, \nu_2 ), \quad \text{if} \quad \beta_0=-3  \\
  ( 0 , \frac{2 \gamma_1^\ast  \alpha_-}{\beta_0-1},\frac{2 \gamma_2 \alpha_-}{1 + \beta_0},0),\quad \text{if} \quad \beta_0 \notin \{\pm1,\pm3\}  
 \end{cases}
 \end{equation}
 \end{enumerate}

\subsection{The associated generalized Hamiltonian and lowering operator}
\label{appa-3}
The generalized Hamiltonian and lowering operator associated with the above parameter combination can be calculated directly using the equations (\ref{H-the-most-general}), (\ref{A-the-most-general}) and (\ref{a-0-constant}). When the $\gamma$ parameters are nonzero, the generalized Hamiltonian (\ref{H-the-most-general}) can be viewed as two interacting quantum harmonic oscillators with additional linear coupling terms.  Labeling the generalized Hamiltonian and the generalized lowering operator with the minimal set of subscripts allowing a direct correspondence with the values of the parameters identifying each element of the above classes, that is, as $\hat{H}_{\beta_0, \beta_3, \gamma_1 , \gamma_2}$ and  $\hat{A}_{\beta_0,\beta_3,\gamma_1,\gamma_2,\alpha_{\mp}},$ we obtain the following lists:

\begin{eqnarray}
 \hat{H}_{1, 1, 0, 0} &=& \hat{N} + \hat{J}_3, \nonumber \\
  \hat{A}_{1,1,0,0,\alpha_+} & =&  \mu_1 \hat{a}_1 + \alpha_+ \hat{J}_-,  \label{HA-plus-basic-1}  \\
\hat{H}_{3, 1, 0 , 0} &=& 3 \hat{N} + \hat{J}_3,  \nonumber \\ 
 \hat{A}_{3,1,0,0,\alpha_+}  &=& \mu_2 \hat{a}_2 +  \alpha_+ \hat{J}_- , \\
\hat{H}_{-1, 1, 0 , 0} &=&  -  \hat{N} + \hat{J}_3,  \nonumber \\ 
 \hat{A}_{-1,1,0,0,\alpha_+}  &=& \nu_2 \hat{a}_2^\dagger +  \alpha_+ \hat{J}_- , \\
 \hat{H}_{-3, 1, 0 , 0}  &=& -3 \hat{N} + \hat{J}_3, \nonumber \\
  \hat{A}_{-3,1,0,0,\alpha_+} & =&  \nu_1 \hat{a}_1^\dagger + \alpha_+ \hat{J}_-, \\
    \hat{H}_{\beta_0, 1, 0 , 0} &=& \beta_0 \hat{N} + \hat{J}_3, \nonumber \\
  \hat{A}_{\beta_0,1,0,0,\alpha_+} & =&  \alpha_+ \hat{J}_- , \quad  \beta_0 \notin \{\pm 1, \pm 3\}. \label{HA-plus-basic-5}
 \end{eqnarray}

\begin{eqnarray}
 \hat{H}_{1, 1, \gamma_1  , 0} &=& \hat{N} + \hat{J}_3 + \gamma_1 \hat{a}_1^\dagger + \gamma_1^\ast \hat{a}_1, \nonumber \\
  \hat{A}_{1,1,\gamma_1,0,\alpha_+} & =&  \mu_1 \hat{a}_1 + \gamma_1 \alpha_+ \hat{a}_2^\dagger + \alpha_+ \hat{J}_- +  \mu_{1} \gamma_1 \hat{I}, \\
\hat{H}_{3, 1, \gamma_1  , 0} &=& 3 \hat{N} + \hat{J}_3 + \gamma_1 \hat{a}_1^\dagger + \gamma_1^\ast \hat{a}_1,  \nonumber \\ 
 \hat{A}_{3,1,\gamma_1,0,\alpha_+}  &=& \mu_2 \hat{a}_2 +  \frac{\gamma_1 \alpha_+}{2} \hat{a}_2^\dagger + \alpha_+ \hat{J}_- , \\
 \hat{H}_{-3, 1, \gamma_1  , 0}  &=& -3 \hat{N} + \hat{J}_3 + \gamma_1 \hat{a}_1^\dagger + \gamma_1^\ast \hat{a}_1, \nonumber \\
  \hat{A}_{-3,1,\gamma_1,0,\alpha_+} & =&  \nu_1 \hat{a}_1^\dagger -  \gamma_1 \alpha_+ \hat{a}_2^\dagger + \alpha_+ \hat{J}_-  -  \gamma_1^\ast \nu_1 \hat{I}, \\
    \hat{H}_{\beta_0, 1, \gamma_1  , 0} &=& \beta_0 \hat{N} + \hat{J}_3 + \gamma_1 \hat{a}_1^\dagger + \gamma_1^\ast \hat{a}_1, \nonumber \\
  \hat{A}_{\beta_0,1,\gamma_1,0,\alpha_+} & =&  \frac{2 \gamma_1 \alpha_+}{1+\beta_0} \hat{a}_2^\dagger +  \alpha_+ \hat{J}_- , \quad  \beta_0 \notin \{\pm 1, \pm 3\}.
 \end{eqnarray}

 \begin{eqnarray}
 \hat{H}_{3, 1, 0,  \gamma_2} &=& 3 \hat{N} + \hat{J}_3 + \gamma_2 \hat{a}_2^\dagger + \gamma_2^\ast \hat{a}_2, \nonumber \\
  \hat{A}_{3,1,0,\gamma_2, \alpha_+} & =&  \gamma_2^\ast \alpha_+  \hat{a}_1 + \mu_2 \hat{a}_2  + \alpha_+ \hat{J}_-  + \gamma_2 \mu_2 \hat{I}, \\
\hat{H}_{-1, 1, 0,\gamma_2} &=&  - \hat{N} + \hat{J}_3 + \gamma_2 \hat{a}_2^\dagger + \gamma_2^\ast \hat{a}_2,  \nonumber \\ 
 \hat{A}_{-1,1,0,\gamma_2, \alpha_+}  &=& - \gamma_2^\ast  \alpha_+  \hat{a}_1 +  \nu_2 \hat{a}_2^\dagger + \alpha_+ \hat{J}_-   -     \gamma_2^\ast  \nu_2  \hat{I}, \\
 \hat{H}_{-3, 1, 0,\gamma_2} &=& -3 \hat{N} + \hat{J}_3 + \gamma_2 \hat{a}_2^\dagger + \gamma_2^\ast \hat{a}_2, \nonumber \\
  \hat{A}_{-3,1,0,\gamma_2, \alpha_+} & =&  - \frac{\gamma_2^\ast \alpha_+}{2} \hat{a}_1  +  \nu_1 \hat{a}_1^\dagger + \alpha_+ \hat{J}_- , \\
    \hat{H}_{\beta_0, 1, 0,\gamma_2} &=& \beta_0 \hat{N} + \hat{J}_3 + \gamma_2 \hat{a}_2^\dagger + \gamma_2^\ast \hat{a}_2, \nonumber \\
  \hat{A}_{\beta_0,1,0,\gamma_2, \alpha_+} & =&  \frac{2 \gamma_2^\ast }{\beta_0 -1} \hat{a}_1  +  \alpha_+ \hat{J}_-    \quad  \beta_0 \notin \{\pm 1, \pm 3\}. 
 \end{eqnarray}

\begin{eqnarray}
 \hat{H}_{3, 1, \gamma_1,  \gamma_2} &=& 3 \hat{N} + \hat{J}_3 + \gamma_1 \hat{a}_1^\dagger + \gamma_1^\ast \hat{a}_1 +  \gamma_2 \hat{a}_2^\dagger + \gamma_2^\ast \hat{a}_2, \nonumber \\
 \hat{A}_{3,1,\gamma_1,\gamma_2, \alpha_+} & =&  \gamma_2^\ast \alpha_+  \hat{a}_1 + \mu_2 \hat{a}_2  + \frac{\gamma_1 \alpha_+}{2} \hat{a}_2^\dagger + \alpha_+ \hat{J}_-  +  
 \left( \gamma_2 \mu_2 +  \frac{\gamma_1 \gamma_2^\ast \alpha_+}{2}  \right)  \hat{I}, \label{A-17} \\
 \hat{H}_{-3, 1, \gamma_1,\gamma_2} &=& -3 \hat{N} + \hat{J}_3 + \gamma_1 \hat{a}_1^\dagger + \gamma_1^\ast \hat{a}_1 +  \gamma_2 \hat{a}_2^\dagger + \gamma_2^\ast \hat{a}_2,  \nonumber \\
 \hat{A}_{-3,1,\gamma_1,\gamma_2, \alpha_+} & =&  - \frac{\gamma_2^\ast \alpha_+}{2} \hat{a}_1  +  \nu_1 \hat{a}_1^\dagger - \gamma_1 \alpha_+ \hat{a}_2^\dagger + \alpha_+ \hat{J}_-   -    \left( \gamma_1^\ast  \nu_1 - \frac{\gamma_1 \gamma_2^\ast \alpha_+ }{2} \right)  \hat{I}, \\
   \hat{H}_{\beta_0, 1, \gamma_1,\gamma_2} &=& \beta_0 \hat{N} + \hat{J}_3 +  \gamma_1 \hat{a}_1^\dagger + \gamma_1^\ast \hat{a}_1 +  \gamma_2 \hat{a}_2^\dagger + \gamma_2^\ast \hat{a}_2, \nonumber \\
  \hat{A}_{\beta_0,1,\gamma_1,\gamma_2, \alpha_+} & =&  \frac{2 \gamma_2^\ast \alpha_+}{\beta_0 -1} \hat{a}_1  +   \frac{2 \gamma_1 \alpha_+}{\beta_0 + 1} \hat{a}_2^\dagger  
  +  \alpha_+ \hat{J}_-  +  \frac{4 \gamma_1 \gamma_2^\ast \alpha_+}{\beta_0^2-1} \hat{I}, \quad  \beta_0 \notin \{\pm 1, \pm 3\}, 
 \end{eqnarray}
 
\begin{eqnarray}
 \hat{H}_{1, -1, 0, 0} &=& \hat{N} - \hat{J}_3, \nonumber \\
  \hat{A}_{1,-1,0,0,\alpha_-} & =&  \mu_2 \hat{a}_2 + \alpha_- \hat{J}_+,  \label{HA-minus-basic-1}\\
\hat{H}_{3, -1, 0 , 0} &=& 3 \hat{N} - \hat{J}_3,  \nonumber \\ 
 \hat{A}_{3,-1,0,0,\alpha_-}  &=& \mu_1 \hat{a}_1 +  \alpha_- \hat{J}_+, \\
\hat{H}_{-1, -1, 0 , 0} &=&  -  \hat{N} - \hat{J}_3,  \nonumber \\ 
 \hat{A}_{-1,-1,0,0,\alpha_-}  &=& \nu_1 \hat{a}_1^\dagger +  \alpha_- \hat{J}_+ , \\
 \hat{H}_{-3, -1, 0 , 0}  &=& -3 \hat{N} - \hat{J}_3, \nonumber \\
  \hat{A}_{-3,-1,0,0,\alpha_-} & =&  \nu_2 \hat{a}_2^\dagger + \alpha_- \hat{J}_+, \\
    \hat{H}_{\beta_0, -1, 0 , 0} &=& \beta_0 \hat{N} - \hat{J}_3, \nonumber \\
  \hat{A}_{\beta_0,-1,0,0,\alpha_-} & =&  \alpha_- \hat{J}_+ , \quad  \beta_0 \notin \{\pm 1, \pm 3\}. \label{HA-minus-basic-5}
 \end{eqnarray}

\begin{eqnarray}
 \hat{H}_{3, -1, \gamma_1  , 0} &=& 3 \hat{N} - \hat{J}_3 + \gamma_1 \hat{a}_1^\dagger + \gamma_1^\ast \hat{a}_1, \nonumber \\
  \hat{A}_{3,-1,\gamma_1,0,\alpha_-} & =&  \mu_1 \hat{a}_1 + \gamma_1^\ast \alpha_-\hat{a}_2  + \alpha_- \hat{J}_+  +  \gamma_1  \mu_{1} \hat{I}, \\
\hat{H}_{-1, -1, \gamma_1  , 0} &=&  - \hat{N} - \hat{J}_3 + \gamma_1 \hat{a}_1^\dagger + \gamma_1^\ast \hat{a}_1,  \nonumber \\ 
 \hat{A}_{-1,-1,\gamma_1,0,\alpha_-}  &=&  \gamma_1^\ast \alpha_-  \hat{a}_2  +  \nu_1 \hat{a}_1^\dagger + \alpha_- \hat{J}_+  -   \gamma_1^\ast \nu_1 , \\
 \hat{H}_{-3, -1, \gamma_1  , 0} &=& -3 \hat{N} - \hat{J}_3 + \gamma_1 \hat{a}_1^\dagger + \gamma_1^\ast \hat{a}_1, \nonumber \\
  \hat{A}_{-3,-1,\gamma_1,0,\alpha_-} & =&  - \frac{\gamma_1^\ast \alpha_-}{2} \hat{a}_2 + \nu_2 \hat{a}_2^\dagger  + \alpha_- \hat{J}_+, \\
    \hat{H}_{\beta_0, -1, \gamma_1  , 0}  &=& \beta_0 \hat{N} -  \hat{J}_3 + \gamma_1 \hat{a}_1^\dagger + \gamma_1^\ast \hat{a}_1, \nonumber \\
  \hat{A}_{\beta_0,-1,\gamma_1,0,\alpha_-} & =&  \frac{2 \gamma_1^\ast  \alpha_-}{\beta_0 - 1} \hat{a}_2 +  \alpha_- \hat{J}_+   , \quad  \beta_0 \notin \{\pm 1, \pm 3\}.
 \end{eqnarray}

 \begin{eqnarray}
 \hat{H}_{1, -1, 0,  \gamma_2} &=&  \hat{N} - \hat{J}_3 + \gamma_2 \hat{a}_2^\dagger + \gamma_2^\ast \hat{a}_2, \nonumber \\
  \hat{A}_{1,-1,0,\gamma_2, \alpha_-} & =&  \mu_2  \hat{a}_2  +  \gamma_2 \alpha_- \hat{a}_1^\dagger + \alpha_- \hat{J}_+  + \gamma_2 \mu_2 \hat{I}, \\
\hat{H}_{3, -1, 0,\gamma_2} &=&  3 \hat{N} - \hat{J}_3 + \gamma_2 \hat{a}_2^\dagger + \gamma_2^\ast \hat{a}_2,  \nonumber \\ 
 \hat{A}_{3,-1,0,\gamma_2, \alpha_-}  &=& \mu_1 \hat{a}_1 +  \frac{\gamma_2 \alpha_-}{2} \hat{a}_1 ^\dagger + \alpha_- \hat{J}_+, \\
 \hat{H}_{-3, -1, 0,\gamma_2} &=& -3 \hat{N} - \hat{J}_3 + \gamma_2 \hat{a}_2^\dagger + \gamma_2^\ast \hat{a}_2, \nonumber \\
  \hat{A}_{-3,-1,0,\gamma_2, \alpha_-} & =&  -  \gamma_2 \alpha_- \hat{a}_1^\dagger  +  \nu_2 \hat{a}_2^\dagger + \alpha_- \hat{J}_+ -\gamma_2^\ast \nu_2  \hat {I}, \\
    \hat{H}_{\beta_0, -1, 0,\gamma_2} &=& \beta_0 \hat{N} - \hat{J}_3 + \gamma_2 \hat{a}_2^\dagger + \gamma_2^\ast \hat{a}_2, \nonumber \\
  \hat{A}_{\beta_0,-1,0,\gamma_2, \alpha_-} & =&  \frac{2 \gamma_2 \alpha_- }{\beta_0 + 1} \hat{a}_1^\dagger  +  \alpha_- \hat{J}_+,    \quad  \beta_0 \notin \{\pm 1, \pm 3\}. 
 \end{eqnarray}

\begin{eqnarray}
 \hat{H}_{3, -1, \gamma_1,  \gamma_2} &=& 3 \hat{N} - \hat{J}_3 + \gamma_1 \hat{a}_1^\dagger + \gamma_1^\ast \hat{a}_1 +  \gamma_2 \hat{a}_2^\dagger + \gamma_2^\ast \hat{a}_2, \nonumber \\
 \hat{A}_{3,-1,\gamma_1,\gamma_2, \alpha_-} & =&  \mu_1 \hat{a}_1  +  \gamma_1^\ast \alpha_-  \hat{a}_2 + \frac{\gamma_2 \alpha_-}{2} \hat{a}_1^\dagger + \alpha_- \hat{J}_+  +  
 \left(\gamma_1 \mu_1 + \frac{\gamma_1^\ast \gamma_2  \alpha_-}{2}  \right)  \hat{I}, \\
 \hat{H}_{-3, -1, \gamma_1,\gamma_2} &=& -3 \hat{N} - \hat{J}_3 + \gamma_1 \hat{a}_1^\dagger + \gamma_1^\ast \hat{a}_1 +  \gamma_2 \hat{a}_2^\dagger + \gamma_2^\ast \hat{a}_2,  \nonumber \\
 \hat{A}_{-3,-1,\gamma_1,\gamma_2, \alpha_-} & =&  - \frac{\gamma_1^\ast \alpha_-}{2} \hat{a}_2  - \gamma_2 \alpha_-  \hat{a}_1^\dagger  +  \nu_2 \hat{a}_2^\dagger + \alpha_- \hat{J}_+   -    \left(\gamma_2^\ast  \nu_2 - \frac{\gamma_1^\ast  \gamma_2  \alpha_- }{2} \right)  \hat{I}, \\
   \hat{H}_{\beta_0, -1, \gamma_1,\gamma_2} &=& \beta_0 \hat{N} -  \hat{J}_3 +  \gamma_1 \hat{a}_1^\dagger + \gamma_1^\ast \hat{a}_1 +  \gamma_2 \hat{a}_2^\dagger + \gamma_2^\ast \hat{a}_2, \nonumber \\
  \hat{A}_{\beta_0,-1,\gamma_1,\gamma_2, \alpha_-} & =&  \frac{2 \gamma_1^\ast \alpha_-}{\beta_0 -1} \hat{a}_2  +   \frac{2 \gamma_2 \alpha_-}{\beta_0 + 1} \hat{a}_1^\dagger  
  +  \alpha_- \hat{J}_+  +  \frac{4 \gamma_1^\ast \gamma_2 \alpha_-}{\beta_0^2-1} \hat{I}, \quad  \beta_0 \notin \{\pm 1, \pm 3\}, 
 \end{eqnarray}

\section{$ \left\{ h (1) \oplus h(1) \right\} \roplus u(2) $ \; Hamiltonian and associated lowering operator }
\label{appb-0}

When $b^2= 4\beta_+ \beta_- + \beta_3^2 =1$ and $\beta_\pm \neq 0,$ according to equations  (\ref{eq-mu-1-2}) and  (\ref{eq-nu-1-2}),  several situations may arise  depending on the values of the parameters involved.  In such a case, it is pertinent to start  the analysis by fixing  the values of the  parameters $\gamma$ and then to analyze the conditions that the remaining parameters must satisfy for the resulting algebraic equation to have nonzero solutions.  

\subsection{Case $b^2=1,$ $\beta_{0} \notin  \{1,3,-1,-3\}$ and $\vec{\gamma}=0$}
\label{appb-1}
In this case, the generalized Hamiltonian  (\ref{H-2-D-gen-A-gen}) becomes

 \begin{equation} 
   \hat{H}_{(\beta_0,\vec{\beta},\vec{0}, \vec{0})} =   \beta_0 \hat{N} + \vec{\beta} \cdot \vec{\hat{J}} + h_0 \hat{I} 
        \label{H-beta-0-beta}
    \end{equation}
   and the lowering operator  (\ref{A-the-most-general} )  reduces to

\begin{equation}
     \hat{A}_{(\vec{0},\vec{0},\vec{\alpha},0)}  = \alpha_3 \left(  \frac{\beta_-}{1+\beta_3}   \hat{J}_+   - \frac{\beta_+}{1-\beta_3}   \hat{J}_-  + \hat{J}_3 \right). \label{A-vec-alpha}
    \end{equation}

 We notice that by performing a  similarity transformation with the help of the unitary operator $\hat{T}_{(\epsilon,b,\beta_3,\theta)},$ described in equation ( \ref{T-unitary-epsilon-b}), these operators  take the form
    
\begin{equation} 
   \hat{H}^{T_{\epsilon}}_{(\beta_0,\beta_3,\vec{0}, \vec{0})} = \beta_{0} \hat{N} + \epsilon \hat{J_3} +  h_0  \hat{I}= \frac{(\beta_0 + \epsilon)}{2}
   \hat{a}_1^\dagger \hat{a}_1 + \frac{(\beta_0 - \epsilon)}{2}
   \hat{a}_2^\dagger \hat{a}_2 +  h_0  \hat{I}  \label{H-beta-0-beta-3}
    \end{equation}
    and
 \begin{equation}
     \hat{A}^{T_{\epsilon}}_{(\vec{0},\vec{0},\alpha_3,0)}  =  \begin{cases}   - \frac{ \alpha_3 e^{i \theta}}{2R} \hat{J}_-, \quad \text{if} \quad \epsilon =1 \\    \frac{ \alpha_3 e^{- i \theta}}{2R} \hat{J}_+, \quad \text{if} \quad \epsilon = - 1, 
\end{cases}    \label{A-vec-alpha-3},
    \end{equation}
    respectively, i.e., they show the structure of a $(\beta_0 + \epsilon): (\beta_0 -\epsilon)$ rate anysotropic Hamiltonian together with its associated $su(2)$ canonical lowering operator.  
\subsection{Case $b^2=1,$ $\beta_{0} \in  \{1,3\}$ and $\vec{\gamma}=0$}
\label{appb-2}
When all the parameters $\gamma $ are equal to $0,$ and $\beta_0 \neq -1$ and $\beta_0 \neq -3,$  we have  $\nu_1=\nu_2=0,$ and therefore we   fall in the case already discussed at the beginning of this article, i.e., we must analyze under what conditions the parameters  $\mu_1$ and $\mu_2$ are nonzero, and we find that it is possible if and only if  $\beta_0=1$ or $\beta_0 =3.$ The expression of the generalized Hamiltonians and its associated lowering operators in this case are shown in equations   ( \ref{H-2D-semi-gen}-- \ref{A-2D-generalized}).   
 
 \subsection{Case $b^2=1,$ $\beta_{0} \in  \{-1,-3\}$ and $\vec{\gamma}=0$}
\label{appb-3} On the other hand, when all the parameters $\gamma $ are equal to $0$  and  $\beta_0 \neq  1$ and $\beta_0 \neq 3,$  we have  $\mu_1=\mu_2=0.$  Then from  (\ref{eq-nu-1-2})  we deduce  $\nu_2 = - \frac{2 \beta_+ \nu_1}{2 + \beta_0 -\beta_3}.$ Hence, the  generalized Hamiltonian  (\ref{H-2-D-gen-A-gen}) becomes

 \begin{equation} 
   \hat{H}_{(\beta_0,\vec{\beta},\vec{0}, \vec{0})} =   \beta_0 \hat{N} +  \vec{\beta} \cdot \vec{\hat{J}} + h_0 \hat{I} 
        \label{H-nu-beta-0-beta}
    \end{equation}
   and the lowering operator  (\ref{A-the-most-general} )  reduces to
\begin{equation}
\hat{A}_{(\vec{0},\vec{\nu},\vec{\alpha},0)}  =   \nu_1 \left(   \hat{a}_1^\dagger  -   \frac{2 \beta_+ }{2 + \beta_0 -\beta_3} \hat{a}_2^\dagger \right)  +  \alpha_3 \left(  \frac{\beta_-}{1+\beta_3}  \hat{J}_+  - \frac{\beta_+}{1-\beta_3}   \hat{J}_-  +  \hat{J}_3 \right). \label{A-nu-vec-alpha}
    \end{equation}
     
\subsection{Case $b^2=1,$  and  $\frac{\gamma_1}{2}= \frac{\gamma_2 \beta_-}{1-\beta_3}$}
\label{appb-4}
In this case, depending on the value of $\beta_0,$ we  distinguish the following subcases:

\begin{itemize}

\item When $\beta_0 =1,$ the  Hamiltonian  (\ref{H-2-D-gen-A-gen}) becomes

\begin{equation}
   \hat{H}_{(1,\vec{\beta},\vec{\gamma_1}, \vec{\gamma_2})} =   \hat{N} + \vec{\beta} \cdot \vec{\hat{J}} +  \frac{ 2 \gamma_2 \beta_-}{1-\beta_3}  \hat{a}_1^\dagger + \frac{ 2 \gamma_2^\ast \beta_+}{1-\beta_3}  \hat{a}_1  + \gamma_2 \hat{a}_2^\dagger + \gamma_2^\ast \hat{a}_2 +   h_0 \hat{I} 
\end{equation}
and the lowering operator  (\ref{A-the-most-general} )  reduces to

\begin{equation} \hat{A}_{(\vec{\mu},\vec{\nu},\vec{\alpha},a_0)}  =   \mu_1 \left(   \hat{a}_1  +  \frac{2 \beta_- }{1 + \beta_3} \hat{a}_2 \right) +  \alpha_3  \gamma_2  \left(  \frac{2 \beta_-}{1 -\beta_3^2}  \hat{a}_1^\dagger - \frac{ 1}{1-\beta_3}  \hat{a}_2^\dagger \right) + \alpha_3 \left(  \frac{\beta_-}{1+\beta_3}   \hat{J}_+   - \frac{\beta_+}{1-\beta_3}   \hat{J}_-  + \hat{J}_3 \right) + \frac{\gamma_2 \mu_1}{\beta_+}.
\end{equation}

\item When $\beta_0 =3,$ the  Hamiltonian  (\ref{H-2-D-gen-A-gen}) becomes

\begin{equation}
   \hat{H}_{(3,\vec{\beta},\vec{\gamma_1}, \vec{\gamma_2})} =  3  \hat{N} + \vec{\beta} \cdot \vec{\hat{J}} +  \frac{ 2 \gamma_2 \beta_-}{1-\beta_3}  \hat{a}_1^\dagger + \frac{ 2 \gamma_2^\ast \beta_+}{1-\beta_3}  \hat{a}_1  + \gamma_2 \hat{a}_2^\dagger + \gamma_2^\ast \hat{a}_2 +   h_0 \hat{I} 
\end{equation}
and the lowering operator  (\ref{A-the-most-general} )  reduces to

\begin{equation} \hat{A}_{(\vec{\mu},\vec{\nu},\vec{\alpha},0)}  =   \mu_1 \left(   \hat{a}_1  +  \frac{2 \beta_- }{\beta_3 -1}  \hat{a}_2 \right) +  \alpha_3  \gamma_2 \left(  \frac{\beta_-}{1 -\beta_3^2}  \hat{a}_1^\dagger - \frac{1}{2(1-\beta_3)}  \hat{a}_2^\dagger \right) + \alpha_3 \left(  \frac{\beta_-}{1+\beta_3}   \hat{J}_+   - \frac{\beta_+}{1-\beta_3}   \hat{J}_-  + \hat{J}_3 \right).  
\end{equation}

\item When $\beta_0 =-3,$ the  Hamiltonian  (\ref{H-2-D-gen-A-gen}) becomes

\begin{equation}
   \hat{H}_{(-3,\vec{\beta},\vec{\gamma_1}, \vec{\gamma_2})} =  -3  \hat{N} + \vec{\beta} \cdot \vec{\hat{J}} +  \frac{ 2 \gamma_2 \beta_-}{1-\beta_3}  \hat{a}_1^\dagger + \frac{ 2 \gamma_2^\ast \beta_+}{1-\beta_3}  \hat{a}_1  + \gamma_2 \hat{a}_2^\dagger + \gamma_2^\ast \hat{a}_2 +   h_0 \hat{I} 
\end{equation}
and the lowering operator  (\ref{A-the-most-general} )  reduces to

\begin{equation} \hat{A}_{(\vec{0},\vec{\nu},\vec{\alpha},a_0)}  =  \nu_1  \hat{a}_1^\dagger + 2  \left( \frac{ \beta_+ \nu_1}{1+\beta_3} + \frac{\gamma_2 \alpha_3}{1-\beta_3^2}  \right) \hat{a}_2^\dagger + \alpha_3 \left(  \frac{\beta_-}{1+\beta_3}   \hat{J}_+   - \frac{\beta_+}{1-\beta_3}   \hat{J}_-  + \hat{J}_3 \right) - 2 \gamma_{2}^\ast \left(\frac{2 \beta_+ \beta_3 \nu_1 + \gamma_2 \alpha_3 }{1- \beta_3^2}\right).
\end{equation}

\item When $\beta_0  \notin \{1,3,-1,-3 \},$ the  Hamiltonian  (\ref{H-2-D-gen-A-gen}) becomes

\begin{equation}
   \hat{H}_{(\beta_0,\vec{\beta},\vec{\gamma_1}, \vec{\gamma_2})} =  \beta_0  \hat{N} + \vec{\beta} \cdot \vec{\hat{J}} +  \frac{ 2 \gamma_2 \beta_-}{1-\beta_3}  \hat{a}_1^\dagger + \frac{ 2 \gamma_2^\ast \beta_+}{1-\beta_3}  \hat{a}_1  + \gamma_2 \hat{a}_2^\dagger + \gamma_2^\ast \hat{a}_2 +   h_0 \hat{I} 
\end{equation}
and the lowering operator  (\ref{A-the-most-general} )  reduces to

\begin{equation} \hat{A}_{(\vec{0},\vec{\nu},\vec{\alpha},0)}  =  \frac{ 2 \alpha_3 \gamma_2 (3 +\beta_0)}{(1-\beta_3^2)\left( (2 + \beta_0)^2 -1  \right)     }  \left( 2 \beta_-  \hat{a}_1^\dagger    -  (1 + \beta_3) \hat{a}_2^\dagger \right) + \alpha_3 \left(  \frac{\beta_-}{1+\beta_3}   \hat{J}_+   - \frac{\beta_+}{1-\beta_3}   \hat{J}_-  + \hat{J}_3 \right).
\end{equation}

 \end{itemize}

\subsection{Case $b^2=1,$  and  $\frac{\gamma_1}{2}= - \frac{\gamma_2 \beta_-}{1+\beta_3}$}
\label{appb-5}
In this case, depending on the value of $\beta_0,$ we  distinguish the following sub-cases:

\begin{itemize}

\item When $\beta_0 =-1,$ the  Hamiltonian  (\ref{H-2-D-gen-A-gen}) becomes

\begin{equation}
   \hat{H}_{(-1,\vec{\beta},\vec{\gamma_1}, \vec{\gamma_2})} =  - \hat{N} + \vec{\beta} \cdot \vec{\hat{J}} -  \frac{  2 \gamma_2 \beta_-}{1+\beta_3}  \hat{a}_1^\dagger - \frac{  2 \gamma_2^\ast \beta_+}{1+\beta_3}  \hat{a}_1 + \gamma_2 \hat{a}_2^\dagger + \gamma_2^\ast  \hat{a}_2 +   h_0 \hat{I} 
\end{equation}
and the lowering operator  (\ref{A-the-most-general} )  reduces to

\begin{equation} \hat{A}_{(\vec{\mu},\vec{\nu},\vec{\alpha},a_0)}  =  \alpha_3  \gamma_2^\ast  \left(  \frac{2 \beta_+}{1 -\beta_3^2}  \hat{a}_1 + \frac{ 1}{1+\beta_3}  \hat{a}_2 \right) +  \nu_1 \left(   \hat{a}_1^\dagger  -  \frac{2 \beta_+ }{1 - \beta_3} \hat{a}_2^\ast \right) +  \alpha_3 \left(  \frac{\beta_-}{1+\beta_3}   \hat{J}_+   - \frac{\beta_+}{1-\beta_3}   \hat{J}_-  + \hat{J}_3 \right) + \frac{\gamma_2^\ast \nu_1}{\beta_-}.
\end{equation}

\item When $\beta_0 =-3,$ the  Hamiltonian  (\ref{H-2-D-gen-A-gen}) becomes

\begin{equation}
   \hat{H}_{(-3,\vec{\beta},\vec{\gamma_1}, \vec{\gamma_2})} =  - 3 \hat{N} + \vec{\beta} \cdot \vec{\hat{J}} -  \frac{  2 \gamma_2 \beta_-}{1+\beta_3}  \hat{a}_1^\dagger - \frac{  2 \gamma_2^\ast \beta_+}{1+\beta_3}  \hat{a}_1 + \gamma_2 \hat{a}_2^\dagger + \gamma_2^\ast  \hat{a}_2 +   h_0 \hat{I} 
   \end{equation}
and the lowering operator  (\ref{A-the-most-general} )  reduces to

\begin{equation} \hat{A}_{(\vec{\mu},\vec{\nu},\vec{\alpha},0)}  =    \alpha_3  \gamma_2^\ast  \left(  \frac{\beta_+}{1 -\beta_3^2}  \hat{a}_1 + \frac{1}{2(1+\beta_3)}  \hat{a}_2 \right) +  \nu_1 \left(   \hat{a}_1^\dagger  +  \frac{2 \beta_+ }{1 + \beta_3}  \hat{a}_2^\dagger \right) + \alpha_3 \left(  \frac{\beta_-}{1+\beta_3}   \hat{J}_+   - \frac{\beta_+}{1-\beta_3}   \hat{J}_-  + \hat{J}_3 \right).  
\end{equation}

\item When $\beta_0 =3,$ the  Hamiltonian  (\ref{H-2-D-gen-A-gen}) becomes

\begin{equation}
   \hat{H}_{(3,\vec{\beta},\vec{\gamma_1}, \vec{\gamma_2})} =  3 \hat{N} + \vec{\beta} \cdot \vec{\hat{J}} -  \frac{  2 \gamma_2 \beta_-}{1+\beta_3}  \hat{a}_1^\dagger - \frac{  2 \gamma_2^\ast \beta_+}{1+\beta_3}  \hat{a}_1 + \gamma_2 \hat{a}_2^\dagger + \gamma_2^\ast  \hat{a}_2 +   h_0 \hat{I} 
\end{equation}
and the lowering operator  (\ref{A-the-most-general} )  reduces to

\begin{equation} \hat{A}_{(\vec{\mu},\vec{0},\vec{\alpha},a_0)}  =  \mu_1  \hat{a}_1 - 2  \left( \frac{ \beta_- \mu_1}{1-\beta_3} + \frac{\gamma_2^\ast \alpha_3}{1-\beta_3^2}  \right) \hat{a}_2 + \alpha_3 \left(  \frac{\beta_-}{1+\beta_3}   \hat{J}_+   - \frac{\beta_+}{1-\beta_3}   \hat{J}_-  + \hat{J}_3 \right) + 2 \gamma_{2} \left(\frac{2 \beta_- \beta_3 \mu_1 - \gamma_2^\ast \alpha_3 }{1- \beta_3^2}\right).
\end{equation}

\item When $\beta_0  \notin \{1,3,-1,-3 \},$ the  Hamiltonian  (\ref{H-2-D-gen-A-gen}) becomes

\begin{equation}
   \hat{H}_{(\beta_0,\vec{\beta},\vec{\gamma_1}, \vec{\gamma_2})} =  \beta_0 \hat{N} + \vec{\beta} \cdot \vec{\hat{J}} -  \frac{  2 \gamma_2 \beta_-}{1+\beta_3}  \hat{a}_1^\dagger - \frac{  2 \gamma_2^\ast \beta_+}{1+\beta_3}  \hat{a}_1 + \gamma_2 \hat{a}_2^\dagger + \gamma_2^\ast  \hat{a}_2 +   h_0 \hat{I} 
   \end{equation}
and the lowering operator  (\ref{A-the-most-general} )  reduces to

\begin{equation} \hat{A}_{(\vec{\mu},\vec{0},\vec{\alpha},0)}  =  \frac{ 2 \alpha_3 \gamma_2^\ast (3 -\beta_0)}{(1-\beta_3^2)\left( (2 - \beta_0)^2 -1  \right)     }  \left( 2 \beta_+  \hat{a}_1  +  (1 - \beta_3) \hat{a}_2 \right) + \alpha_3 \left(  \frac{\beta_-}{1+\beta_3}   \hat{J}_+   - \frac{\beta_+}{1-\beta_3}   \hat{J}_-  + \hat{J}_3 \right).
\end{equation}

 \end{itemize}

\subsection{Case $b^2=1,$    $\frac{\gamma_1}{2} \neq  \frac{\gamma_2 \beta_-}{1-\beta_3}$ and  $\frac{\gamma_1}{2} \neq - \frac{\gamma_2 \beta_-}{1+\beta_3}$}
\label{appb-6}
\begin{itemize}

\item When $\beta_0 =3,$ the  Hamiltonian  (\ref{H-2-D-gen-A-gen}) becomes

\begin{equation}
   \hat{H}_{(3,\vec{\beta},\vec{\gamma_1}, \vec{\gamma_2})} =  3  \hat{N} + \vec{\beta} \cdot \vec{\hat{J}} +   \gamma_1 
   \hat{a}_1^\dagger + \gamma_1^\ast \hat{a}_1  + \gamma_2 \hat{a}_2^\dagger +  \gamma_2^\ast \hat{a}_2 +   h_0 \hat{I}  
   \label{H-2-D-gen-3-gamma-all-diff}
\end{equation}
and the lowering operator  (\ref{A-the-most-general} )  reduces to

\begin{eqnarray} \hat{A}_{(\vec{\mu},\vec{\nu},\vec{\alpha},a_0)} & =&   \left[ \mu_1  \hat{a}_1  + \left( \frac{\alpha_3}{\beta_+}  \left(
\frac{\gamma_1^\ast}{2} - \frac{ \gamma_2^\ast \beta_+}{1-\beta_3}  
\right)    -   \frac{2 \mu_1 \beta_- }{1-\beta_3}  \right)\hat{a}_2 \right] +
 \frac{\alpha_3}{2} \left(
\frac{\gamma_2}{2} + \frac{\gamma_1 \beta_+}{1-\beta_3} \right)  \left[ \frac{2 \beta_-}{1+\beta_3}  \hat{a}_1^\dagger - \hat{a}_2^\dagger   \right]
\nonumber \\ &+&  \alpha_3 \left(  \frac{\beta_-}{1+\beta_3}   \hat{J}_+   - \frac{\beta_+}{1-\beta_3}   \hat{J}_-  + \hat{J}_3 \right) \nonumber \\ &+&
2 \mu_1   \left(
\frac{\gamma_1}{2} - \frac{ \gamma_2  \beta_-}{1-\beta_3}  
\right) + \frac{\gamma_2 \alpha_3}{\beta_+}   \left(
\frac{\gamma_1^\ast}{2} - \frac{ \gamma_2^\ast \beta_+}{1-\beta_3} 
  \right) + \alpha_3   \left(
\frac{\gamma_2}{2} + \frac{ \gamma_1 \beta_+}{1-\beta_3}  
\right)  \left(
\frac{\gamma_2^\ast}{2} - \frac{ \gamma_1^\ast \beta_-}{1+\beta_3}  
\right). \label{A-2-D-gen-3-gamma-all-diff}
  \end{eqnarray}

\item When $\beta_0 =-3,$ the  Hamiltonian  (\ref{H-2-D-gen-A-gen}) becomes

\begin{equation}
      \hat{H}_{(-3,\vec{\beta},\vec{\gamma_1}, \vec{\gamma_2})} = - 3  \hat{N} + \vec{\beta} \cdot \vec{\hat{J}} +   \gamma_1 
   \hat{a}_1^\dagger + \gamma_1^\ast \hat{a}_1  + \gamma_2 \hat{a}_2^\dagger +  \gamma_2^\ast \hat{a}_2 +   h_0 \hat{I} 
\end{equation}
and the lowering operator  (\ref{A-the-most-general} )  reduces to

\begin{eqnarray} \hat{A}_{(\vec{\mu},\vec{\nu},\vec{\alpha},a_0)} & =& 
 \frac{\alpha_3}{2} \left(
\frac{\gamma_2^\ast}{2} - \frac{\gamma_1^\ast \beta_-}{1+\beta_3} \right)  \left[ \frac{2 \beta_+}{1- \beta_3}  \hat{a}_1 + \hat{a}_2   \right]  +  \left[ \nu_1  \hat{a}_1^\dagger  +  \left( \frac{\alpha_3}{\beta_-}  \left(
\frac{\gamma_1}{2} +  \frac{ \gamma_2  \beta_-}{1+\beta_3}  
\right)    +   \frac{2 \nu_1 \beta_+}{1+\beta_3}  \right)\hat{a}_2^\dagger \right] 
\nonumber \\ &+&  \alpha_3 \left(  \frac{\beta_-}{1+\beta_3}   \hat{J}_+   - \frac{\beta_+}{1-\beta_3}   \hat{J}_-  + \hat{J}_3 \right) \nonumber \\ &-&
2 \nu_1   \left(
\frac{\gamma_1^\ast}{2} + \frac{ \gamma_2^\ast  \beta_+}{1+\beta_3}  
\right) - \frac{\gamma_2^\ast \alpha_3}{\beta_-}   \left(
\frac{\gamma_1}{2} + \frac{ \gamma_2  \beta_-}{1+\beta_3} 
  \right) + \alpha_3   \left(
\frac{\gamma_2^\ast}{2} - \frac{ \gamma_1^\ast \beta_-}{1+\beta_3}  
\right)  \left(
\frac{\gamma_2}{2} + \frac{ \gamma_1  \beta_+}{1-\beta_3}  
\right).
  \end{eqnarray}

\item When $\beta_0  \notin \{1,3,-1,-3 \},$ the  Hamiltonian  (\ref{H-2-D-gen-A-gen}) becomes

\begin{equation}
     \hat{H}_{(\beta_0,\vec{\beta},\vec{\gamma_1}, \vec{\gamma_2})} =  \beta_0  \hat{N} + \vec{\beta} \cdot \vec{\hat{J}} +   \gamma_1 
   \hat{a}_1^\dagger + \gamma_1^\ast \hat{a}_1  + \gamma_2 \hat{a}_2^\dagger +  \gamma_2^\ast \hat{a}_2 +   h_0 \hat{I} \label{gen-gen-hamiltonian-beta-0-diff}
\end{equation}
and the lowering operator  (\ref{A-the-most-general} )  reduces to

\begin{eqnarray} \hat{A}_{(\vec{\mu},\vec{\nu},\vec{\alpha},a_0)}  &=&  \frac{2  \alpha_3}{ 1-\beta_0}
\left( \frac{\gamma_2^\ast}{2}  - \frac{\gamma_1^\ast \beta_-}{1+\beta_3} \right) \left( \frac{2 \beta_+}{1-\beta_3}  \hat{a}_1 + \hat{a}_2     \right)   \nonumber \\&+  &\frac{2  \alpha_3}{1+\beta_0}
\left( \frac{\gamma_2}{2}  +  \frac{\gamma_1 \beta_+}{1-\beta_3} \right) \left( \frac{2 \beta_-}{1+\beta_3}  \hat{a}_1^\dagger - \hat{a}_2^\dagger     \right) \nonumber \\&+&   \alpha_3 \left(  \frac{\beta_-}{1+\beta_3}   \hat{J}_+   - \frac{\beta_+}{1-\beta_3}   \hat{J}_-  + \hat{J}_3 \right) \nonumber \\ &+& \frac{8 \alpha_3}{1-\beta_0^2} \left( \frac{\gamma_2^\ast}{2}  - \frac{\gamma_1^\ast \beta_-}{1+\beta_3} \right)  \left( \frac{\gamma_2}{2}  +  \frac{\gamma_1 \beta_+}{1-\beta_3} \right). \label{gen-gen-lowering-beta-0-diff}
\end{eqnarray}

 \end{itemize}

\end{document}